\documentclass[aps, prd, twocolumn, superscriptaddress, 10pt]{revtex4-2}

\usepackage[utf8]{inputenc}

\usepackage{mathtools}
\usepackage{amsfonts}
\usepackage{mathrsfs}
\usepackage{bbm}
\usepackage{slashed}

\usepackage{color}
\usepackage{array}
\usepackage{graphicx}

\usepackage{caption}
\usepackage{subcaption}

\usepackage{float}
\usepackage{placeins}
\usepackage{booktabs}
\usepackage{makecell}
\usepackage{adjustbox}
\usepackage{tabularx}

\usepackage{xspace}
\usepackage{siunitx}
\usepackage{hyperref}
\usepackage[nameinlink]{cleveref}

\usepackage{xifthen}
\usepackage{xcolor}
\hypersetup{
	colorlinks,
	linkcolor={red!75!black},
	citecolor={blue!75!black},
	urlcolor={blue!75!black}
}

\usepackage{etoolbox}
\apptocmd{\sloppy}{\hbadness 10000\relax}{}{}

\usepackage{silence}
\graphicspath{{./figures/}}

\newcommand{\gettitle}{On Gauge Consistency In Gauge-Fixed Yang-Mills Theory}

\newcommand{\getHeidelbergAffiliation}{\affiliation{Institut f\"ur Theoretische Physik, Universit\"at Heidelberg, Philosophenweg 16, 69120 Heidelberg, Germany}}
\newcommand{\getDarmstadtAffiliation}{\affiliation{Institut f\"ur Kernphysik, Technische Universit\"at Darmstadt, Schlossgartenstra{\ss}e 2, 64289 Darmstadt, Germany}}
\newcommand{\getEMMIAffiliation}{\affiliation{ExtreMe Matter Institute EMMI, GSI, Planckstr. 1, 64291 Darmstadt, Germany}}

\hypersetup{
	pdftitle={\gettitle},
	pdfauthor={Pawlowski, Schneider, Wink},
	pdfkeywords={Yang-Mills theory}
		{correlations functions} {confinement} 
		{functional renormalisation group}
		{gauge symmetry},
	bookmarksopen=true,
	bookmarksopenlevel=2,
	bookmarksnumbered=true
}

\begin{document}

\title{\gettitle}

\author{Jan M. Pawlowski}
\getHeidelbergAffiliation
\getEMMIAffiliation

\author{Coralie S. Schneider}
\getHeidelbergAffiliation

\author{Nicolas Wink}
\getHeidelbergAffiliation
\getDarmstadtAffiliation

\begin{abstract}
We investigate BRST invariance in Landau gauge Yang-Mills theory with functional methods. To that end, we solve the coupled system of functional renormalisation group equations for the  momentum-dependent ghost and gluon propagator, ghost-gluon, and three- and four-gluon vertex dressings. The equations for both, transverse and longitudinal correlation functions are solved self-consistently: all correlation functions are fed back into the loops. Additionally, we also use the Slavnov-Taylor identities for computing the longitudinal correlation functions on the basis of the above results. Then, the gauge consistency of the solutions is checked by comparing the respective longitudinal correlation functions. We find good agreement of these results, hinting at the gauge consistency of our setup. 
	\vspace*{35pt}
\end{abstract}

\maketitle

%%%%%%%%%%%%%%%%%%%%%%%%%%
\section{Introduction}%%%%
%%%%%%%%%%%%%%%%%%%%%%%%%%
\label{sec:YM:Introduction}%%
%%%%%%%%%%%%%%%%%%%%%%%%%%
%%%%%%%%%%%%%%%%%%%%%%%%%%
Infrared QCD is a strongly correlated system that is governed by confinement and spontaneous chiral symmetry breaking, whose understanding and resolution require numerical non-perturbative first principle approaches. Functional diagrammatic approaches such as the functional renormalisation group (fRG) and Dyson-Schwinger equations (DSE), or $n$-particle irreducible approaches (nPI) potentially offer both, analytic access to the mechanisms behind the infrared dynamics of QCD as well as its quantitative numerical resolution. For recent reviews on fRG and DSE approaches to QCD see~\cite{Fischer:2018sdj, Dupuis:2020fhh}.

The diagrammatic nature of functional approaches is best implemented within gauge-fixed QCD, for a recent discussion of gauge invariant alternatives see~\cite{Dupuis:2020fhh} and references therein. A specifically well-suited gauge fixing is the Landau gauge, in particular for numerical applications. The latter require truncations to the infinite hierarchy of coupled loop equations for correlation functions in functional approaches. Then, the strongly correlated nature of infrared QCD begs the question of whether the truncations commonly used for explicit numerical solutions transport the underlying gauge symmetry of QCD: are the correlation functions computed \textit{gauge consistent}. Naturally, the gauge consistency of the correlation functions is essential for confinement, both its manifestation in gauge-fixed approaches as well as its understanding.

In the present work, we put forward a systematic approach towards the evaluation of the above question: First, one computes transverse and longitudinal correlation functions in QCD with functional approaches. Then, the gauge consistency of the results is tested by inserting them into the Slavnov-Taylor identities (STIs). The STIs encode the changes of gauge-fixed correlation functions under gauge or BRST-transformations. 

The evaluation of the latter test comes with an intricacy. The STIs also constitute a set of functional relations for correlation functions. These relations can be derived from the BRST identity of the effective action, the BRST master equation, similarly to deriving functional relations from the fRG-flow of the effective action or the functional DSE for the latter. In the case of the STIs, longitudinal correlation functions are related to combinations of transverse and longitudinal correlation functions. In summary, fRG equations, DSEs, and STIs represent different resummation schemes for correlation functions. While their exact solutions agree, any approximations thereof will not coincide in general. To exemplify this situation, let us emphasise that one can easily guarantee the STIs for a finite set of correlation functions in a given order of the truncation if simply using the STIs for the computation of the longitudinal correlation functions. Evidently, this resolution of the STIs does not guarantee the gauge consistency of the transverse correlation functions. This is best seen in the Landau gauge, where the set of functional relations for transverse correlation functions is decoupled from the system of longitudinal ones. Instead, gauge consistency should rather be evaluated by the \textit{smallness} of the deviations of correlation functions within \textit{all} functional relations at hand: fRG flows, DSEs, STIs and $n$PI-relations. For a detailed discussion of these subtleties, see \cite{Dupuis:2020fhh, Fischer:2008uz}.

In the present work, we study Landau gauge Yang-Mills theory within a systematic vertex expansion with the functional renormalisation group. Longitudinal correlation functions are also computed from their respective Slavnov-Taylor identities, which follow from the BRST master equation for the effective action. This task is facilitated with the code framework \textit{QMeS}-Derivation \cite{Pawlowski:2021tkk, github:QMeS}, that allows for the derivation of fRG and DSE equations as well as (m)STIs. We use an advanced truncation for these computations, even though fully quantitative computations are still more advanced, see in particular \cite{Cyrol:2016tym, Huber:2020keu}. The respective results are also compared with the corresponding lattice results. In \Cref{sec:YM:NumericalResults} we discuss the consequences for the gauge consistency of the present results as well as extensions of the current computations. \Cref{sec:YM:Conclusion} contains a brief conclusion to the present work. The Appendices \Cref{app:YM:Implementation}, \Cref{app:YM:FRGYM} and \Cref{app:YM:mSTI} contain some technical details of the computation and the diagrammatic functional equations within the truncation used throughout this work are shown in \Cref{app:YM:Diags}.

%%%%%%%%%%%%%%%%%%%%%%%%%%
\section{Slavnov-Taylor Identities in Functional Approaches}%%%%
\label{sec:YM:Flows}%%
%%%%%%%%%%%%%%%%%%%%%%%%%%

We consider $SU(N_c)$ Yang-Mills theory in Euclidean space-time within the Landau gauge. The explicit results are obtained in the physical (QCD) case $N_c=3$, but they readily extend to general $N_c$. The approximation used here carries the large $N_c$ scaling which is present in the dominant tensor structure, for a respective discussion see \cite{Corell:2018yil}. The gauge-fixed classical action, including the ghost term and BRST source terms, is given by 
\begin{align}
S= &\,
\int_x \Biggl[\frac{1}{4}F^a_{\mu\nu}F^a_{\mu\nu}+\partial_\mu \bar{c}^a D^{ab}_\mu c^b+B^a\partial_\mu A_\mu^a-\frac{\xi}{2}B^aB^a\nonumber\\[4pt]
&\hspace{-.2cm}-Q_{A,\mu}^a D^{ab}_\mu c^b -\frac{1}{2}g f^{abc}Q_c^ac^bc^c + 
	Q_{\bar{c}}^a B^a  \Biggr]\,, 
\label{eq:Sclass}\end{align}
where we have introduced the shorthand notation $\int_x = \int d^4 x$. The covariant derivative $D$ in the adjoint representation and the field strength tensor are defined as,
\begin{align}\nonumber 
D^{ab}_\mu c^b  = &\, \partial_\mu c^a + g f^{acb}A^c_\mu c^b\,,\\[1ex]
F_{\mu\nu}^a = &\, \partial_\mu A_\nu^a-\partial_\nu A_\mu^a + g f^{abc}A_\mu^bA_\nu^c\,, 
\end{align}
and the Landau gauge is implemented with $\xi\rightarrow 0$. The action \labelcref{eq:Sclass} depends on the superfield $\Phi$, 
\begin{align}\label{eq:Phi}
	\Phi=(A_\mu, c, \bar c, B)\,,\qquad  Q =(Q_A, Q_c,Q_{\bar c})\,, 
\end{align}
and we have already introduced source terms the BRST transformations (Becchi, Rouet, Stora, Tyutin)~\cite{Becchi:1975nq,  Tyutin:1975qk} of the fields  $\delta_{\text{BRST}}\Phi_a = \mathfrak{s}\Phi_a\delta \lambda$ in a general covariant gauge, coupled to the currents $Q_{\Phi_a}$. The respective transformations are given by 
\begin{align}
\mathfrak{s}A_\mu^a &= D^{ab}_\mu c^b\,,\nonumber\\[10pt]
\mathfrak{s}c^a &= \frac{1}{2} g f^{abc}c^bc^c\,,\nonumber\\[10pt]
\mathfrak{s}\bar{c}^a &= B^a\,,\nonumber\\[10pt]
\mathfrak{s} B^a &= 0\,,
\label{eq:YM:YMBRSTtrafo}
\end{align}
with the Nakanishi-Lautrup field $B^a$. Evidently, we do not need $Q_B$, as the respective BRST transformation vanishes identically. Moreover, with $\mathfrak{s}^2=0$ these additional terms are also BRST invariant.  Our general conventions and notation follows~\cite{Pawlowski:2021tkk}.

%%%%%%%%%%%%%%%%%%%%%%%%%%%
\subsection{Functional Renormalisation Group}
 
In the functional renormalisation group approach, an infrared cutoff is introduced by adding momentum-dependent mass terms $\Delta S_k$ to the classical action $S[\Phi,Q]$ defined in \labelcref{eq:Sclass}, 
\begin{align}
\Delta S_k = \int \left(\frac{1}{2}(R_A)^{ab}_{\mu\nu}\, A^a_\mu A^b_\nu + (R_c)^{ab}\,\bar{c}^ac^b\right)\,,
\label{eq:YM:cutoff}
\end{align}
with the infrared cutoff scale $k$. The regulators $R=(R_A,R_c)$ suppress quantum fluctuations for momenta $p^2 \lesssim k^2$ for the respective fields, and vanish for $p^2 \gtrsim k^2$. The specific form of the regulators used in the present work can be found in \Cref{app:YM:reg}. The infrared regularised classical action $S+\Delta S_k$ is then used in the path integral representation of the generating functional $Z[J,Q]$. Here, the super-current $J$ includes currents for all component fields $J=(J_A, J_c, J_{\bar c}, J_B)$. 

The functional flow equation is derived for the scale-dependent effective action  
\begin{align}\label{eq:G}
\Gamma_k[\Phi,Q]= \int J\cdot \Phi -\log Z[J,Q] - \Delta S_k[\Phi]\,, 
\end{align}
the modified Legendre transform of the Schwinger functional $\log Z$. Its logarithmic scale derivative provides us with the one-loop exact  fRG equation,
\begin{align}
\partial_t \Gamma_k = \int_x\left[\frac{1}{2}(\dot{R}_A)^{ab}_{\mu\nu}\,(G_{AA})_{\mu\nu}^{ab}-(\dot{R}_c)^{ab} (G_{c\bar{c}})^{ab}\right](x,x)\,,
\label{eq:YM:fRG}
\end{align}
with $\dot R=\partial_t R$,  the (negative) RG time $t = \ln(k/\Lambda)$, and the full ghost and gluon propagators 
\begin{align}
	G_{AA} = \left[\frac{1}{\Gamma^{(2)}+R}\right]_{AA} \,,\qquad 
	G_{c\bar c}= \left[ \frac{1}{\Gamma^{(2)}+R} \right]_{c\bar c}\,. 
	\label{eq:YM:propeq}
\end{align}
Here, $\Lambda$ is a reference scale, typically chosen to be the initial cutoff scale deep in the ultraviolet (UV). If the initial scale is chosen sufficiently large, the effective action tends towards the local UV effective action $\Gamma_\Lambda$ that consists out of all UV-relevant terms. For Yang-Mills theory this UV effective action includes all terms in the classical action as well as a mass term for the gluon. The latter term originates in the breaking of gauge invariance via the regulator term for $k\neq 0$. In turn, for $k\to 0$, this breaking disappears and we are left with the full BRST invariant effective action. For a full derivation and further discussions, see e.g.~\cite{Pawlowski:2021tkk, LectureNotesprep} and the QCD-related reviews \cite{Litim:1998nf, Berges:2000ew, Pawlowski:2005xe, Gies:2006wv, Rosten:2010vm, Braun:2011pp,  Dupuis:2020fhh}.

At this point we would like to remark on some implicit assumptions within the derivation of the flow equation \labelcref{eq:YM:fRG}: it implies that the only $k$-dependence of the \textit{renormalised} generating functional originates in the cutoff term. Then, in particular, the renormalisation procedure is assumed to be $k$-independent, for a detailed discussion see e.g.~\cite{Pawlowski:2005xe, Rosten:2010vm}. While the self-consistency of this assumptions is readily shown for the wave function renormalisations and vertices, the absence of a mass renormalisation for the gluon is less trivial and potentially affects the underlying BRST invariance. This question is of utmost importance for the interpretation of Yang-Mills theory in the presence of the infrared cutoff term as a deformation of Yang-Mills theory, rather than a massive extension of Yang-Mills theory, for the Curci-Ferrari (CF) model see~e.g.~\cite{Tissier:2010ts, Tissier:2011ey, Serreau:2012cg, Reinosa:2017qtf, Pelaez:2021tpq}. In the CF model, masses for ghost and gluons are part of the gauge fixing. In contrast, one may also simply add a gluonic mass term after the gauge fixing, and the two procedures agree in the Landau gauge. Note also, that the latter procedure is similar to adding the regulator term to the gauge-fixed action in the fRG approach, a subtle difference being the mass renormalisation and the related classical BRST symmetry. 

Both procedures have to be contrasted to massive extensions of Yang-Mills theory, where the mass is added to the classical action without a gauge fixing, bound to introduce non-localities.

We close with the remark that the massless limit of massive extensions of Yang-Mills theory is an intricate one as it defines a flow in theory space. Instead, the removal of the momentum-local infrared regulator in Yang-Mills theory is smooth, as the cutoff term can be interpreted as the local deformation of Yang-Mills theory. Still, even for local deformations the correct description of the massless limit is intricate as we shall see later, see in particular \Cref{sec:YM:FunRel}.

\subsection{STI \& mSTI}\label{sec:YM:STI+mSTI}

Classical BRST symmetry with the infinitesimal transformations \labelcref{eq:YM:YMBRSTtrafo} leads to symmetry identities on the quantum level, that can be formulated in terms of a master equation, see e.g.~\cite{Zinn-Justin:1974ggz, Zinn-Justin:1999euj},
\begin{align}
\int_x \left( \frac{\delta \Gamma}{\delta Q_{A,\mu}^a} \frac{\delta \Gamma}{\delta A_\mu^a} +\frac{\delta \Gamma}{\delta Q_c^a} \frac{\delta \Gamma}{\delta c^a}+\frac{\delta \Gamma}{\delta Q_{\bar{c}}^a} \frac{\delta \Gamma}{\delta \bar{c}^a}\right)= 0\,, 
\label{eq:YM:STI}
\end{align}
and integrating out the (Gau{\ss}ian) Nakanishi-Lautrup field $B$ converts \Cref{eq:YM:STI} into 
\begin{align}
\int_x \left( \frac{\delta \Gamma}{\delta Q_{A,\mu}^a} \frac{\delta \Gamma}{\delta A_\mu^a} +\frac{\delta \Gamma}{\delta Q_c^a} \frac{\delta \Gamma}{\delta c^a}-\frac{1}{\xi} (\partial_\nu A^a_\nu)\partial_\mu \frac{\delta \Gamma}{\delta Q_{A,\mu}^a}\right)= 0\,.
\label{eq:YM:STIwo}
\end{align}
In the derivation of \labelcref{eq:YM:STIwo} one also uses the fact that the anti-ghost field only appears linearly in the action. 

Finally, the introduction of the cutoff term \labelcref{eq:YM:cutoff} leads to a modification of the symmetry identities \labelcref{eq:YM:STI} or \labelcref{eq:YM:STIwo}, the \textit{modified Slavnov-Taylor identity} (mSTI). It is derived similarly to the flow equation \labelcref{eq:YM:fRG}: there the right hand side can be interpreted as the equation which governs the violation of scale invariance. For the mSTI, the additional term in comparison to \labelcref{eq:YM:STI} or \labelcref{eq:YM:STIwo} originates from the lack of BRST invariance of the regulator term. As for the flow equation, it is a  one-loop exact equation. In the presence of the Nakanishi-Lautrup field $B$, the mSTI takes the concise form, 
\begin{align}
&\int_x  \Gamma_{Q_i}  \Gamma_{\Phi_i} =\int_{x,y}\, \left( R\, G\right)_{\Phi_i \Phi_j}\, \Gamma_{Q_{j} \Phi_i}
\label{eq:YM:mSTI}
\end{align}
with the notation  
\begin{align}
	\Gamma_{Q_1\cdots Q_n\Phi_{n+1}\cdots \Phi_{n+m}} 
	= \frac{ \delta\Gamma[\Phi,Q]}{ \delta {Q_1}\cdots \delta {Q_n}  \delta\Phi_{n+1} \cdots \delta{\Phi_{n+m}}}\,,
	\end{align}
for mixed derivatives of the effective action with respect to BRST currents and fields. The integral $\int_{x,y}$ on the right-hand side of \labelcref{eq:YM:mSTI} constitutes a space-time trace of the product of operators $(R\,G)(x,y)$ and $\Gamma_{Q \Phi}(y,x)$. On both sides of \labelcref{eq:YM:mSTI}, a sum over species of fields as well as internal and Lorentz indices is implied. For example, the left-hand side of \labelcref{eq:YM:mSTI} is simply that of \labelcref{eq:YM:STI}, if re-instating all indices. 

As already mentioned above, the right-hand side of \labelcref{eq:YM:mSTI} originates in the breaking of BRST invariance induced by the regulator term, similarly to the right-hand side of the flow equation, which manifests the breaking of scale invariance induced by  the regulator term. We emphasise, that \labelcref{eq:YM:mSTI} is derived with the same implicit assumption of the independence of the UV renormalisation procedure of the cutoff term. We will discuss the self-consistency of this assumption in \Cref{sec:YM:FunRel}. 

For $k\rightarrow 0$ the mSTI reduces to the STI. Thus, satisfying the mSTI at all scales $k$, guarantees gauge invariance of observables at $k=0$. Modified Slavov-Taylor identities, Ward identities and Nielsen identities have been studied intensively in the fRG approach as they are key to quantitatively reliable approximations in gauge theories. For a partial use for momentum-dependent correlation functions in QCD see \cite{Cyrol:2016tym, Cyrol:2017ewj}, for a full derivation and further discussions in the present context, see e.g.~\cite{Pawlowski:2021tkk, LectureNotesprep}. For generic fRG literature on modified symmetry identities see e.g.\ \cite{Bonini:1993sj,  Ellwanger:1994iz, Becchi:1996an, DAttanasio:1996tzp, Reuter:1997gx, Freire:2000bq, Igarashi:1999rm, Igarashi:2001ey, Pawlowski:2003sk, Pawlowski:2005xe, Igarashi:2009tj, Donkin:2012ud, Lavrov:2012xz, Sonoda:2013dwa, Safari:2015dva, Cyrol:2016tym, Safari:2016gtj, Igarashi:2016gcf, Cyrol:2017ewj, Asnafi:2018pre, Igarashi:2019gkm, Barra:2019rhz, Lavrov:2020exa, Pawlowski:2020qer}, a complete list of references can be found in the recent review see \cite{Dupuis:2020fhh}.

\begin{figure*}[t]
	\centering
	\begin{subfigure}[t]{0.45\textwidth}
		\centering
		\includegraphics[width=\linewidth]{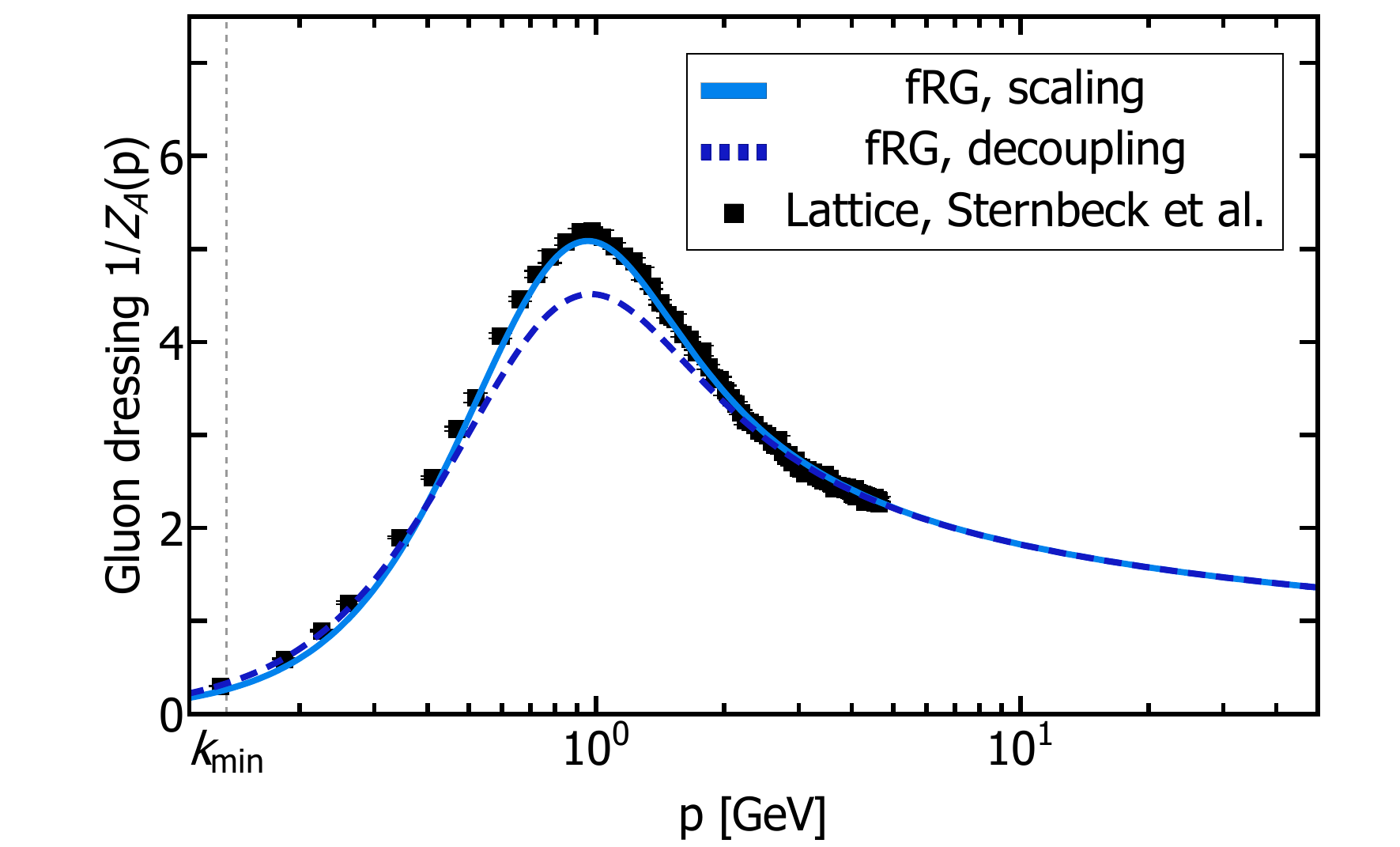}
		\caption{Gluon dressing $1/Z_A(p)$.}
		\label{fig:gluon_dressing}
	\end{subfigure}%
	\hspace{0.05\textwidth}
	\begin{subfigure}[t]{0.45\textwidth}
		\centering
		\includegraphics[width=\linewidth]{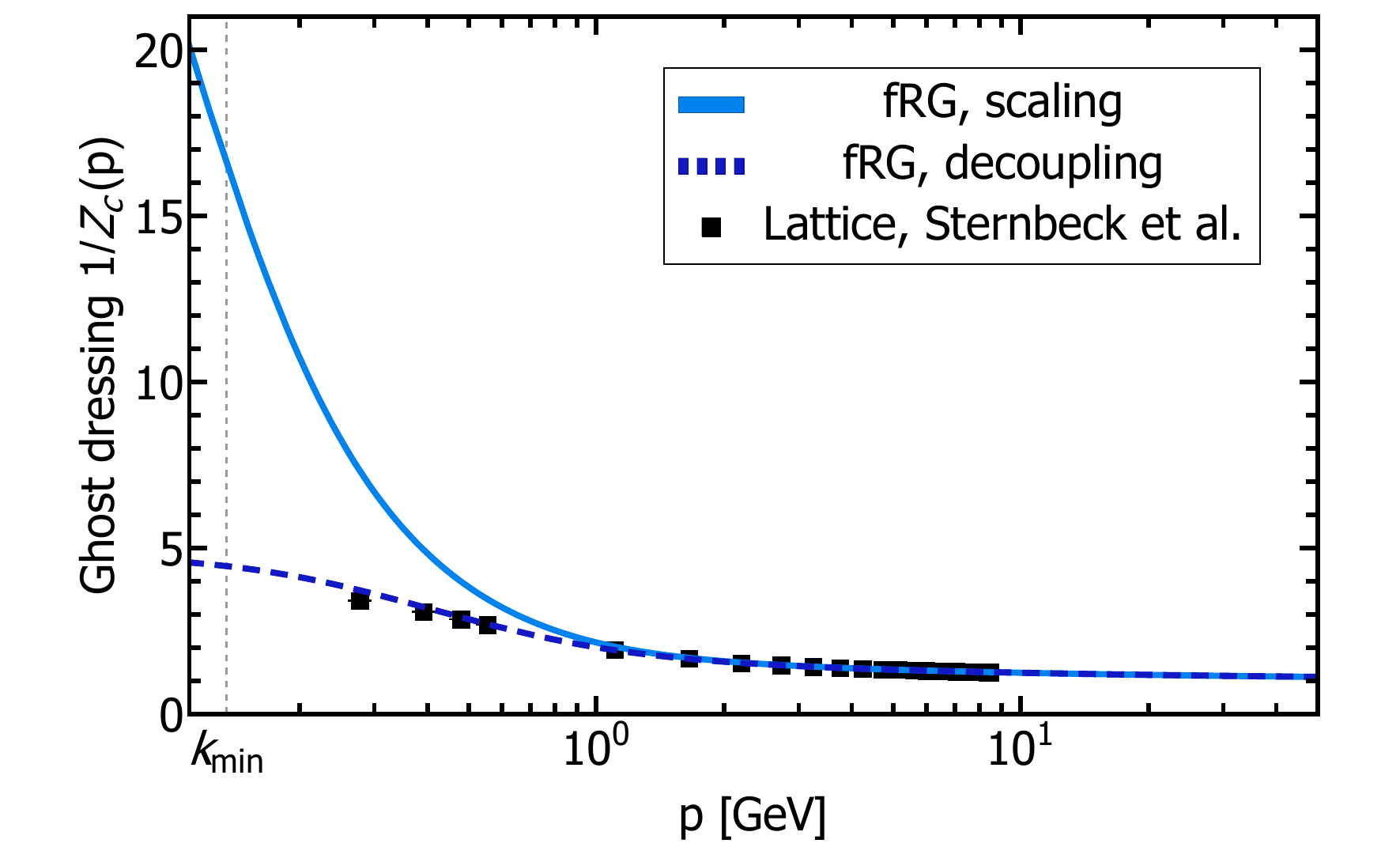}
		\caption{Ghost dressing $1/Z_c(p)$.}
		\label{fig:ghost_dressing}
	\end{subfigure}
	\caption{Propagator dressings form the fRG for scaling (solid) and decoupling (dashed) solutions as a function of momentum $p$ in comparison to the lattice results (squares) of~\cite{Sternbeck:2006cg}. The global normalisation of the lattice results was fixed by our scaling solution, c.f.~\Cref{app:YM:scalesetting}.\hspace*{\fill}}
	\label{fig:YM:propdressings}
\end{figure*}
%
%%%%%%%%%%%%%%%%%%%%%%%%%%%%%%%%
\subsection{Vertex Expansion and Truncations}
\label{sec:YM:VertexExpansion}
We expand the effective action in terms of $\Phi$ and $Q$ vertices. At vanishing BRST current, $Q=0$, this entails, 
\begin{align}
\Gamma_k[\Phi,0] = \sum_{n=1}^{\infty}\int \frac{1}{n!}\Gamma_{\Phi_1\cdots\Phi_n}\Phi_n\dots\Phi_1\,,
\label{eq:YM:vertexexpansion}
\end{align}
with the superfield $\Phi$ introduced \labelcref{eq:Phi}, the 1PI correlation functions $\Gamma_{\Phi_1\cdots\Phi_n}$, and the normalisation $\Gamma_k[0,0]=0$. The prefactor $1/n!$ is a vector-factorial, where each component corresponds to the factorial of the number of fields of the same species in the summand. Including BRST and mixed vertices is done analogously. Inserting the vertex expansion \labelcref{eq:YM:vertexexpansion} into the fRG equation \labelcref{eq:YM:fRG}, one sees readily, that a full solution of the theory requires the complete set of 1PI correlation functions. Specifically, the flow of an $n$-point correlation function $\Gamma_{\Phi_1\cdots\Phi_n}$ depends on correlation functions $\Gamma_{\Phi_1\cdots\Phi_m}$ with $2\leq m\leq n+2$. This leads to an infinite tower of coupled integral-differential equations. For most practical purposes this tower has to be truncated. Similar considerations apply to all closed functional master equations and in particular to the set of DSEs, whose towers of coupled integral equations also satisfy $2\leq m\leq n+2$. For a complete survey on results for Landau gauge correlation functions we refer to the reviews \cite{Dupuis:2020fhh, Alkofer:2000wg, Fischer:2006ub, Binosi:2009qm, Maas:2011se, Boucaud:2011ug, Huber:2018ned, 
Pelaez:2021tpq}. 

\begin{figure*}
	\centering
	\begin{minipage}[t]{0.45\textwidth}
		\begin{figure}[H]
			\includegraphics[width=\linewidth]{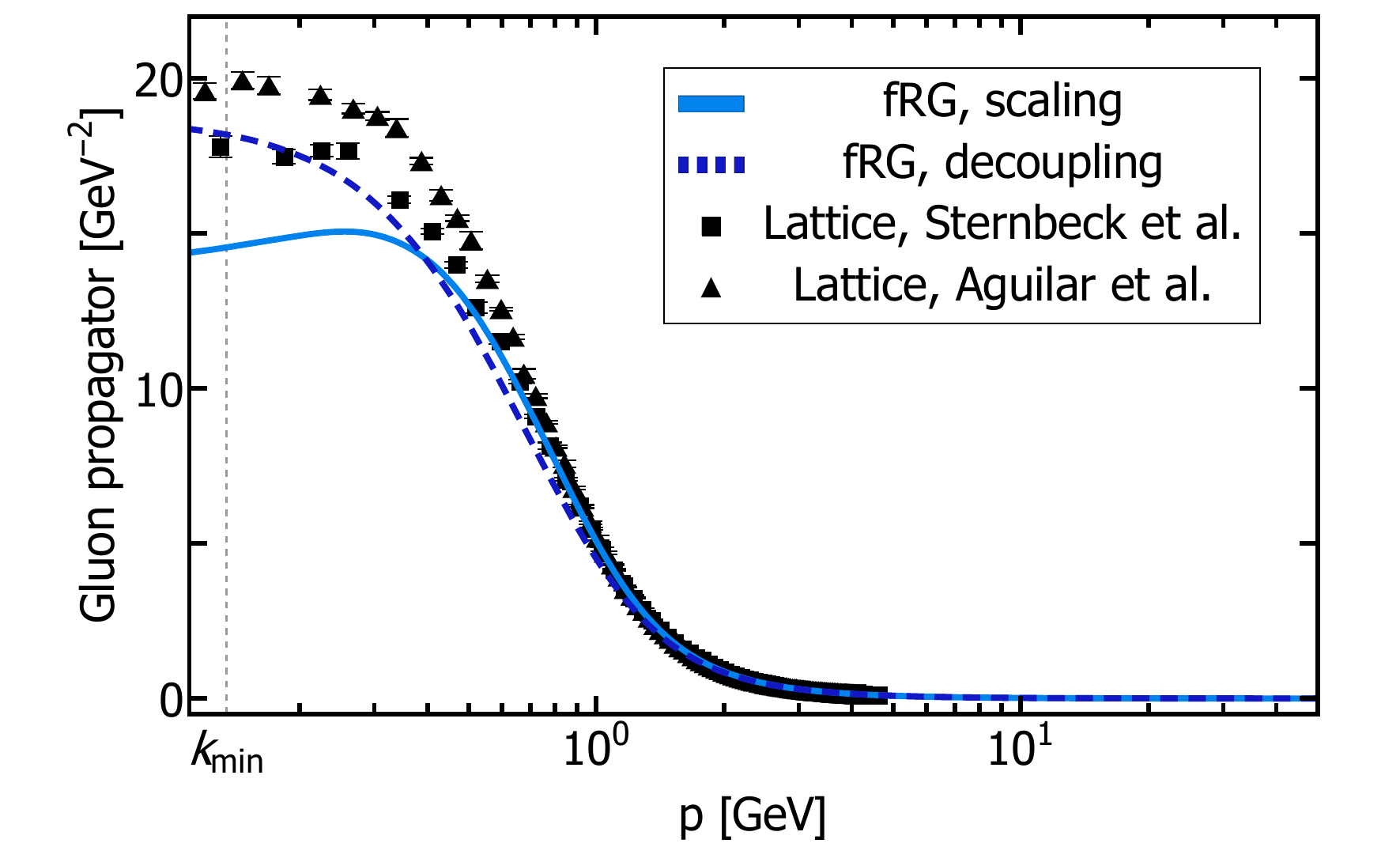}
			\caption{Scaling (solid) and decoupling (dashed) gluon propagator as a function of momentum $p$ in comparison to the lattice data (squares) from~\cite{Sternbeck:2006cg} and (triangles) from~\cite{Aguilar:2021okw}. The global normalisation of the lattice data was fixed by our scaling solution, c.f.~\Cref{app:YM:scalesetting}.\hspace*{\fill}}
			\label{fig:gluon_prop}
		\end{figure}
	\end{minipage}
	\hspace{0.05\textwidth}
	\begin{minipage}[t]{0.45\textwidth}
		\begin{figure}[H]
			\includegraphics[width=\linewidth]{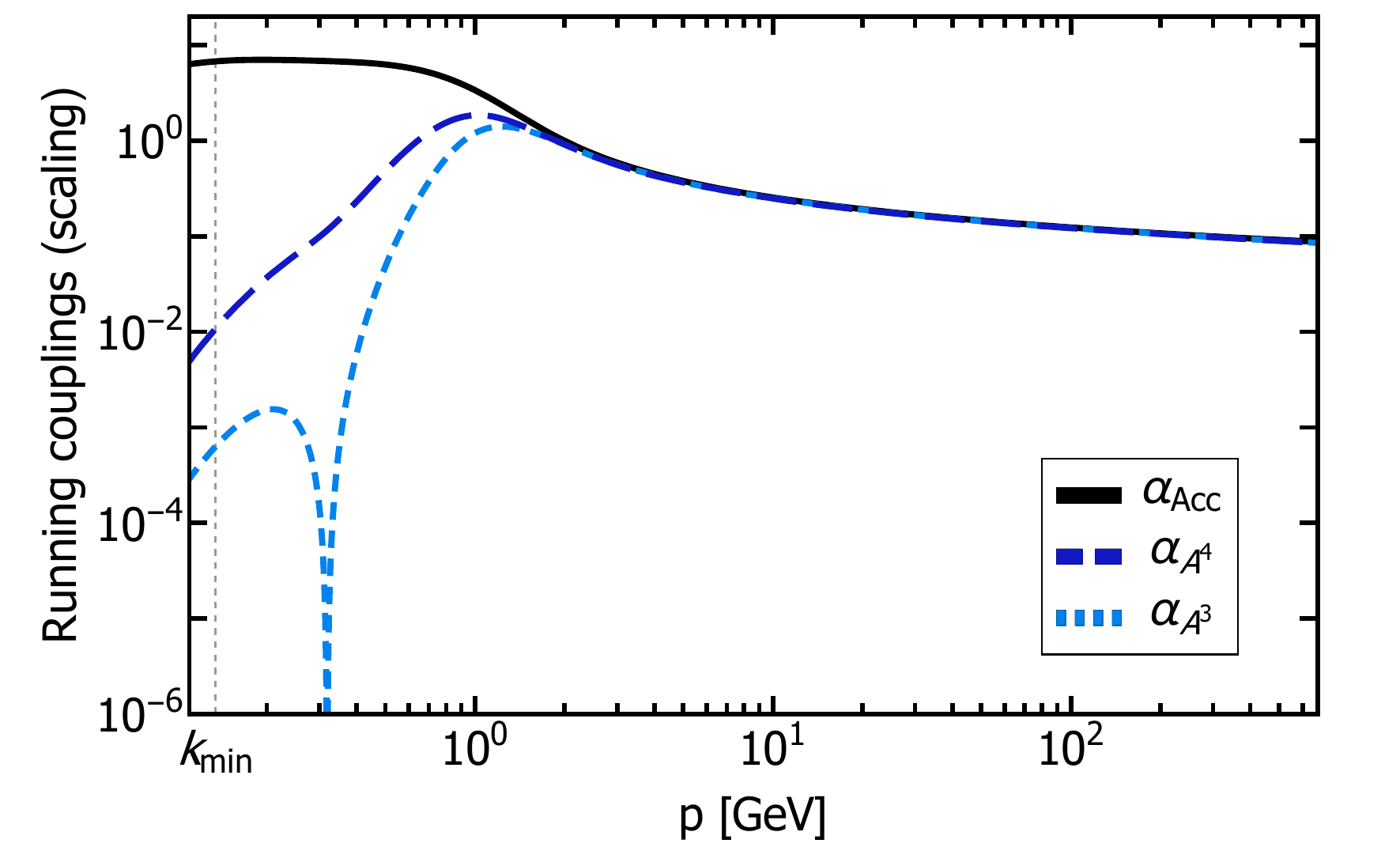}
			\caption{Running couplings as defined in \labelcref{eq:YM:couplingsrelation} obtained from different Yang-Mills vertex dressings as a function of momentum $p$.\hspace*{\fill}}
			\label{fig:YM:couplings}
		\end{figure}
	\end{minipage}
\end{figure*}

For this work, we restrict ourselves to the truncation shown in \Cref{fig:YM:flowequations} in the appendix. The gluon and ghost propagators are obtained from their respective two-point functions via \labelcref{eq:YM:propeq}. The two-point functions read 
\begin{align}
\Gamma_{AA,\mu\nu}^{ab}(p) =&\, \Pi^\perp_{\mu\nu}(p)Z_A(p) p^2\delta^{ab}+\Pi^\parallel_{\mu\nu}(p)\Gamma_{AA}^\parallel(p)\delta^{ab}\,,\nonumber\\[10pt]
\Gamma_{\bar{c}c}^{ab}(p) = & \,Z_c(p)p^2\delta^{ab}\,,
\label{eq:propsplit}\end{align}
with the transverse and longitudinal projection operators,
\begin{align}
	\Pi^{\perp}_{\mu \nu}(p) = \,\left(\delta_{\mu\nu}-\frac{p_\mu p_\nu}{p^2}\right)\,,\qquad \Pi^\parallel_{\mu \nu}(p) =\, \frac{p_\mu p_\nu}{p^2}\,.
	\label{eq:YM:ProjOps}
\end{align}  
The ghost and gluon dressing functions obtained from our computation are depicted in \Cref{fig:YM:propdressings}. The respective gluon propagator is shown in \Cref{fig:gluon_prop}. There, the present results are compared to lattice data. For recent results see in particular~\cite{Cyrol:2016tym} (fRG) and~\cite{Huber:2020keu, Aguilar:2021lke, Horak:2021pfr, Aguilar:2021uwa, Horak:2022myj} (DSE).

The longitudinal scalar part $\Gamma_{AA}^\parallel$ in \labelcref{eq:propsplit} contains the gauge fixing part that diverges in the Landau gauge limit $\xi\to 0$ and a regular contribution, that originates from the breaking of BRST invariance for finite $k$ and vanishes in the limit $k\to 0$. For $k\neq 0$ we have  
\begin{align}\label{eq:Gammaparal}
	\Gamma_{AA}^\parallel(p) = \frac{1}{\xi} p^2 + \Gamma_{AA,\textrm{reg}}^\parallel \,.
\end{align}
We remark that in the Landau gauge limit the gluon propagator is transverse and does not depend on $\Gamma_{AA,\textrm{reg}}^\parallel$, even though the latter contribution does not vanish. This is one of the properties that singles out the Landau gauge not only as the technically least difficult Lorenz gauge, but also suggests the best convergence of the results for correlation functions in a systematic vertex expansions: for any other choice of $\xi$ the regular part $\Gamma_{AA,\textrm{reg}}^\parallel$ feeds back into the dynamics of the system, even though its integrated impact has to vanish at vanishing cutoff scale. A more detailed discussion including the consistency of different functional relations (fRG, DSE, mSTI) is deferred to \Cref{sec:YM:FunRel}.

Utilising the projection operators \labelcref{eq:YM:ProjOps}, we can split correlation functions into their transverse and longitudinal parts,
\begin{align}
\Gamma_{\Phi_1\cdots\Phi_n} =&\, \Gamma_{\Phi_1\cdots\Phi_n}^\perp +\Gamma_{\Phi_1\cdots\Phi_n}^\parallel\,, 
\end{align}
where $\Gamma_{\Phi_1\cdots\Phi_n}^\perp$ is the completely transverse part of the correlation function, structurally given by $\Gamma_{\Phi_1\cdots\Phi_n}^\perp = (\Pi^\perp)^n\Gamma_{\Phi_1\cdots\Phi_n}$.  The longitudinal part simply is the complement, and hence is built up from correlation functions with at least one longitudinal leg. 

We will explain this splitting at the example of the ghost-gluon vertex. Its complete basis is spanned by only two tensor structures:  the classical one, proportional to the anti-ghost momentum, and a longitudinal non-classical one, which is proportional to the gluon momentum, 
\begin{align}\label{eq:ghostgluon}
\Gamma_{A\bar{c}c,\mu}^{abc}(p,q) &= i f^{abc}\Bigl(\lambda_{A \bar{c} c,\text{cl}}(p,q) q_\mu +\lambda_{A \bar{c} c,\text{ncl}}(p,q)p_\mu\Bigr)\,.
\end{align}
However, we can rewrite this in terms of the projection operator defined in \labelcref{eq:YM:ProjOps}, which is transverse in the gluon momentum $p$,
\begin{align}
\Gamma_{A\bar{c}c,\mu}^{abc}(p,q) =&\, i f^{abc}\Bigl(\lambda_{A \bar{c} c}(p,q)\, \Pi^{\perp}_{\mu \nu}(p) q_\nu \nonumber\\[4pt]
&\hspace{2cm}+ \bar \lambda_{A \bar{c} c,1}(p,q)\,p_\mu\Bigr)\,, 
\label{eq:YM:ghostgluonsplit}
\end{align}
with 
\begin{align}
\lambda_{A \bar{c} c}(p,q) &= \lambda_{A \bar{c} c,\text{cl}}(p,q)\,,  \nonumber\\[1ex]
\bar \lambda_{A \bar{c} c,1}(p,q) &=  \frac{p\cdot q}{p^2} \lambda_{A \bar{c} c,\text{cl}}(p,q) + \lambda_{A \bar{c} c,\text{ncl}}(p,q)\,.
\label{eq:YM:nonclDressing}
\end{align}

The singularity of such a split with projection operators at $p=0$ is reflected in the prefactor $p\cdot q/p^2$ of the classical dressing in $\bar \lambda_{A \bar{c} c,1}(p,q)$. It is matched by the respective one in the transverse projection operator multiplying $\lambda_{A \bar{c} c}(p,q)$. Therefore, for non-singular dressings in \labelcref{eq:ghostgluon} we only have a parameterisation singularity.

\begin{figure*}[t]
	\centering
	\begin{subfigure}[t]{0.45\textwidth}
		\centering
		\includegraphics[width=\linewidth]{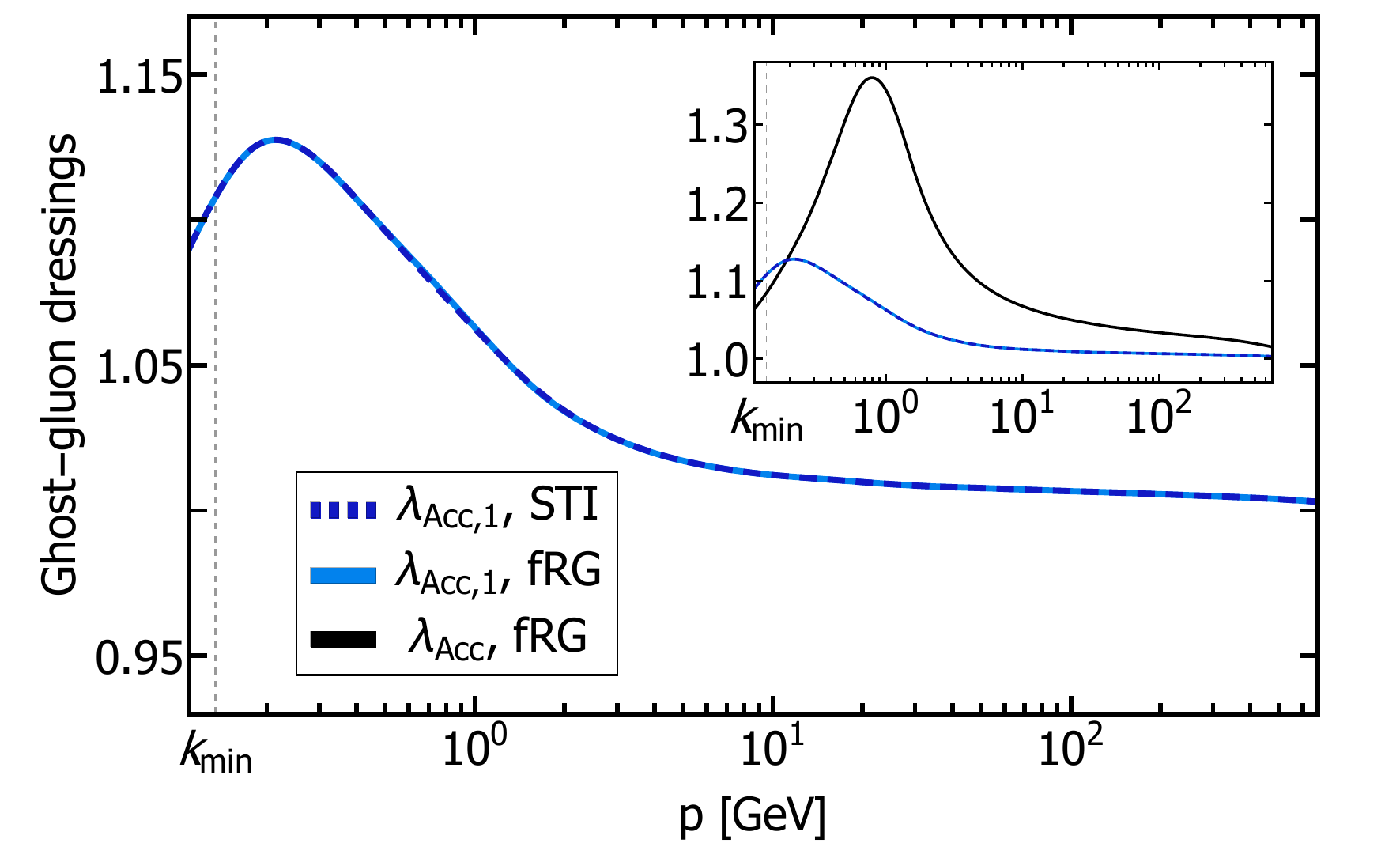}
		\caption{Ghost-gluon vertex dressings.}
		\label{fig:ghost_gluon}
	\end{subfigure}%
	\hspace{0.05\textwidth}
	\begin{subfigure}[t]{0.45\textwidth}
		\centering
		\includegraphics[width=\linewidth]{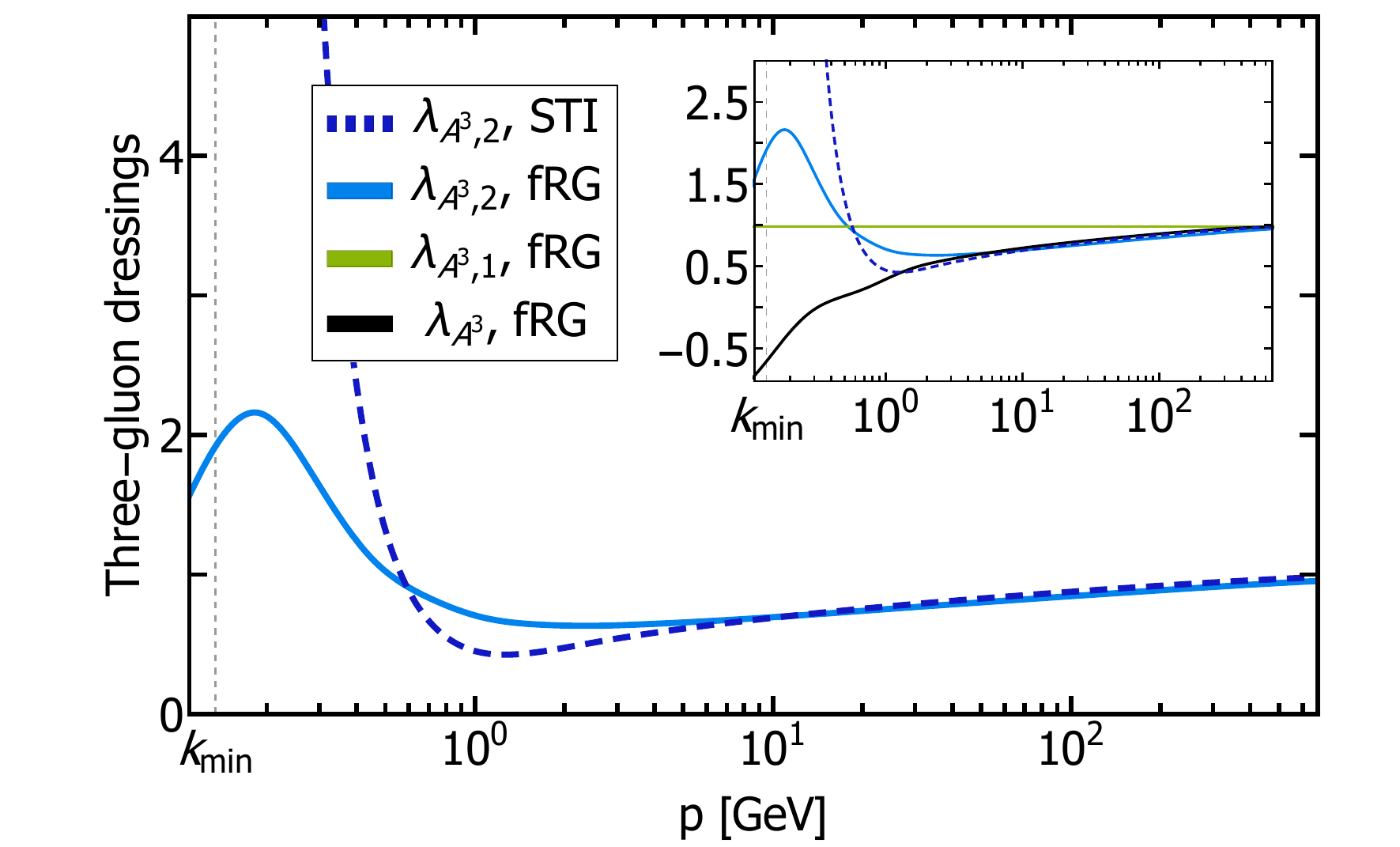}
		\caption{Three-gluon vertex dressings.}
		\label{fig:three_gluon}
	\end{subfigure}
	\caption{Dressings of the three-point functions as a function of momentum $p$. Results from the fRG are shown as solid lines, while STI results are shown as dashed lines. The transverse component is black and the longitudinal ones are depicted in green and blue. Please note that the transverse components are only visible in the insets.\hspace*{\fill} }
	\label{fig:YM:longcouplingsfrg}
\end{figure*}
In the present work, we consider relations between the dressings at the symmetric point $p^2=q^2 = -2 p\cdot q$. In the ultraviolet limit, the non-classical dressing vanishes and we are left with the classical one. Hence, for the present purpose it is convenient to simply use the longitudinal projection of the classical tensor structure, $\Pi_{\mu\nu}^\parallel(p) q_\nu$, for the longitudinal part. This leads us to,   
\begin{align}
	\Gamma_{A\bar{c}c,\mu}^{abc}(p,q) =&\Bigl(\lambda_{A \bar{c} c}(p,q)\, \Pi^{\perp}_{\mu \nu}(p)  \nonumber\\[4pt]
	&\hspace{.5cm}+  \lambda_{A \bar{c} c,1}(p,q)\,\Pi^{\parallel}_{\mu \nu}(p) \Bigr)\,\, i f^{abc} q_\nu\,, 
	\label{eq:YM:ghostgluonsplitP}
\end{align}
see also \cite{Cyrol:2016tym}. The number index in $\lambda_{A \bar{c} c,1}$ indicates the number of longitudinally projected gluons of the classical tensor structure of the vertex, and no index corresponds to the dressing of the fully transverse tensor structure. The dressing $\lambda_{A \bar{c} c,1}$ is given by  
\begin{align}\label{eq:lambda1}
	\lambda_{A \bar{c} c,1}(p,q) = \lambda_{A \bar{c} c,\text{cl}}(p,q) + \frac{p^2}{p\cdot q} \lambda_{A \bar{c} c,\text{ncl}}(p,q)\,.
	\end{align}
We have in particular $\bar \lambda_{A \bar{c} c,1}(p,q) = -2 \lambda_{A \bar{c} c,1}(p,q)$ at the symmetric point. Moreover, for all momenta except those with $p\cdot q =0$, the dressing in \labelcref{eq:lambda1} has the ultraviolet limit $\lambda_{A \bar{c} c,1}\to \lambda_{A \bar{c} c,\textrm{cl}}$ which facilitates the following discussions. In contradistinction, we find $\bar \lambda_{A \bar{c} c,1}\to (p\cdot q)/p^2 \lambda_{A \bar{c} c,\textrm{cl}}$. In conclusion, the transverse ghost-gluon dressing is equivalent to the classical one, whereas the longitudinal dressing is a combination of the classical and non-classical dressing. For a comparison of the ghost-gluon dressings in \labelcref{eq:YM:ghostgluonsplitP}, see \Cref{fig:ghost_gluon} and \Cref{fig:YM:accnonclcouplingsfrg} in the \Cref{app:YM:FRGYM}. For recent results see in particular~\cite{Cyrol:2016tym} (fRG) and~\cite{Huber:2020keu, Aguilar:2021okw, Eichmann:2021zuv} (DSE). 

We now proceed by parametrising the BRST and gluonic vertices in terms of transverse and longitudinal projections analogously to the parameterisation of the ghost-gluon vertex discussed above. With the anti-ghost shift symmetry of the effective action, we can relate the BRST vertices $\Gamma_{cQ_A}$ and $\Gamma_{AcQ_A}$ to the ghost two-point and the ghost-gluon vertex. A derivation thereof can be found in \Cref{app:YM:shiftsym} and we arrive at 
\begin{align}
\Gamma^{ab}_{c \,Q_A,\mu}(p) =& -i Z_c(p) p_\mu \delta^{ab}\,,\nonumber\\[10pt]
\Gamma^{abc}_{Ac\,Q_A,\mu\nu}(p,q) =&\, -f^{abc}\Bigr(\Pi^{\perp}_{\mu \nu}(p)\lambda_{A \bar{c} c}(p,-q-p)\nonumber\\[4pt]
&\hspace{.5cm}+ \Pi^{\parallel}_{\mu \nu}(p)\lambda_{A \bar{c} c,1}(p,-q-p)\Bigl)\,.
\label{eq:YM:BRSTVertices1}
\end{align}
Notably, these identities even hold diagramatically true for the flows of the dressings. The other BRST two-point function and vertex are parameterised as,
\begin{align}
\Gamma^{ab}_{A Q_{\bar{c},}\mu}(p) &= -i p_\mu \frac{1}{\xi} \delta^{ab}\,,\nonumber\\[4pt]
\Gamma_{cc\,Q_c}^{abc}(p,q) &= -i f^{abc}\lambda_{Q_ccc}(p,q)\,.
\label{eq:YM:BRSTVertices2}
\end{align}

\begin{figure}[t]
 	\includegraphics[width=0.98\linewidth]{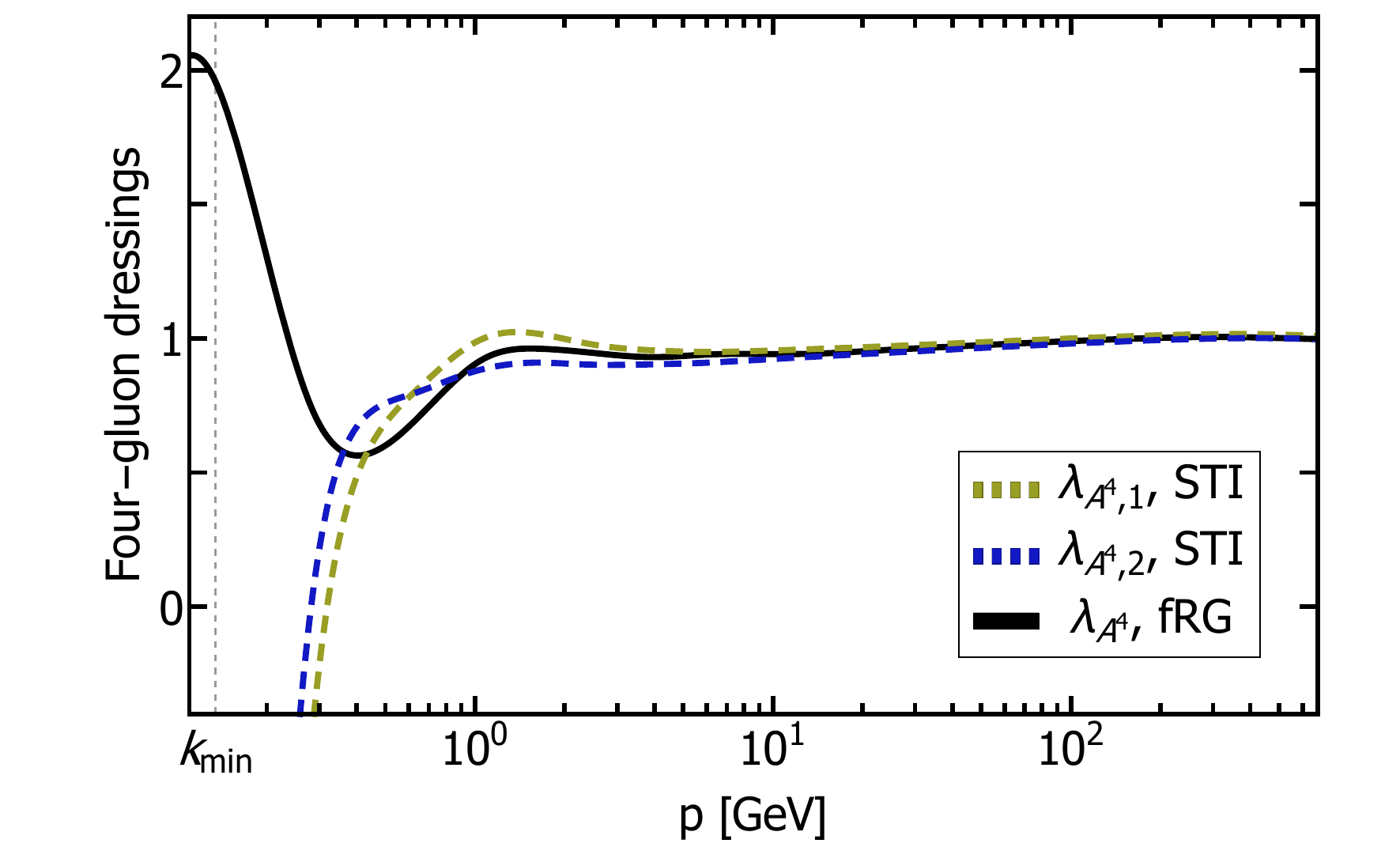}
	\caption{Dressings of the four-point function as a function of momentum $p$. The result from the fRG is shown as a solid line, while STI results are shown as dashed lines. The transverse component is black and the longitudinal ones are depicted in green and blue.\hspace*{\fill}}
	\label{fig:fourgluon}
\end{figure}

In the present work we approximate the three- and four-gluon vertices with their fully dressed classical tensor structures, noted as  $\tau_{A^3,\text{cl}}$ and $\tau_{A^4,\text{cl}}$. The purely transverse parts are obtained by contracting the classical tensor structure with three and four transverse projection operators respectively. The complements now contain mixed vertices with longitudinal and transverse legs as well as a purely longitudinal part. The parts with at least three longitudinal legs do not contribute since they do not feed back into the diagrams and are therefore not included in our truncation. Thus we parameterise the gluonic vertices as,
\begin{align}\label{eq:3-4}
\Gamma_{A^3}(p,q) = &\left(\lambda_{A^3}\left(\Pi^\perp\right)^3+\lambda_{A^3,1}\left(\Pi^\perp\right)^2\Pi^\parallel\right.\nonumber\\[4pt]
&\left.+\lambda_{A^3,2}\Pi^\perp\left(\Pi^\parallel\right)^2\right) \tau_{A^3,\text{cl}}+\cdots \,\nonumber\\[10pt]
\Gamma_{A^4}(p,q,r)   =&\left(\lambda_{A^4}\left(\Pi^\perp\right)^4+\lambda_{A^4,1}\left(\Pi^\perp\right)^3\Pi^\parallel\right.\nonumber\\[4pt]
&\left.+\lambda_{A^4,2}\left(\Pi^\perp\right)^2\left(\Pi^\parallel\right)^2\right) \tau_{A^4,\text{cl}}+\cdots \,
\end{align}
where indices and momenta are dropped and the permutations are implicit for the sake of simplicity. The dots indicate terms proportional to $(\Pi^\parallel)^n\tau_{A^3,\text{cl}}$ and $(\Pi^\parallel)^n\tau_{A^4,\text{cl}}$ with $n\geq 3$, i.e. with at least $3$ longitudinally projected gluon legs, as well as further, non-classical,  tensor structures. We emphasise that for $p\cdot q= 0$ or $p\cdot r=0$ the parameterisation in \labelcref{eq:3-4} may lead to parameterisation singularities for the dressings, that are then seen in the respective projections of the diagrams in the functional relations. For further details on the parameterisation, see \Cref{app:YM:tensors} and e.g.\ \cite{Eichmann:2021zuv} for a similar parametrisation. The longitudinal four-gluon vertices are approximated by their STI-values, see \labelcref{eq:YM:4glSTI}. For a comparison of the dressings in \labelcref{eq:3-4} see \Cref{fig:three_gluon} and \Cref{fig:fourgluon}. For recent results see in particular~\cite{Cyrol:2016tym} (fRG) and~\cite{Eichmann:2014xya, Huber:2020keu, Aguilar:2021lke, Eichmann:2021zuv} (DSE). 

The remaining dressings are obtained from their respective fRG equations, that are depicted in \Cref{app:YM:Diags} in \Cref{fig:YM:flowequations}. They are evaluated at a symmetric point for three(four)-point functions,
\begin{align}
p_i\cdot p_j = \begin{cases}
\;\;\;\;p^2  &\text{for $i=j$,}\\[10pt]
-\frac{1}{n-1} p^2& \text{otherwise, where $n=3(4)$.}
\end{cases}
\label{eq:YM:symmPoint}
\end{align} 
For simplicity, the vertex dressings feeding back into the fRG equations are evaluated at the average momentum configuration $\lambda_i(\bar{p})$ with,
\begin{align}
\bar{p}^2 =  \frac{1}{n} \sum_{i=1}^{n}p_i^2\,,
\label{eq:YM:avmomconfig}
\end{align}
which has been shown to be a good approximation, see e.g.~\cite{Huber:2015fna}.
From the transverse dressings we can derive momentum dependent running couplings,
\begin{align}
\alpha_{A\bar{c}c}(p) &= \frac{1}{4\pi}\frac{\lambda_{A \bar{c} c}(p)^2}{Z_A(p)Z_c(p)^2}\,,\nonumber\\[10pt]
\alpha_{A^3}(p) &= \frac{1}{4\pi}\frac{\lambda_{A^3}(p)^2}{Z_A(p)^3}\,,\nonumber\\[10pt]
\alpha_{A^4}(p) &= \frac{1}{4\pi}\frac{\lambda_{A^4}(p)}{Z_A(p)^2}\,.
\label{eq:YM:couplingsrelation}
\end{align}
These vertex couplings are perturbatively two-loop degenerate (two-loop universality), leading to 
\begin{align}
\alpha_{A\bar{c}c}(p)\approx \alpha_{A^3}(p)\approx \alpha_{A^4}(p)\,,\quad\text{for} \quad p\gtrsim  5\,\text{GeV}.
\label{eq:YM:couplingsdegeneracy}
\end{align}
This property is reflected well in our truncation, see \Cref{fig:YM:couplings}. We emphasise that  \labelcref{eq:YM:couplingsdegeneracy} is non-trivial, as our initial condition only feature constant vertices. These initial conditions are chosen or tuned such that \labelcref{eq:YM:couplingsdegeneracy} is fulfilled. Satisfying \labelcref{eq:YM:couplingsdegeneracy} over several orders of magnitude is already a significant indication for gauge consistency. For a detailed discussion of this aspect see~\cite{Cyrol:2016tym} and more information on the tuning procedure is given in \Cref{sec:YM:transverse}. 

%%%%%%%%%%%%%%%%%%%%%%%%%%%%%%%%%%%%
\subsection{Functional Relations and Consistency Constraints}
\label{sec:YM:FunRel}
%%%%%%%%%%%%%%%%%%%%%%%%%%%%%%%%%%%%

The computation of functional relations is facilitated in Landau gauge Yang-Mills theory, if contrasted with general covariant gauges: in the Landau gauge the set of all transverse functional relations is closed, i.e.
\begin{subequations}
	\label{eq:FunRelP}
\begin{align}
\Gamma_{(n)}^\perp = \text{funRel}^\perp_{(n)}[\{\Gamma_{(2\leq m\leq n+2)}^\perp\}]\,,
\label{eq:YM:closedtransversal}
\end{align}
where $\{\Gamma_{(2\leq m\leq n+2)}^\perp\}$ indicates the set of all transverse $m$-point correlation functions with $2 \leq m\leq n+2$.  In contradistinction, the longitudinal correlation functions satisfy 
\begin{align}\label{eq:YM:long}
\Gamma_{(n)}^\parallel = \text{funRel}^\parallel_{(n)}[\{\Gamma_{(2 < m\leq n+2)}^\parallel\},\{\Gamma_{(2\leq m\leq n+1)}^\perp\}]\, .
\end{align}
\end{subequations}
Note that the longitudinal two-point function $\Gamma_{(2)}^\parallel$ does not feed back into the longitudinal relations \labelcref{eq:YM:long} and \labelcref{eq:YM:mSTISpace}. \Cref{eq:YM:closedtransversal} follows from the fact that the gluon propagator $G_A$ as well as $(G\,\dot R\, G)_{AA}$ are transverse in the Landau gauge and so are all internal legs. Also, the propagator only depends on $\Gamma_{(2)}^\perp$. The structure \labelcref{eq:FunRelP} applies to fRG equations and  DSEs, for more details see \cite{Fischer:2008uz, Cyrol:2016tym, Dupuis:2020fhh, LectureNotesprep}. 

The  mSTIs in \labelcref{eq:YM:mSTI} involve longitudinal and transverse correlation functions. They are similar to \labelcref{eq:YM:long}, reading 
\begin{align}
\Gamma_{(n)}^\parallel = \text{mSTI}_{(n)}[\{\Gamma_{(2 < m\leq n+2)}^\parallel\},\{\Gamma_{(2\leq m\leq n+1)}^\perp\} ]\,. 
\label{eq:YM:mSTISpace}
\end{align}
Note that \labelcref{eq:YM:mSTISpace} constitutes yet another tower of functional relations for longitudinal correlation functions. Naturally, as for truncated towers of fRG equations and DSEs, its solution for a given truncation will in general differ from both, the solution of the longitudinal fRG equations and the longitudinal DSEs (or any other tower of longitudinal functional relations). Moreover, for a finite set of longitudinal and transverse correlation functions, one may simply use the (m)STI for computing the longitudinal correlation functions. Such a scheme seemingly builds in gauge consistency by construction. Naturally, the longitudinal correlation functions will then in general not satisfy the respective fRG equations or DSEs. In conclusion, it is not the violation of the mSTI alone, which comprises the information about gauge consistency, but the combination of all functional relations.

However, even though \labelcref{eq:YM:mSTISpace} is very similar to the fRG equations and DSEs for the longitudinal dressings, there is an important structural difference which is most apparent in the Landau gauge limit: while \labelcref{eq:YM:closedtransversal} and \labelcref{eq:YM:long} only depend on the transverse part of the regulator, the mSTI \labelcref{eq:YM:mSTISpace} depends on the longitudinal part for regulators whose longitudinal part diverges with $1/\xi$. A natural class of regulators with this property is that of \textit{RG-adapted} or spectrally adapted regulators, see \cite{Pawlowski:2005xe, Gies:2002af}. In this case, the regulators are proportional to the full dispersion of the respective field, which is detailed in \Cref{app:YM:reg}. This choice ensures, that the effective action in the presence of the regulator has the underlying RG-invariance of the full theory at vanishing regulator, hence the name \textit{RG-adapted}. Specifically, the longitudinal part of the gluon regulator reads 
\begin{align}
	(R_A)^{\parallel, ab}_{\mu\nu}(p) =&\,\delta^{ab} \Pi^\parallel_{\mu \nu}(p)\Gamma^{\parallel}_{AA}(p) \, r\left(\frac{p^2}{k^2}\right) \,.
	\label{eq:YM:longregs}
\end{align}
The class of regulators with \labelcref{eq:YM:longregs} leads to 
\begin{align}\label{eq:RGTL}
	\lim_{\xi\to 0} 	p_\mu (G_{AA} R_A)_{\mu\nu} = p_\nu \frac{r_A}{1+r_A}(p^2/k^2)\,,
\end{align}
and hence one of the internal lines in the mSTI diagrams has a longitudinal part. In turn, for gluon regulators with a regular longitudinal part we have 
\begin{align}\label{eq:RGT}
	p_\mu (G_{AA} R_A)_{\mu\nu} \propto \xi p_\nu\,,
\end{align}
and hence all internal lines are transverse in the Landau gauge. Then, the mSTI has the same structure as \labelcref{eq:YM:long} in terms of contributing correlation functions. 

Interestingly, the fRG equations and DSEs (in the presence of a regulator), \labelcref{eq:FunRelP}, are the same in the Landau gauge limit for regulators with either  \labelcref{eq:RGTL} or \labelcref{eq:RGT}. However, it can be shown, that the contributions of the longitudinal part of the propagator \labelcref{eq:RGTL}  do not cancel in the mSTI. In particular, the effective gluon mass $m_k^2$ from the STI only receives contributions from the longitudinal part of $(R_k G)_{AA}$. In conclusion, only for gluon regulators whose longitudinal part contains the gauge fixing term $1/\xi p^2 r(p^2/k^2)$, the effective mSTI gluon mass and that from the fRG flow equation or DSE agree, and the implicit assumption of $k$-independence of the renormalisation procedure is self-consistent. In short, only for RG-adapted regulators we do not implicitly introduce a gluon mass (counter) term on the level of the classical action. 

This intricacy emphasises the importance of gauge consistency of the regularisation procedure. The comparison of functional mSTI and fRG/DSE always allows us to choose a longitudinal regulator $R_A^\parallel$ such that all functional relations are compatible. However, this also implies that gauge consistency of truncated fRG flows or DSEs should rather be checked with a set of correlation functions and not a single one.   

Throughout this work we will use \textit{RG-adapted} regulators \labelcref{eq:YM:longregs}, which are also \textit{gauge consistent}, as discussed above. A more detailed study of this intricacy and the general regulator (shape) dependence is still under investigation and will be published elsewhere. 

We conclude this section with two remarks. Firstly, for regular correlation functions, one can use the STIs for also extracting a part of the respective transverse correlation functions. Loosely speaking, regularity implies  
\begin{align}\label{eq:regular}
	|\partial_{p_i} \Gamma_{(n)}| < \infty\,, 
\end{align}
in which case the longitudinal and transverse parts are linked. However, in the present case of Yang-Mills theories, it can be shown that regular correlation functions are at odds with confinement. In turn, in perturbation theory, one can show that the regularity assumption holds true. Indeed, it is also related to the identification of running couplings \labelcref{eq:YM:couplingsrelation}. This adds yet another layer of complexity to the current situation: for perturbative and semi-perturbative momenta regularity holds true and we should see the respective relations between longitudinal and transverse correlation functions. 

In summary, this leaves us with the necessity as well as a large variety of non-trivial consistency checks. We also emphasise that great care is needed in the interpretation of violations of the mSTI or alternatively of other functional relations for longitudinal correlation functions within truncations.

%%%%%%%%%%%%%%%%%%%%%%%%%%%%
\subsection{Confinement}
Confinement describes the effect that no coloured particles can be detected directly by experiments, i.e.\ the absence of coloured states from the physical asymptotic spectrum of our theory, and is in one-to-one correspondence to a \textit{dynamical} mass gap $m_\textrm{gap}^\textrm{\tiny YM}$ in the physical spectrum of Yang-Mills theory. It is fundamentally linked to the non-Abelian gauge group of QCD, and hence it is most cleanly studied in Yang-Mills theory in the absence of dynamical quarks that complicate the matter. 

In Landau gauge Yang-Mills theory confinement is in one-to-one correspondence to a mass gap (finite correlation length $\xi_\textrm{conf}$) in the gluon propagator, as has been shown in \cite{Braun:2007bx, Fister:2013bh}. There, it has been also shown that the confinement-deconfinement temperature $T_\textrm{conf}$ is proportional to the (inverse) screening length, $T_\textrm{conf}\propto 1/\xi_\textrm{conf}$. Moreover, it also sets the scale of the glueball spectrum, $m_\textrm{glueball}\propto 1/\xi_\textrm{conf}$, and in conclusion $m^\textrm{\tiny YM}_\textrm{gap}\propto 1/\xi_\textrm{conf}$. Finally, it also sets the saturation scale of the effective charge, see \cite{Aguilar:2009nf}. 

\begin{figure}[t]
	\includegraphics[width=0.98\linewidth]{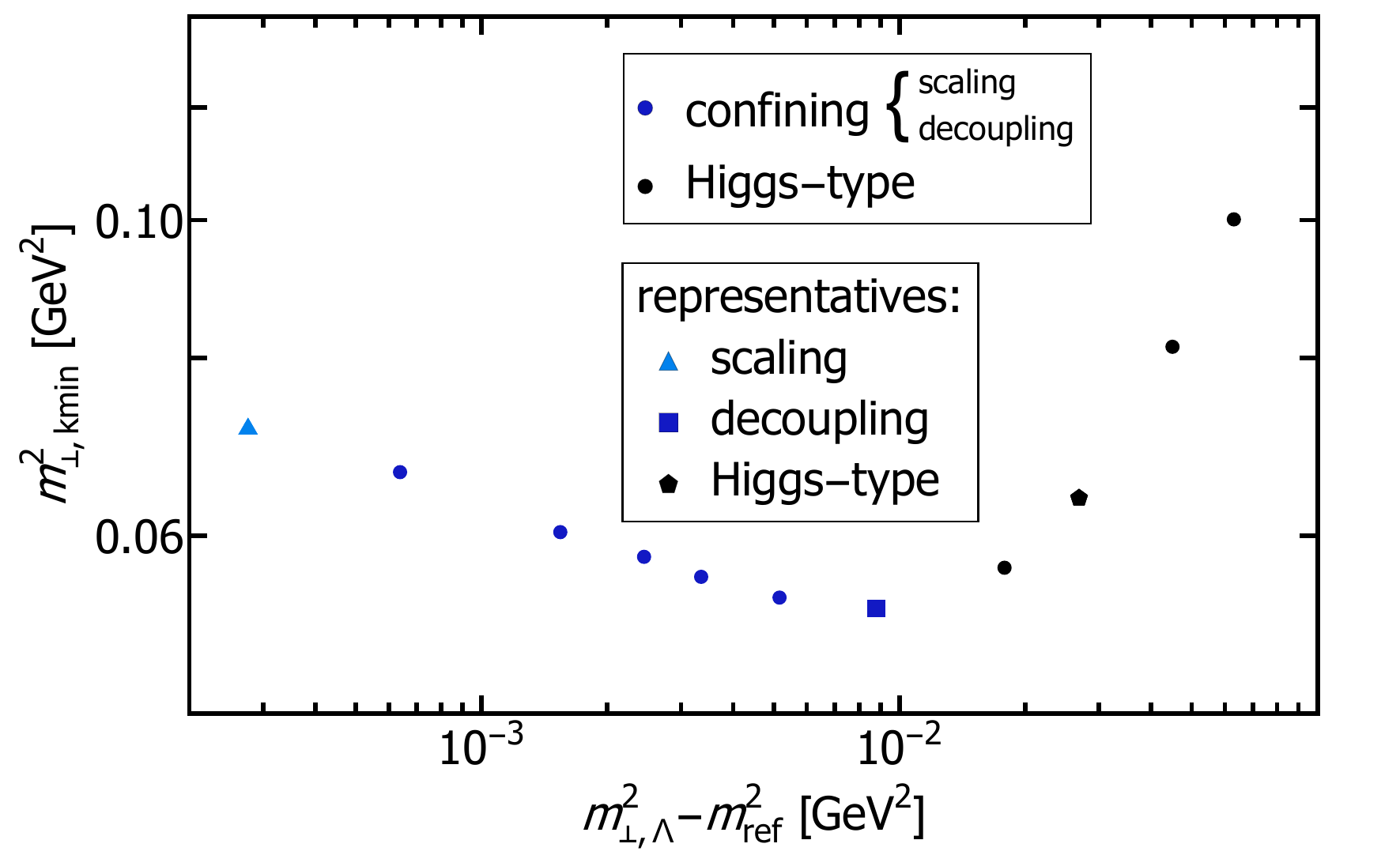}
	\caption{Effective gluon mass $m_\perp^2$, defined in \labelcref{eq:mPerp}, at $k=k_{\mathrm{min}}$ from different fRG solutions as a function of the difference $m_{\bot,\Lambda}^2-m_{\mathrm{ref}}^2$ of the mass parameters at the initial cutoff scale $\Lambda$. More details on the effective gluon mass at a finite cutoff scale are deferred to \Cref{app:YM:reg}. The reference mass $m_{\mathrm{ref}}^2$ indicates the initial mass of the extrapolated UV value for the mass parameter of the true scaling solution, see also~\Cref{app:YM:reg}; and the light blue triangle indicates our approximate scaling solution. The black markers correspond to Higgs-type and the blue markers to confining solutions. The dark blue square is used as a representative decoupling solution. The black pentagon is used as a representative Higgs-type solution. \hspace*{\fill}}
	\label{fig:gluon_mass_tuning}
\end{figure}

In conclusion, while the mass gap $1/\xi_\textrm{conf}$ in the Landau gauge gluon propagator is a property of a gauge-fixed correlation function, it is directly linked to the \textit{dynamical} mass gap $m_\textrm{gap}^\textrm{\tiny YM}$ in Yang-Mills theory. At small momenta, the mass gap manifests itself as an effective gluon mass parameter $m_\perp^2$ in the gluon propagator or transverse gluon two-point function, 
\begin{align}\label{eq:mPerp}
	m_\perp^2 = \Gamma_{AA}^\perp(p=0)\,, 
\end{align}
see also \labelcref{eq:mPerpPar}. The effective gluon mass $m_\perp^2 $ is bounded from below by the screening length $\xi_\textrm{conf}$ (subject to an appropriate normalisation of the gluon propagator). 

Evidently, the dynamics of Yang-Mills theory for momenta below the confining mass gap is dominated by the infrared aspects of the gauge fixing at hand. Even if restricting ourselves to the Landau gauge, we may encounter different IR solutions, see e.g.~\cite{Fischer:2008uz}. An investigation of the different types of solutions within our set-up can be found in \Cref{sec:YM:Discussiondec}. In the present fRG set-up the mass parameter \labelcref{eq:mPerp} is approximated by its counterpart at the lowest infrared cutoff scale considered, $ k_{\min}= 135$\,MeV. The full range of solutions in terms of the effective mass parameter $m^2_{\perp,k_{\min}}$ is depicted in \Cref{fig:gluon_mass_tuning}: right of the minimum the theory approaches massive Yang-Mills theory, while left of the minimum confining Yang-Mills theory is found. The Landau gauge solution found in lattice simulation is best approximated for effective gluon masses about the minimum. In the remainder of this section we discuss the different branches of the effective gluon mass curve depicted in \Cref{fig:gluon_mass_tuning}. 

One possible mechanism for confinement has been proposed by Kugo and Ojima \cite{Kugo:1979gm}. The Kugo-Ojima confinement scenario can generally be broken down into three criteria. First, if we have a global, non-perturbative BRST charge, i.e.\ unbroken BRST symmetry, it can be used to construct the physical state space $\mathcal{H}_{phys}$. Second, if the global colour charge is unbroken, $\mathcal{H}_{phys}$ only contains colourless states. And third, the cluster decomposition principle has to be violated in the total state space $\mathcal{H}$, but not in $\mathcal{H}_{phys}$. 

For the latter, it has already been argued in \cite{Nakanishi:1990qm} that this is indeed fulfilled in the covariant operator formulation of QCD. From the second criterion, we can deduce direct implications for our infrared (IR) Green's functions. It can be shown that this requires an IR enhancement of the ghost propagator. This yields a unique IR renormalisation condition. We shall refer to this as the \textit{scaling} solution. This enhancement is accompanied by a dynamical creation of an effective gluon mass,
\begin{align}
&\lim\limits_{p\rightarrow 0} Z_c(p^2) \propto (p^2)^\kappa\,,\nonumber\\[10pt]
&\lim\limits_{p\rightarrow 0} Z_A(p^2) \propto (p^2)^{-2\kappa}\,,
\label{eq:YM:scalingexponentsscaling}
\end{align}
with $1/2 < \kappa < 1$, see \cite{vonSmekal:1997ohs, vonSmekal:1997ern} and the reviews \cite{Alkofer:2000wg, Fischer:2006ub}. This scaling behaviour can be tuned in the present fRG set-up, and we approximate it by the leftmost gluon mass parameter in  \Cref{fig:gluon_mass_tuning}, represented by the blue triangle. While only an approximation to full scaling even for $p\to 0$, this solution shows scaling for the physical momenta considered, $p\gtrsim k_\textrm{min}$.  The scaling exponents obtained from our results can be found in \labelcref{eq:YM:scalingexponents}. 

The Kugo-Ojima confinement scenario rests upon at least one non-trivial assumption: the existence of a global BRST charge for the standard Landau gauge BRST symmetry with the \textit{local} or infinitesimal BRST transformations \labelcref{eq:YM:YMBRSTtrafo}, for a detailed discussion see also~\cite{Fischer:2008uz}. First of all, the BRST charge may not exist or may be broken. Second of all, specific realisations of the Landau gauge may include infrared modifications such as in the modified Gribov-Zwanziger scenario. For a review on the latter see~\cite{Vandersickel:2012tz}, for recent progress within a BRST-invariant formulation see e.g.~\cite{Capri:2015ixa, Pereira:2016fpn, Capri:2016aif, Capri:2016jgn, Dudal:2017jfw, Capri:2021pye}. 

All these scenarios are signalled by the absence of the Kugo-Ojima scaling: while the gluon propagator still exhibits a mass gap, it settles at a finite value in the infrared, which is well-described by an effective gluon mass in the infrared. We rush to add that there is no trace of this mass in the ultraviolet which is the qualitative difference to massive Yang-Mills theory. The respective ghost exhibits a finite infrared enhancement but shows no scaling. Still, for these infrared solutions of Yang-Mills theories in the Landau gauge the infrared dynamics is decoupled, and they have been called \textit{decoupling} or \textit{massive} solutions for the above reasons. In particular, they are realised within lattice simulations in the Landau gauge as well as in many functional implementations. An appealing scenario discussed in functional approaches is the dynamical emergence of the gluonic mass gap via the Schwinger mechanism \cite{Schwinger:1962tn, Schwinger:1962tp, Jackiw:1973tr, Eichten:1974et, Poggio:1974qs}, for more recent works see e.g.~\cite{, Cornwall:1981zr, Cornwall:1988ad, Aguilar:2011xe, Aguilar:2021uwa}, or gluon condensation, see~\cite{Horak:2022aqx}. In the present work we consider the decoupling solution with the smallest dynamically created effective gluon mass $m^2_{\perp,k_{\mathrm{min}}}$ indicated by the blue square data point in \Cref{fig:gluon_mass_tuning}. In the present approximation this solution resembles best that seen on the lattice. 

Lastly, we consider solutions with an \textit{explicit} effective gluon mass. This explicit mass is present in massive extensions of Yang-Mills theory as studied in e.g.~\cite{Tissier:2010ts, Tissier:2011ey, Serreau:2012cg, Reinosa:2017qtf} in the Curci-Ferrari gauge (CF model), for a recent review see \cite{Pelaez:2021tpq}. These works aim at modelling Yang-Mills theory as the CF model with a specific mass parameter, which allows to use perturbative expansion techniques.  Alternatively, explicit mass solutions can be generated by a Higgs-type mechanism. In the approximation of the present paper these two physically distinct cases are not distinguished and we call them \textit{Higgs-type} solutions, see also \cite{Cyrol:2016tym}. While this type of solution is realised for asymptotically large UV masses $m^2_{\bot,\Lambda}$ in massive extensions of  Yang-Mills theory, the transition from the \textit{Higgs-type} branch to the confining decoupling branch  happens around the minimum in \Cref{fig:gluon_mass_tuning}, but is not determined yet. We remark, that the mass parameter used in the CF model for modelling Yang-Mills theory is also within this regime. It trivially equals that seen on the lattice as the CF mass is adjusted such that the lattice data are reproduced best. 

The IR behaviour of the decoupling and Higgs-type solution is characterized by
\begin{align}
&\lim\limits_{p\rightarrow 0} Z_c(p^2) \propto 1\,,\nonumber\\[10pt]
&\lim\limits_{p\rightarrow 0} Z_A(p^2) \propto (p^2)^{-1}\,.
\end{align}

More details in general and on the investigation of the different solutions can be found in \cite{Fischer:2006ub, Fischer:2008uz, Cyrol:2016tym, Reinosa:2017qtf}. 

We close this section with the discussion of a rather interesting aspect of the results depicted in  \Cref{fig:gluon_mass_tuning}. They demonstrate a smooth deformation or tuning from the scaling solution towards decoupling solutions in a vertex expansion. The present approximation assumes regular vertices, i.e. the longitudinal ones are directly related to the transverse ones. With such longitudinal vertices, the STIs cannot hold true for small momenta as regular vertices entail a vanishing transverse gluon mass gap, which is at odds with smooth deformations. This suggests that also the vertices in the scaling solution exhibit not only the scaling irregularities but further ones, see~e.g.~\cite{Eichmann:2021zuv}.

%%%%%%%%%%%%%%%%%%%%%%%%%%
\section{Numerical Results}%%%%
\label{sec:YM:NumericalResults}%%
%%%%%%%%%%%%%%%%%%%%%%%%%%

In this work, we solve the transverse sector of Yang-Mills correlation functions self-consistently in the Landau gauge. Here, self-consistency refers to the fact that all correlation functions computed are fed back to the fRG loops. We also compute the longitudinal dressings and the BRST dressings both from the fRG and the mSTI. We proceed by comparing the results obtained with both approaches. All momentum-dependent correlation functions are evaluated at the symmetric point. As before, for a complete survey on results for Landau gauge correlation functions we refer to the reviews \cite{Dupuis:2020fhh, Alkofer:2000wg, Fischer:2006ub, Binosi:2009qm, Maas:2011se, Boucaud:2011ug, Huber:2018ned, 
	Pelaez:2021tpq}. 

%%%%%%%%%%%%%%%%%%%%%%%%%%%%%%%%%%%%%%%%%%%%%%%%%%%%%%%%

\subsection{Transverse correlation functions}
\label{sec:YM:transverse}
As stated in \Cref{sec:YM:FunRel}, the transverse sector of Yang-Mills theory is closed in Landau gauge. Thus, in a given approximation to the transverse sector, we deal with a closed and finite system of momentum-dependent self-consistent coupled differential equations. The mSTI for the gluon propagator, together with regularity, leads to a non-zero transverse gluon mass $m^2_{\perp,\Lambda}$ at the cutoff. Its exact value is uniquely fixed by demanding a scaling, decoupling or Higgs-type solution. 

The constant initial values of the vertex dressings are chosen such that the respective momentum-dependent couplings are perturbatively degenerate at $k=k_{\mathrm{min}}$: they satisfy the two-loop exact relation from \labelcref{eq:YM:couplingsdegeneracy} for perturbative momenta, thus rendering only the initial value of the ghost-gluon dressing a free parameter. It is chosen to be $\lambda_{A \bar{c} c,\Lambda}(p) = 1$ at the cutoff. 

Although the initial gluon mass parameter as well as the initial vertex dressings are uniquely fixed by the two aforementioned conditions, their exact values are determined via a fine-tuning procedure: one chooses a set of initial conditions for the vertex dressings and varies the initial gluon mass parameter thereby obtaining the different types of solutions and finally, the scaling solution, see e.g. \Cref{fig:gluon_mass_tuning}. For the respective  scaling solution the difference from the coupling relations \labelcref{eq:YM:couplingsdegeneracy} is used for retuning the initial vertex dressings.  This process is iterated until the relations \labelcref{eq:YM:couplingsdegeneracy} are satisfied. The main difficulty in the selection or adaptation of the initial parameters after each step in the process is the fact, that the flow equations are a set of coupled differential equations. This means varying only one initial condition influences the outcome of the computation non-linearly, especially the scaling mass parameter is very sensitive to such a retuning. This also implies that the existence of a set of constant, i.e. momentum-independent, initial conditions leading to the momentum-dependent relations \labelcref{eq:YM:couplingsdegeneracy} within a truncation over several orders of magnitude is a highly non-trivial result. This has first been achieved in \cite{Cyrol:2016tym} within the fRG, for a respective DSE work see \cite{Huber:2020keu}, for fRG results in full QCD see \cite{Cyrol:2017ewj}. 
%
%For other relevant work in the fRG see~e.g.~\cite{Ellwanger:1996wy, Ellwanger:1995qf, Bergerhoff:1997cv, Pawlowski:2003hq, Fischer:2004uk, Fischer:2006vf, Gies:2006wv, Fischer:2008uz, Marhauser:2008fz, Fischer:2009tn, Braun:2010cy, Eichhorn:2010zc, Fister:2011uw, Cyrol:2017qkl, Corell:2018yil, Asnafi:2018pre, Dupuis:2020fhh, Horak:2022aqx} and in \commentnw{CHANGE!!!} DSEs see~e.g.~\cite{Hauck:1996sm, vonSmekal:1997ohs, vonSmekal:1997ern, Hauck:1998fz, Alkofer:2000wg, Alkofer:2003jr, Schleifenbaum:2004id, Alkofer:2004it, Maas:2005xh, Maas:2005ym, Fischer:2005ui, Fischer:2006ub, Fischer:2006vf, Huber:2007kc, Alkofer:2008jy, Alkofer:2008dt, Kellermann:2008iw, Fischer:2008uz, Fischer:2009tn, Huber:2010tvj, Maas:2011se, Boucaud:2011ug, Huber:2012kd, Cyrol:2014kca, Eichmann:2014xya, Huber:2015ria, Huber:2016tvc, Huber:2017txg, Huber:2018ned, Horak:2021pfr, Fischer:2020xnb, Napetschnig:2021ria, Eichmann:2021zuv, Ferreira:2022wzf, Horak:2022myj, Aguilar:2006gr, Aguilar:2008xm, Binosi:2012sj, Aguilar:2013xqa, Aguilar:2021okw, Aguilar:2013vaa, Athenodorou:2016oyh, Aguilar:2019uob, Binosi:2014kka, Aguilar:2011xe, Aguilar:2021uwa, Aguilar:2009nf}. 
%
The initial conditions used for our approximation of the scaling solution can be found in \Cref{app:YM:tuning}. The vertex dressings of the three(four)-point functions generally depend on two(three) momentum variables. To reduce the numerical effort we approximate the momentum dependencies of the dressings to one variable, which we chose to be the average momentum $\bar{p}$ at the symmetric point, see \labelcref{eq:YM:avmomconfig}. The symbolic equations for these quantities can be found in figure \Cref{fig:YM:flowequations} in the appendix and were derived with \textit{QMeS}-Derivation \cite{Pawlowski:2021tkk, github:QMeS}. More details on the projection procedure of the equations are given in \Cref{app:YM:tensors}.

The results for the transverse correlation functions obtained from this setup are shown in \Cref{fig:YM:propdressings}, \Cref{fig:gluon_prop} and \Cref{fig:YM:couplings}. The vertex dressings are depicted in \Cref{fig:YM:longcouplingsfrg} and \Cref{fig:fourgluon}. We adapted our scales to that of the lattice data \cite{Sternbeck:2006cg}, and more details regarding the scale setting and global normalisation can be found in \Cref{app:YM:scalesetting}.

The couplings are degenerate for scales $p \gtrsim 2$\,GeV in agreement with the perturbative STIs. In this region, the correlation functions from the scaling and decoupling solution also agree very well with the lattice result. At smaller momenta, the results are not comparable due to the different gauge fixing procedures on the lattice and in functional computations, see e.g.~\cite{Maas:2019ggf, Li:2021wol}, as well as the increasing systematic error of the functional computations for momenta $p\ll 1$\,GeV in the deep infrared. Beyond the deep infrared regime, we find that the scaling solution matches the lattice results at intermediate scales $p\approx1\,$GeV much better than the decoupling solution obtained from the fRG, see \Cref{fig:gluon_dressing} (dressing) and \Cref{fig:gluon_prop} (propagator). This is to be expected, as the approximation used here is more amiable to the scaling solution. In turn, the decoupling solutions in the present fRG approximation lack  the mechanism of dynamical mass generation leading to the necessary irregularities in the vertices, see \cite{Cyrol:2016tym}. 

A comparison of the transverse dressings from the fRG obtained from our setup to vertex dressings from lattice computations \cite{Cucchieri:2006tf, Cucchieri:2008qm, Maas:2016pr} and to DSE dressings from \cite{Huber:2020keu} are depicted in \Cref{app:YM:FRGYM} in \Cref{fig:YM:latticedressings}, \Cref{fig:YM:dsepropdressings},  \Cref{fig:YM:dseprop} and \Cref{fig:YM:dsedressings}.

\begin{figure*}[t]
	\centering
	\begin{subfigure}[t]{0.45\textwidth}
		\centering
		\includegraphics[width=\linewidth]{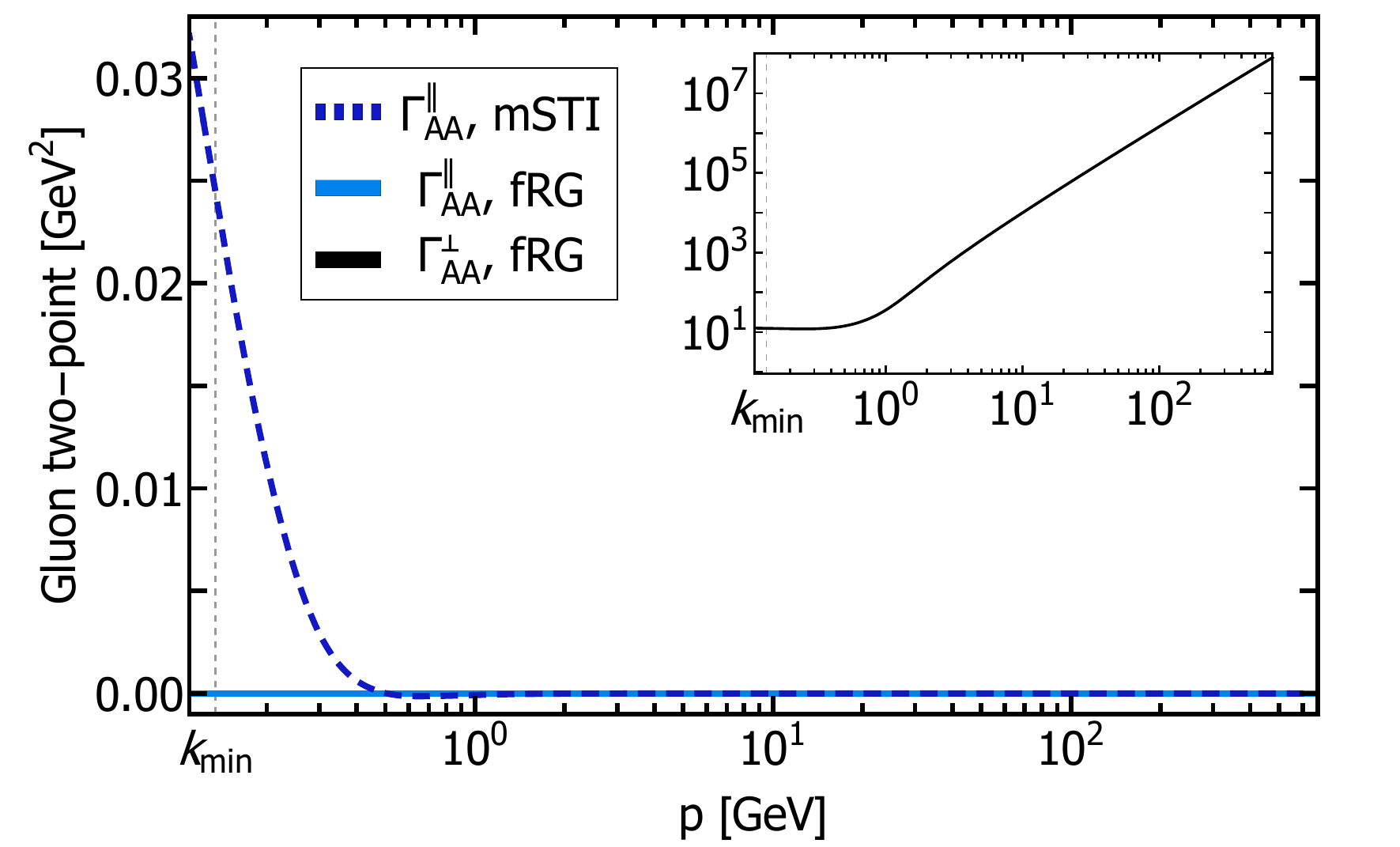}
		\caption{Gluon two-point function.}
		\label{fig:gluon_two-point}
	\end{subfigure}
	\hspace{0.05\textwidth}
	\begin{subfigure}[t]{0.45\textwidth}
		\centering
		\includegraphics[width=\linewidth]{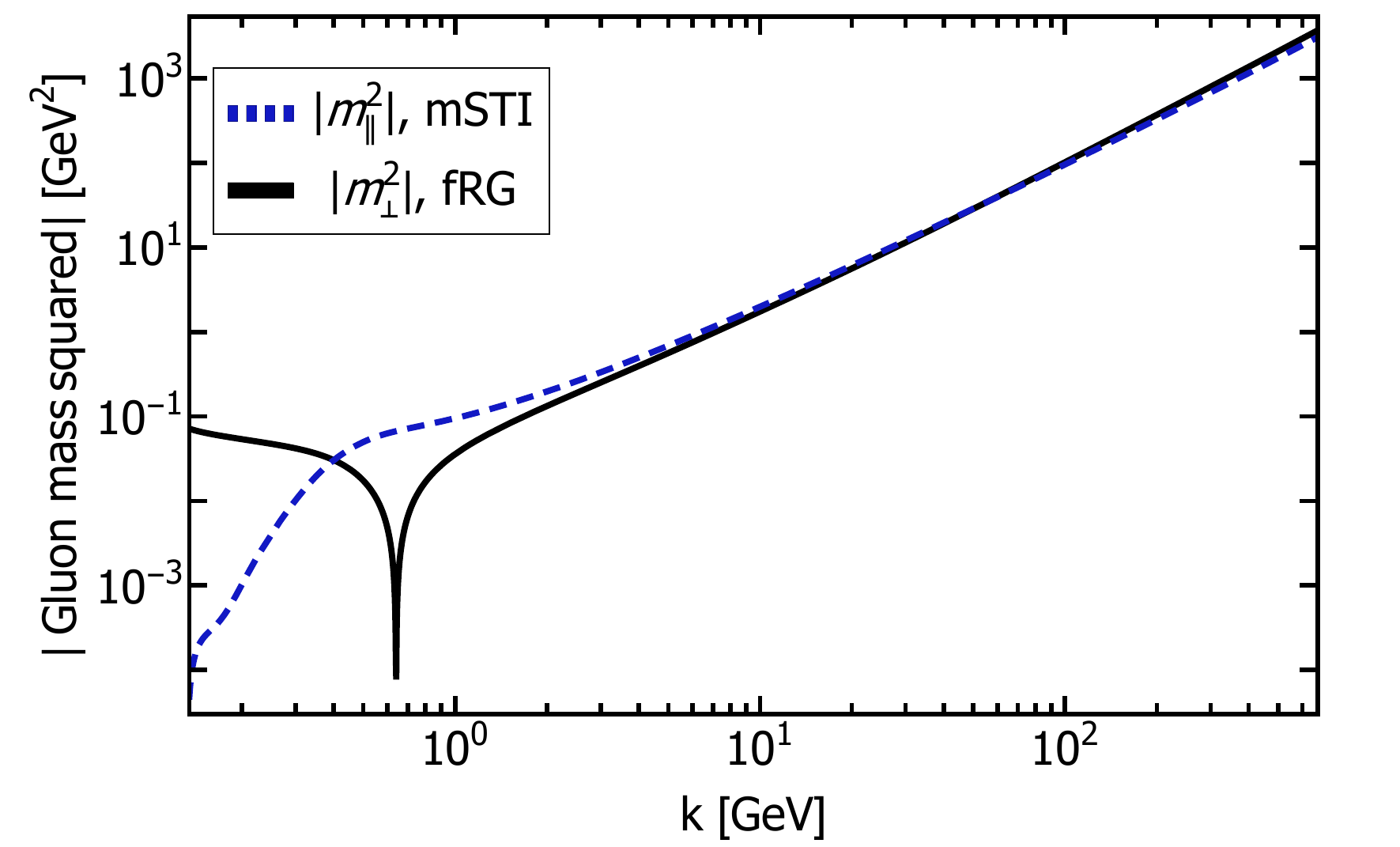}
		\caption{Absolute value of effective gluon mass squared.}
		\label{fig:gluon_mass_flow}
	\end{subfigure}%
	\caption{The gluon two-point function (\labelcref{fig:gluon_two-point}) as a function of momentum $p$ and the corresponding effective mass (\labelcref{fig:gluon_mass_flow}) as a function of RG-scale $k$. Results from the fRG are shown as solid lines, while mSTI results are shown as dashed lines. The transverse component is black and the longitudinal ones are depicted in blue. Please note that the transverse two-point function is only displayed in the inset. \hspace*{\fill}}
	\label{fig:YM:gluonlongmassplot}
\end{figure*}
%
%%%%%%%%%%%%%%%%%%%%%%%%%%%%%%%%%%%%%%%%%%%%%%%%%%%%%%%%%%%%%%%%%%%%%
\subsection{Longitudinal \& BRST correlation functions}
The BRST symmetric Yang-Mills action additionally contains source terms $Q_i$ coupled to the respective BRST transformations of the fields. These vertices are fully dressed, see \Cref{sec:YM:VertexExpansion}, and this dressing carries the non-trivial deformation of the classical BRST transformations in the quantised theory. Then (quantum) BRST symmetry, and thus the mSTIs, relate the longitudinal parts of vertex functions to themselves and the transverse parts, see \labelcref{eq:YM:mSTISpace}.

Evidently, for a self-consistent check of the (m)STIs we also need to compute the longitudinal and BRST dressings from the fRG. The symbolic equations were also derived using \textit{QMeS-Derivation} and are depicted in \Cref{fig:YM:flowequations} in the \Cref{app:YM:Diags}. Again, the average momentum approximation at the symmetric point \labelcref{eq:YM:avmomconfig} was used to simplify the momentum-dependence of the vertex dressings.

Diagrammatically, the fRG equations of the transverse and longitudinal quantities are identical, the vertices contributing in the diagrams are however different. An illustrative and important example for these differences is given by the transverse and longitudinal gluon mass. In \Cref{app:YM:twopointmSTI} we derive the transverse and longitudinal one-loop effective gluon mass from the fRG and mSTI.

As a first application we discuss the three-gluon dressing with one longitudinal leg. It vanishes at the symmetric point,
\begin{align}
\lambda_{A^3,1} \propto \mathcal{P}^{abc}_{A^3,1,\mu\nu\rho}(p,q) \Gamma_{A^3,\mu\nu\rho}^{abc}(p,q) &= 0\,, 
\label{eq:YM:AAA1nonrunning}
\end{align}
rendering this dressing RG-invariant, $\dot{\lambda}_{A^3,1}  = 0$. We are led to 
\begin{align}
\lambda_{A^3,1,k}(p)\equiv \lambda_{A^3,1,\Lambda} \equiv\lambda_{A^3,\Lambda}\,.
\end{align}
The same holds true for the three-gluon dressing with three longitudinal legs.

We use the same initial values for the longitudinal dressings as for the transverse ones. As the longitudinal gluon two-point function does not feed back into the flows, its initial value is chosen such that $\Gamma^{\parallel}_{AA,k_{\mathrm{min}}}(p) = 0$. 

The fine-tuning procedure described in \Cref{sec:YM:transverse} also applies to the longitudinal sector: the initial values of the longitudinal dressings and of the longitudinal gluon mass at the cutoff have to be tuned such that they satisfy the respective STIs perturbatively at $k_{\text{min}}$. In this work, we refrain from tuning the longitudinal sector and rather choose regular vertices, where the initial values of the respective longitudinal and transverse vertex dressings agree. We thus expect a small deviation of the STIs for perturbative momenta. This deviation is directly related to the functional nature of the the mSTIs and the fRG equations that are solved within a truncation. 

Longitudinal fRG equations are coupled to the transverse sector and to themselves, see \Cref{sec:YM:FunRel}. In contrast to those in the transverse sector, the equations decouple hierarchically due to the non-running of the three-gluon vertex dressing with one longitudinal leg. Hence we can solve them successively. The projection procedure for the derivation of longitudinal fRG equations can be found in \Cref{app:YM:tensors}. 

The longitudinal system is solved self-consistently, the input being the transverse dressings. Moreover, we compute the longitudinal four-gluon dressings from the STI \labelcref{eq:YM:4glSTI} as these relations are computationally far simpler accessible. They are depicted in \Cref{fig:fourgluon}. The results for the longitudinal fRG dressings are shown in \Cref{fig:YM:longcouplingsfrg} and \Cref{fig:YM:gluonlongmassplot}. A discussion is postponed to \Cref{sec:YM:Discussion}.

We also compute the BRST dressings from the fRG. Here, shift symmetry allows for the identification of the $Z_{cQ_A}$, $\lambda_{AcQ_A}$  and $\lambda_{AcQ_A,1}$ with the ghost and transverse and longitudinal ghost-gluon dressing, see \Cref{app:YM:shiftsym}, thus one only has to compute the remaining dressing, $\lambda_{Q_ccc}$. Furthermore, we show in \Cref{sec:YM:AccmSTIsection} that this dressing is also related to the longitudinal ghost-gluon dressing.

%%%%%%%%%%%%%%%%%%%%%%%%%%%%%%%%%%%%%
\subsection{Modified Slavnov-Taylor Identitites}

In this section we insert the fRG results into the (m)STI equations and recompute longitudinal correlation functions. Hence, we solve the STIs within our truncation, which offers a non-trivial check of gauge consistency of our solution. A discussion and comparison of the correlation functions obtained from the fRG and from the mSTI can be found in \Cref{sec:YM:Discussion} and a comparison of thereof for different types of solutions is given in \Cref{sec:YM:Discussiondec}.

For the derivation procedure and the diagrammatic contributions, see \Cref{app:YM:mSTI}. All (m)STIs are evaluated at the symmetric point \labelcref{eq:YM:symmPoint} and the dressings are computed on the average momentum configuration \labelcref{eq:YM:avmomconfig}.

Numerically, it is not feasible to integrate the equations down to $k=0$, and we keep a small, yet finite, cutoff $k_{\text{min}}=0.135\,$GeV. Consequently, we do not check the gauge consistency of our results with the STIs but rather with the mSTIs at $k=k_{\text{min}}$. In the following we will refer to expressions of the general form 
\begin{align}
	\label{eq:shortmSTI}
\textrm{STI $-$ Diagrams} \overset{!}{=} 0\,, 
\end{align}
as mSTIs. In \labelcref{eq:shortmSTI}, Diagrams stands for the one-loop diagrammatic corrections of the STI due to the BRST symmetry breaking cutoff terms \labelcref{eq:YM:mSTI}. The symbolic equations are depicted in \Cref{app:YM:Diags} in \Cref{fig:YM:mSTIequations}. 

As briefly discussed before, the mSTIs carry the underlying gauge invariance of the theory. Hence, the offer non-trivial gauge consistency checks of the correlation functions of a specific solution (e.g. scaling, decoupling or Higgs-type) of the theory obtained numerically within a truncation. Deviations of the mSTIs can be traced back to three sources: numerical precision, the truncation, and the chosen solution. 

The first can be further split into infrared cutoff effects, that might break the mSTIs for momenta around the IR cutoff scale, $p \approx k_{\text{min}}$ and overall numerical precision. Accordingly, a quantitative check of the mSTIs is only possible at momentum scales, that are at least one order of magnitude larger than the cutoff $k_{\text{min}}$. 

Truncation artefacts manifest themselves in a twofold way: effects related to dropped vertices and sub-leading tensor structures should only result in a deviation in the infrared and the semi-perturbative regime, $p\ll 5\,$GeV. 

Nonetheless, even the most sophisticated truncation in the fRG does not lead to correlation functions that fully satisfy the mSTIs: both constitute different resummations and can only agree within approximations that lead to the same full resummations in both classes of functional equations. Thus, one expects an overall deviation of the mSTIs on all momentum scales. Lastly, deviations in the mSTIs might occur due to the chosen solution of Yang-Mills theory, i.e. the scaling, the decoupling or a Higgs-type solution. Generally, we expect the mSTIs to be best fulfilled by the true scaling solution. Moving away from that (along the x-axis in \Cref{fig:gluon_mass_tuning}) deviations of the mSTIs should occur at larger and larger momentum scales.

This entails, that an interpretation of the mSTI results is a difficult task as a clear disentanglement of all of the mentioned sources necessitates further computations of the different types of solutions within different truncations, and with different IR cutoffs and numerical precision.

%%%%%%%%%%%%%%%%%%%%%%%%%
\subsubsection{Gluon Two-Point and Gluon Mass STI}
\begin{figure}[t]
	\includegraphics[width=0.98\linewidth]{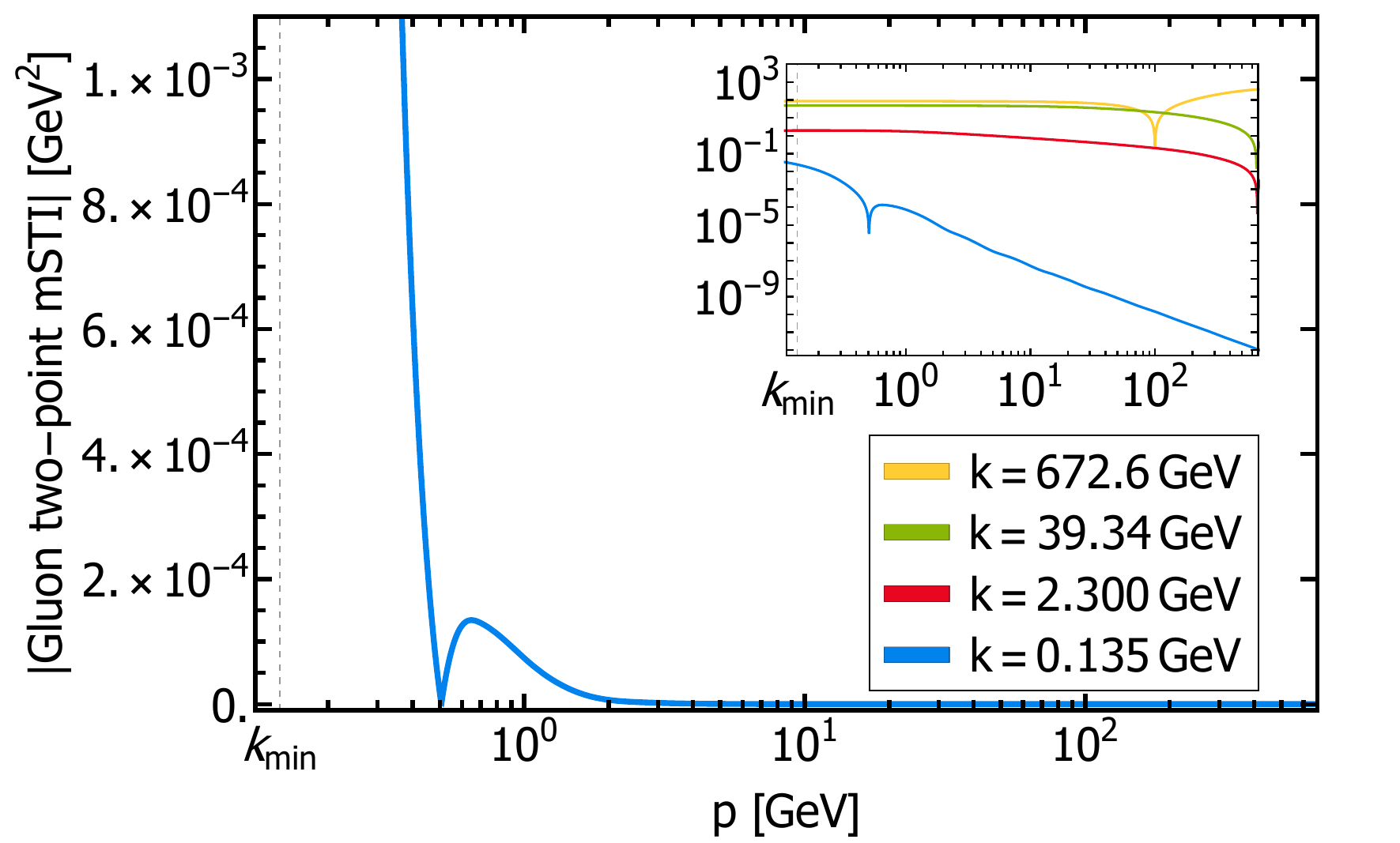}
	\caption{Absolute value of the gluon two-point mSTI as a function of momentum $p$ for different values of $k$.\hspace*{\fill}}
	\label{fig:gluon_two_point_mSTI}
\end{figure}

%%%%%%%%%%%%%%%%%%%%%%%%%
The STI for the longitudinal gluon two-point function is simply given by vanishing quantum corrections of the longitudinal two-point function,
\begin{align}\label{eq:STI2gluon}
\Gamma^{\parallel}_{AA,\text{reg}} (p) = 0\,, 
\end{align}
where the full longitudinal gluon two-point function consists out of the loop contribution $\Gamma^{\parallel}_{AA,\text{reg}}$ and the gauge fixing, see \labelcref{eq:Gammaparal}. This also implies a vanishing longitudinal gluon mass,
\begin{align}
m_\parallel^2 = \Gamma^{\parallel}_{AA} (p =0) = 0\,.
\label{eq:YM:massSTI}
\end{align}
Both, the full longitudinal gluon two-point function $ \Gamma^{\parallel}_{AA} (p)$ and the longitudinal gluon mass $m_\parallel^2 $ are depicted in \Cref{fig:YM:gluonlongmassplot}. In \Cref{fig:gluon_two_point_mSTI} we depict the the deviation of the two-point function mSTI as defined in \labelcref{eq:shortmSTI}, the cutoff dependence of transverse (fRG) and longitudinal (mSTI) masses is depicted in \Cref{fig:YM:gluonlongmassplot}. Technical details can be found in \Cref{app:YM:reg}. A discussion of these findings is deferred to \Cref{sec:YM:Discussion}. 
\begin{figure*}[t]
	\centering
	\begin{subfigure}[t]{0.45\textwidth}
		\centering
		\includegraphics[width=\linewidth]{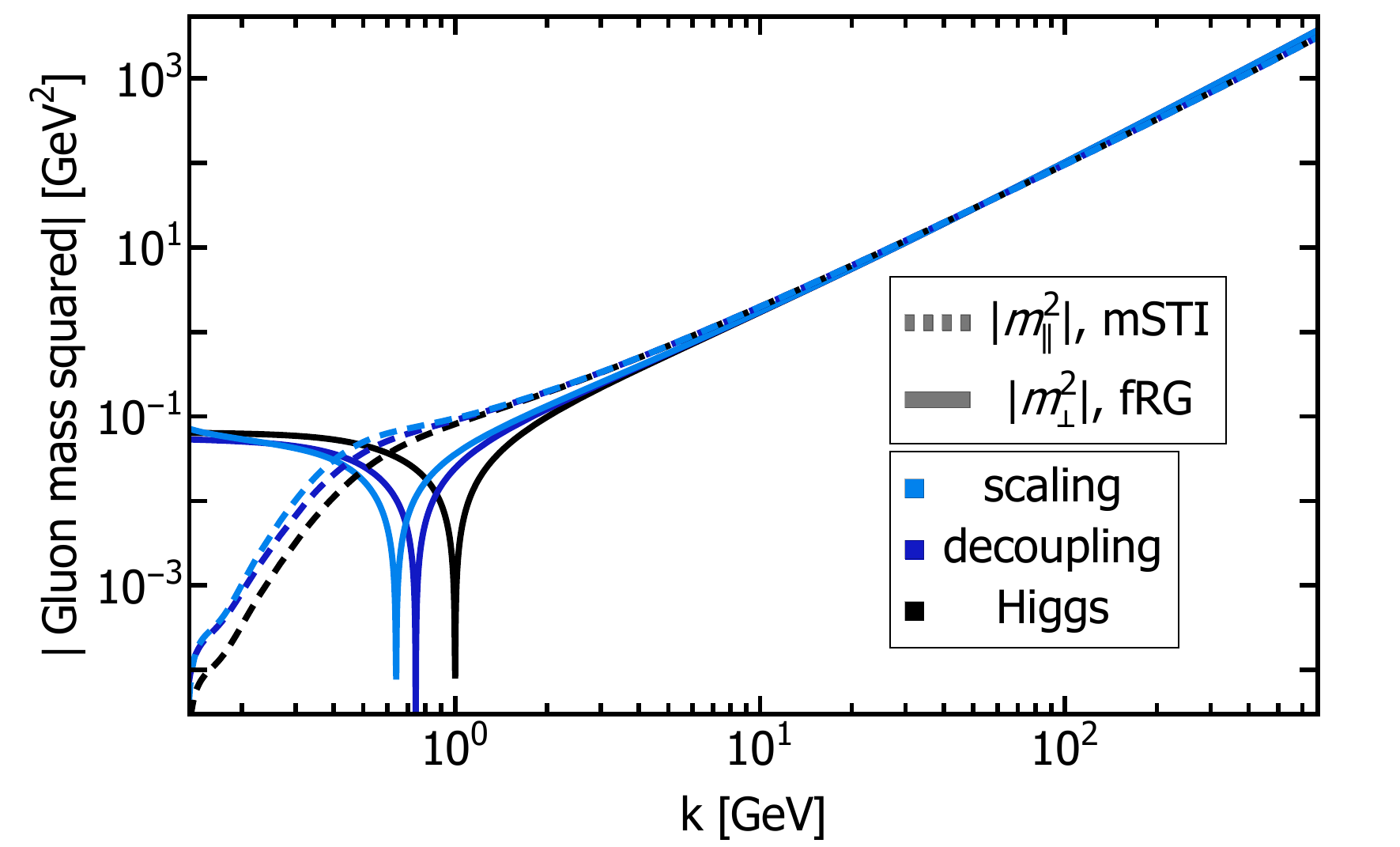}
		\caption{Absolute value of individual contributions. fRG results are shown with solid lines and mSTI results with dashed lines.\hspace*{\fill}}
		\label{fig:gluon_mass_flow_abs}
	\end{subfigure}%
	\hspace{0.05\textwidth}
	\begin{subfigure}[t]{0.45\textwidth}
		\centering
		\includegraphics[width=\linewidth]{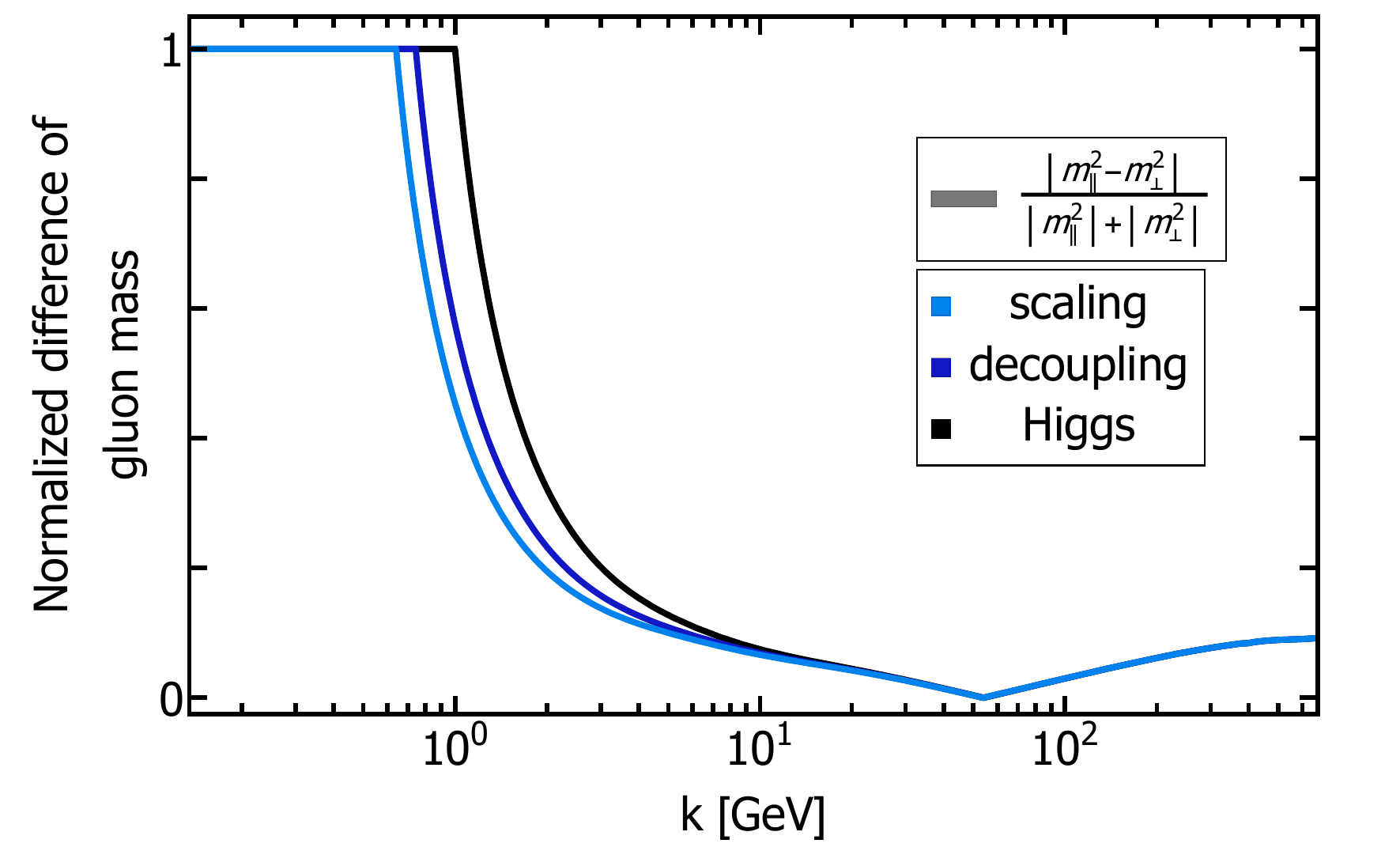}
		\caption{Normalized difference.}
		\label{fig:gluon_mass_flow_relative}
	\end{subfigure}
	\caption{Comparison of the flow of the effective transverse (fRG) and longitudinal gluon mass (mSTI) as a function of the cutoff scale $k$. The transverse part is taken from the fRG. The colors differentiate the different solutions (scaling, decoupling, Higgs-type).\hspace*{\fill}}
	\label{fig:YM:massMeasuredec}
\end{figure*}
%

%%%%%%%%%%%%%%%%%%%%%%%%%
\subsubsection{Ghost-Gluon STI}
\label{sec:YM:AccmSTIsection}
%%%%%%%%%%%%%%%%%%%%%%%%
The STI for the ghost-gluon vertex relates the longitudinal part of ghost-gluon vertex to the BRST projected vertex $Q_ccc$. In our truncation we are led to  
\begin{align}\label{eq:STIGhostGluon}
2 p_\mu \Gamma_{A\bar{c}c,\mu} (p,q) Z_c(p)+  q^2 \Gamma_{ccQ_c}(p,-p-q) Z_c(q)=0\,.
\end{align}
Thus we can derive the STI for the respective dressings at the symmetric point, 
\begin{align}
\lambda_{A \bar{c} c,1}(p) -\lambda_{Q_ccc}(p) =0\,.
\label{eq:YM:AccSTI}
\end{align}
A formulation of the ghost-gluon STI in terms of the classical and non-classical tensor basis is presented in \Cref{app:YM:AccmSTI}. The respective dressings are depicted in \Cref{fig:YM:longcouplingsfrg} and \Cref{fig:YM:accnonclcouplingsfrg}.

%%%%%%%%%%%%%%%%%%%%%%%%%
\subsubsection{Three-Gluon STI}
%%%%%%%%%%%%%%%%%%%%%%%%%

\begin{figure*}[t]
	\centering
	\begin{subfigure}[t]{0.45\textwidth}
		\centering
		\includegraphics[width=\linewidth]{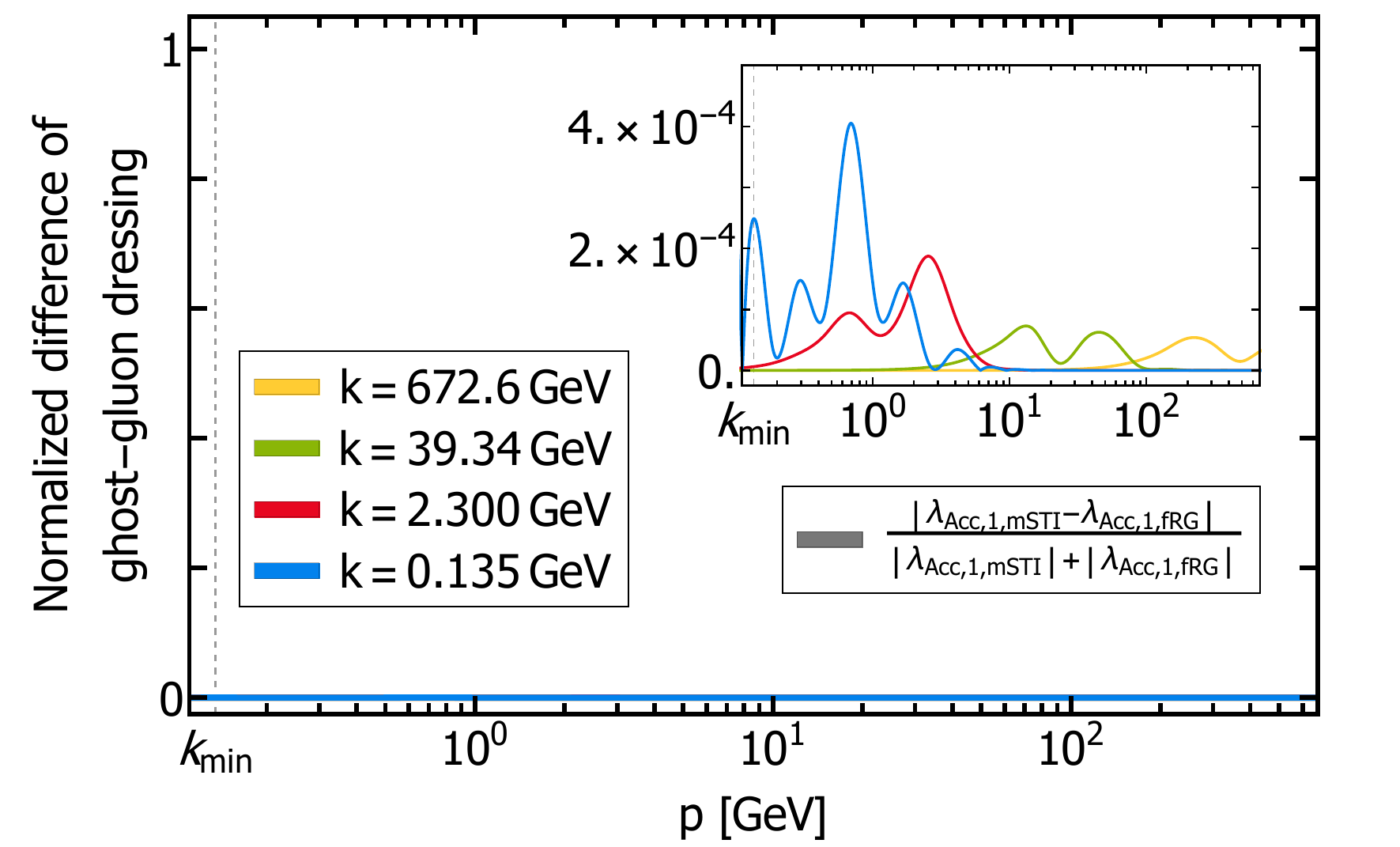}
		\caption{Ghost-gluon vertex dressing. The result is numerically compatible with zero, which becomes evident in the inset.\hspace*{\fill}}
		\label{fig:ghost_gluon_mSTI}
	\end{subfigure}%
	\hspace{0.05\textwidth}
	\begin{subfigure}[t]{0.45\textwidth}
		\centering
		\includegraphics[width=\linewidth]{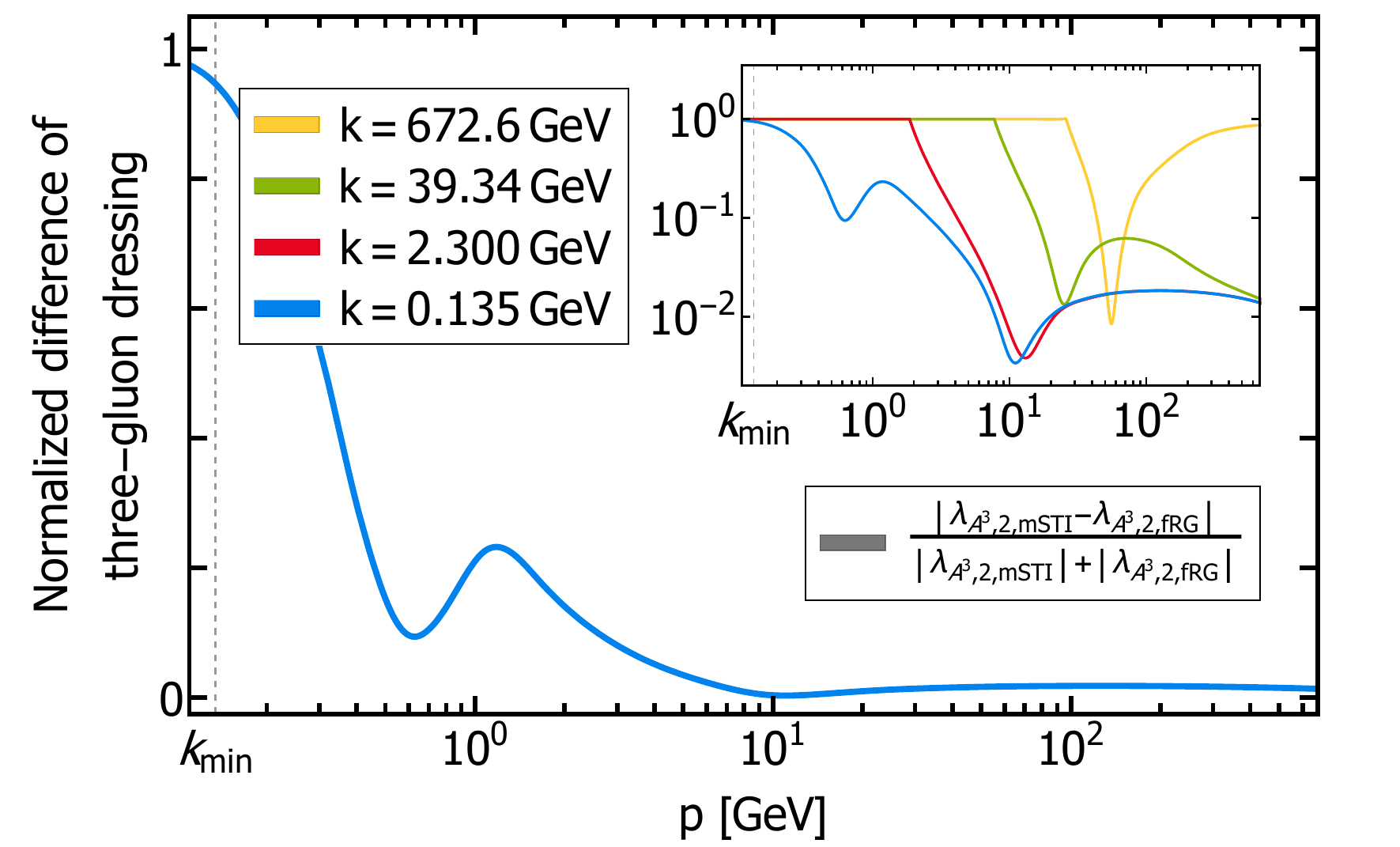}
		\caption{Three-gluon vertex dressing. The difference gets maximal in the IR, a logarithmic version is shown in the inset.\hspace*{\fill}}
		\label{fig:three_gluon_mSTI}
	\end{subfigure}
	\caption{Normalized difference of the longitudinal three-point dressings between the fRG and the mSTI at different RG-scales $k$ as a function of momentum $p$.\hspace*{\fill}}
	\label{fig:YM:ThreePointmSTIPlots}
\end{figure*}

The STI for the three-gluon vertex with two longitudinal legs at the symmetric point is given by,
\begin{align}
&\lambda_{A^3,2}(p) - \frac{Z_A(p)}{Z_c(p)} \lambda_{Q_ccc}(p) = 0\,.
\end{align}
After applying the symmetric point ghost-gluon STI in our truncation, \labelcref{eq:YM:AccSTI}, and dividing by propagator dressings, one obtains the relation for the couplings,
\begin{align}
\frac{\lambda^2_{A^3,2}(p)}{4\pi Z^3_A(p)} = \frac{\lambda^2_{A\bar{c}c,1}(p)}{4 \pi Z_A(p)Z^2_c(p)} \,.
\end{align}
Using the ghost-gluon STI to simplifly the equation is a valid approximation as is shown in \Cref{sec:YM:DiscussionAcc} and \Cref{fig:YM:ThreePointmSTIPlots}. Assuming regularity, i.e. agreement of transverse and longitudinal projections of correlation functions (at least for large momenta), the above relation is similar to the well-known perturbative relation presented in \labelcref{eq:YM:couplingsdegeneracy}.

The dressings of the different longitudinal and transverse projection of the classical tensor structure of the three gluon vertex are shown in \Cref{fig:YM:longcouplingsfrg}. We remark, that no STI is obtained for the projection with two transverse and one longitudinal leg at the symmetric point, since this projection renders zero.

%

%%%%%%%%%%%%%%%%%%%%%%%%
\subsubsection{Four-Gluon STI}
%%%%%%%%%%%%%%%%%%%%%%%%%
The four-gluon STI relates different longitudinal four-gluon dressings to longitudinal and transverse three-gluon and ghost-gluon dressings. We have again used the symmetric point ghost-gluon STI \labelcref{eq:YM:AccSTI} and arrive at the four-gluon STIs at the symmetric point. The one for the four-gluon vertex with one longitudinal leg, $\lambda_{A^4,1}(p)$, is given by, 
\begin{align}
&\lambda_{A^4,1}(p) - \frac{1}{Z_c(p)}\lambda_{A \bar{c} c}(\frac{\sqrt{10}}{3}p)\lambda_{A^3}(\frac{\sqrt{10}}{3}p) = 0\,,
\end{align}
The STI for the four-gluon vertex with two longitudinal legs, $\lambda_{A^4,2}(p)$, is given by, 
\begin{align}\nonumber 
&\lambda_{A^4,2}(p) - \frac{32}{33  Z_c(p)}\lambda_{A \bar{c} c,1}\left(\frac{\sqrt{10}}{3}p\right)\lambda_{A^3}\left(\frac{\sqrt{10}}{3}p\right)\\[1ex] 
&-\frac{1}{33  Z_c(p)}\lambda_{A \bar{c} c}\left(\frac{\sqrt{10}}{3}p\right)\lambda_{A^3,1}\left(\frac{\sqrt{10}}{3}p\right) = 0\,, 
\label{eq:YM:4glSTI}
\end{align}
It is worth mentioning that the momentum argument of the dressings in the STIs at the symmetric point are quite different from the ones usually present in the literature, as e.g.~in~\labelcref{eq:YM:couplingsdegeneracy}. Using the STI for the three-gluon dressing, the STIs for the four-gluon dressing with one and two longitudinal legs are equivalent to the form in \labelcref{eq:YM:couplingsdegeneracy} assuming regularity of vertices for large momenta.

%%%%%%%%%%%%%%%%%%%%%%%%%%%%%%%%%%%
\subsection{Discussion}\label{sec:YM:Discussion}
%%%%%%%%%%%%%%%%%%%%%%%%%%%%%%%%%%%

We now compare the longitudinal results of different correlation functions obtained from the fRG and mSTI approach by studying the normalised difference of the respective quantities.

%%%%%%%%%%%%%%%%%%%%%%%%%%%%%%%%%%%
\subsubsection{Longitudinal Gluon Mass}%%%%%%%%%%%%%%%
\label{sec:YM:DiscussionMass}%%%%%%%%%%%%

The longitudinal mSTI mass, $m^2_{\parallel}$, and the effective transverse fRG mass, $m^2_\perp$ exhibit the same running for large $k$ which is shown in \Cref{fig:YM:gluonlongmassplot}. This is in good agreement with the one-loop running, which is the same for both masses, for further details, see \Cref{app:YM:twopointmSTI}. At about cutoff scales $k\approx5\,$GeV, the two mass start deviating from each other. This is triggered by the difference in the IR of the longitudinal and transverse vertices, cf. \Cref{fig:YM:longcouplingsfrg}. 

We remark, that the values of the longitudinal mSTI and transverse fRG masses are finite and approximately agree at $k_{\mathrm{min}}$, while the longitudinal fRG mass is zero. This is due to the regularity of the vertices in the present approximation at finite cutoff. In the scaling solution, irregular vertices emerge at $k\to 0$, while the decoupling solution does not incorporate the irregular vertices required for confinement in the current approximation. The STI, \labelcref{eq:YM:massSTI}, however, renders the longitudinal gluon mass zero at $k=0$. Evidently, this deviation at least partially stems from our truncation in the vertex sector in the fRG equations, see also the discussion in \Cref{sec:YM:DiscussionAA}. To circumvent this discrepancy, we have extracted the longitudinal mSTI mass at $p=4\,k_\textrm{min}$, in order to avoid the influence of any IR cutoff effects that are present for $p\lesssim k_\textrm{min}$. For more details, see \Cref{app:YM:reg}.

%%%%%%%%%%%%%%%%%%%%%%%%%%%%%%%%%%%
\subsubsection{Longitudinal Gluon Two-Point Function}%%%%%%%%%%%%%%%
\label{sec:YM:DiscussionAA}%%%%%%%%%%%%%
%%%%%%%%%%%%%%%%%%%%%%%%%%%%%%%%%%%

The solution of the longitudinal two-point fRG equation constitutes the simplest case possible since it does not at all feed back into fRG diagrams in the Landau gauge. We also remark that the longitudinal propagator does feed back to the mSTI, but only its classical gauge fixing part. In summary this entails, that the longitudinal gluon two-point function is obtained by simply integrating the flow equation, and choosing the initial condition for $\Gamma^{\parallel}_{AA,\mu\nu,\text{reg},\Lambda}(p)$ such, that one obtains zero for the evaluated integral. This procedure not only solves the fRG equation but also satisfies the STI at $k_{\mathrm{min}}$, hence disregarding the small breaking of the STI for small momentum scales. 

By comparing the longitudinal gluon two-point function from the fRG and from the mSTI, one can see that they do not agree well on all scales $k$, see \Cref{fig:YM:gluonlongmassplot} and \Cref{fig:gluon_two_point_mSTI}. They are only identical at $k=k_{\mathrm{min}}$ for $p \gtrsim 0.5\,$GeV. For small momenta one can see a deviation of order $\mathcal{O}(10^{-2}\,\text{GeV}^2)$. Albeit small, the deviation is qualitatively similar to the deviation in the three-gluon mSTI, \Cref{fig:YM:ThreePointmSTIPlots}. Since the longitudinal three-gluon dressing enters the mSTI equation for the gluonic two-point function, but not vice versa, one can therefore conclude, that both deviations are due to the truncation of the fRG equation for the longitudinal three-gluon and four-gluon dressing.

\begin{figure*}[t]
	\centering
	\begin{subfigure}[t]{0.45\textwidth}
		\centering
		\includegraphics[width=\linewidth]{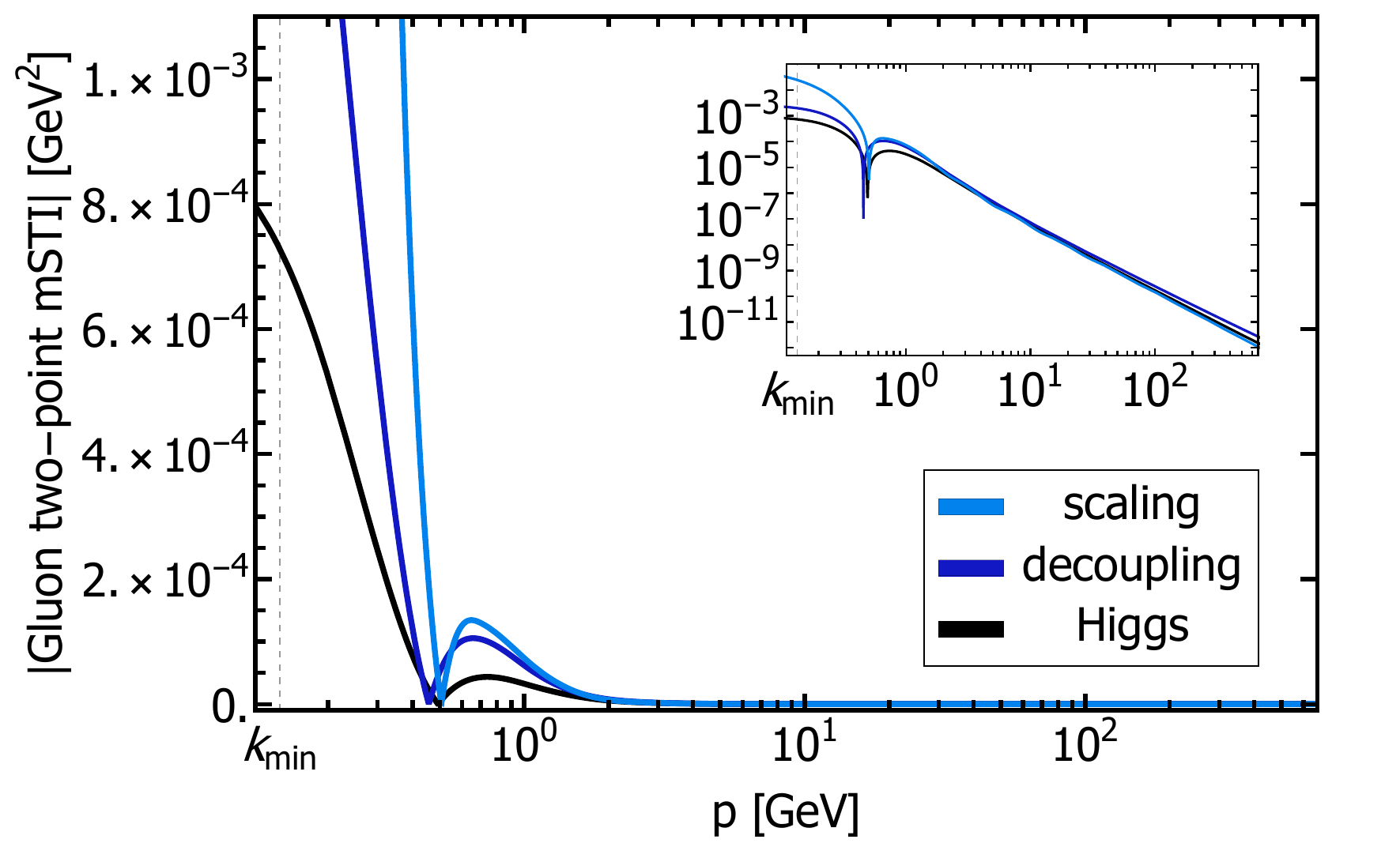}
		\caption{Absolute value of gluon two-point STI. The inset shows a logarithmic scale.\hspace*{\fill}}
		\label{fig:gluon_two-point_mSTI}
	\end{subfigure}%
	\hspace{0.05\textwidth}
	\begin{subfigure}[t]{0.45\textwidth}
		\centering
		\includegraphics[width=\linewidth]{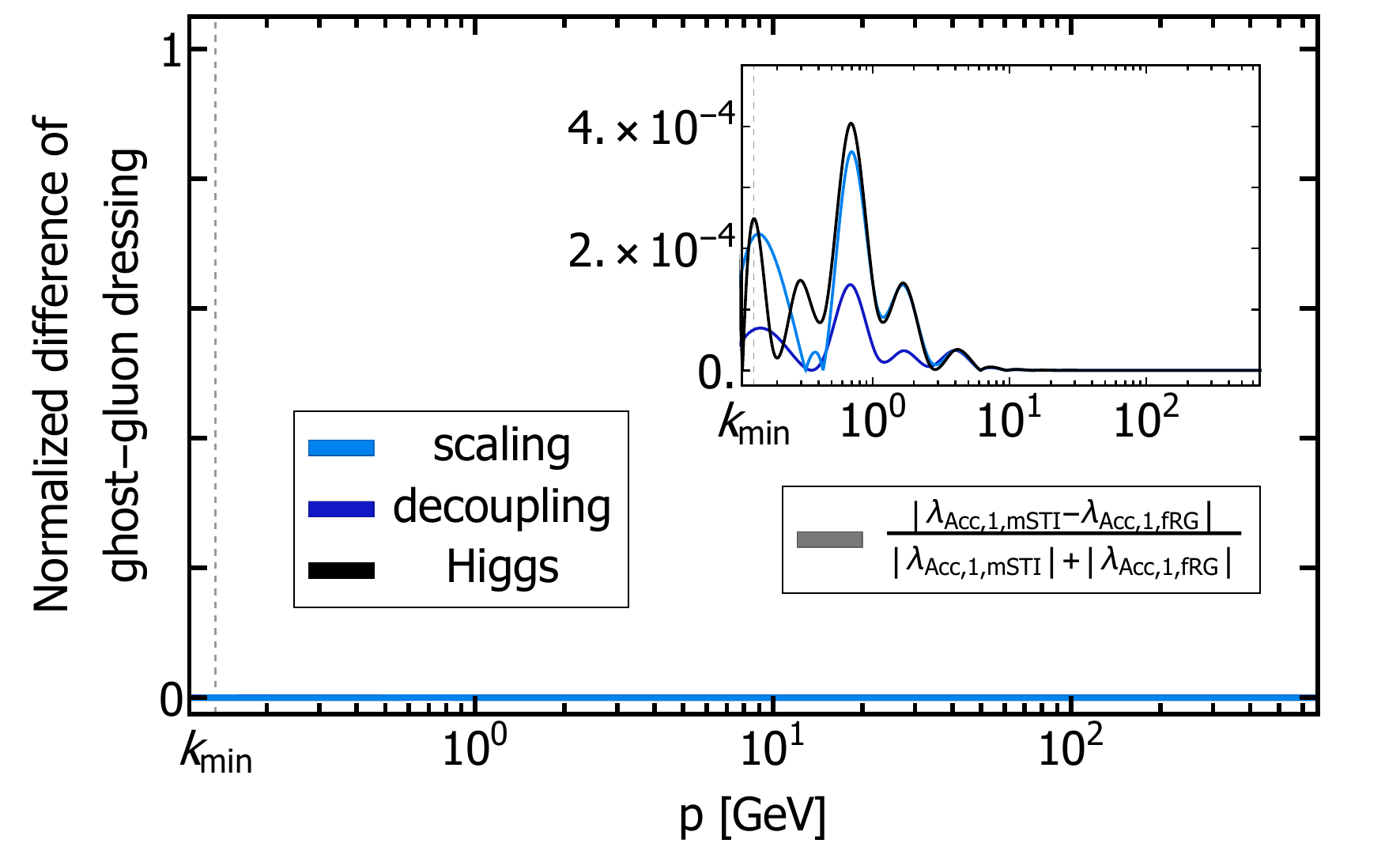}
		\caption{Normalized difference of the ghost-gluon STI. The result is numerically compatible with zero, which is evident from the inset, where the full data range is displayed.\hspace*{\fill}}
		\label{fig:ghost_gluon_mSTI_relative}
	\end{subfigure}
	\caption{Results for the STI of the gluon two-point and the ghost-gluon vertex dressing as a function of momentum $p$. The colors differentiate IR scenarios.\hspace*{\fill}}
	\label{fig:YM:MassPlot}
\end{figure*}
%

%%%%%%%%%%%%%%%%%%%%%%%%%%%%%%%%%%%
\subsubsection{Longitudinal Ghost-Gluon Dressing}%%%%%%%%%%%%%%%
\label{sec:YM:DiscussionAcc}%%%%%%%%%%%%%
%%%%%%%%%%%%%%%%%%%%%%%%%%%%%%%%%%%

In a general covariant gauge, only $\lambda_{A \bar{c} c}$ requires renormalisation. The non-classical dressing $\lambda_{A \bar{c} c, ncl}$ vanishes for large momenta. The non-classical dressing obtained from the fRG and mSTI is depicted in \Cref{fig:YM:accnonclcouplingsfrg}.

We can see that our results fulfil this property perturbatively since the non-classical dressing is approximately zero for large momenta, $p\gg 5\,$GeV. In the infrared, it is however non-trivial. It is worth mentioning that the longitudinal and transverse ghost-gluon dressings do not agree well at any momentum scale $p$.

From  \Cref{fig:YM:longcouplingsfrg},  \Cref{fig:YM:ThreePointmSTIPlots}, and \Cref{fig:YM:accnonclcouplingsfrg} in the appendix, we can see that the mSTI and fRG dressing agree perfectly for all momenta $p$ implying that there is a non-trivial cancellation of diagrams within the two functional approaches. Indeed, \Cref{fig:YM:ThreePointmSTIPlots} merely depicts the numerical accuracy of our computation.

%The (m)STI already holds true at the level of fRG equations, c.f. \Cref{fig:YM:flowequations}: The diagrams contributing to the flow of the longitudinal ghost-gluon dressing that contain three-gluon vertices cancel non-trivially. The remaining diagrams are equivalent if the ghost-gluon STI is fulfilled at the cutoff $k=\Lambda$. 

Generally, this non-trivial result implies BRST symmetry being conserved by the respective fRG computation and thus strongly hints at gauge consistency of our setup.

%%%%%%%%%%%%%%%%%%%%%%%%%%%%%%%%%%%
\subsubsection{Longitudinal Three-Gluon Dressing}%%%%%%%%%%%%%%%
\label{sec:YM:DiscussionAAA}%%%%%%%%%%%%%
%%%%%%%%%%%%%%%%%%%%%%%%%%%%%%%%%%%
The results for the different three-gluon dressings are shown in \Cref{fig:YM:longcouplingsfrg}. As demonstrated in \labelcref{eq:YM:AAA1nonrunning} the three-gluon dressing with one longitudinal leg, $\lambda_{A^3,1}$, does not run. One can see that the dressing with two longitudinal legs from the fRG agrees with the transverse fRG dressing, $\lambda_{A^3}$, and the longitudinal STI dressing for large momenta $p\gg 5\,$GeV. The longitudinal STI dressing and the transverse fRG dressing however agree up to even smaller scales $p\gg 1.5\,$GeV. 

In the infrared one can observe that the general behaviour of the longitudinal fRG and STI dressings agree. The STI dressing diverges for $p\rightarrow 0$ whereas the fRG dressing forms a maximum at $p \approx 0.2\,$GeV. The minimum of both dressings is given at approximately the same momentum, $p\approx 1\,$GeV.

In contrast to that, the transverse dressing gets smaller and even becomes negative for $p\rightarrow 0$. Thus we can observe a clear splitting between the transverse and longitudinal sector in both approaches which is due to the different contributions of longitudinal and transverse vertices in transverse and longitudinal functional equations. 

The slight deviation in the UV can be explained by the logarithmic finetuning procedure: we have chosen the same initial values for the longitudinal and transverse dressings. The contributions on the level of correlation functions to the flows are however different. By choosing different (constant) initial values for the longitudinal dressings at the cutoff it is nevertheless possible to generate a better agreement of the dressings in the UV at $k=k_{\mathrm{min}}$.

As stated in the previous section, the three-gluon mSTI shows a deviation at $p\approx5\,$GeV. This might be due to the fact that we did not include a full tensor basis of the gluonic sector and even approximated the longitudinal four-gluon dressings with their STI values in the fRG equation of the three-gluon vertex.

For momenta $p>5\,$GeV, the three-gluon mSTI is however fulfilled, where the small deviation for very large momenta stems from the tuning procedure, as described before.
\begin{figure*}[t]
	\centering
	\begin{subfigure}[t]{0.45\textwidth}
		\centering
		\includegraphics[width=\linewidth]{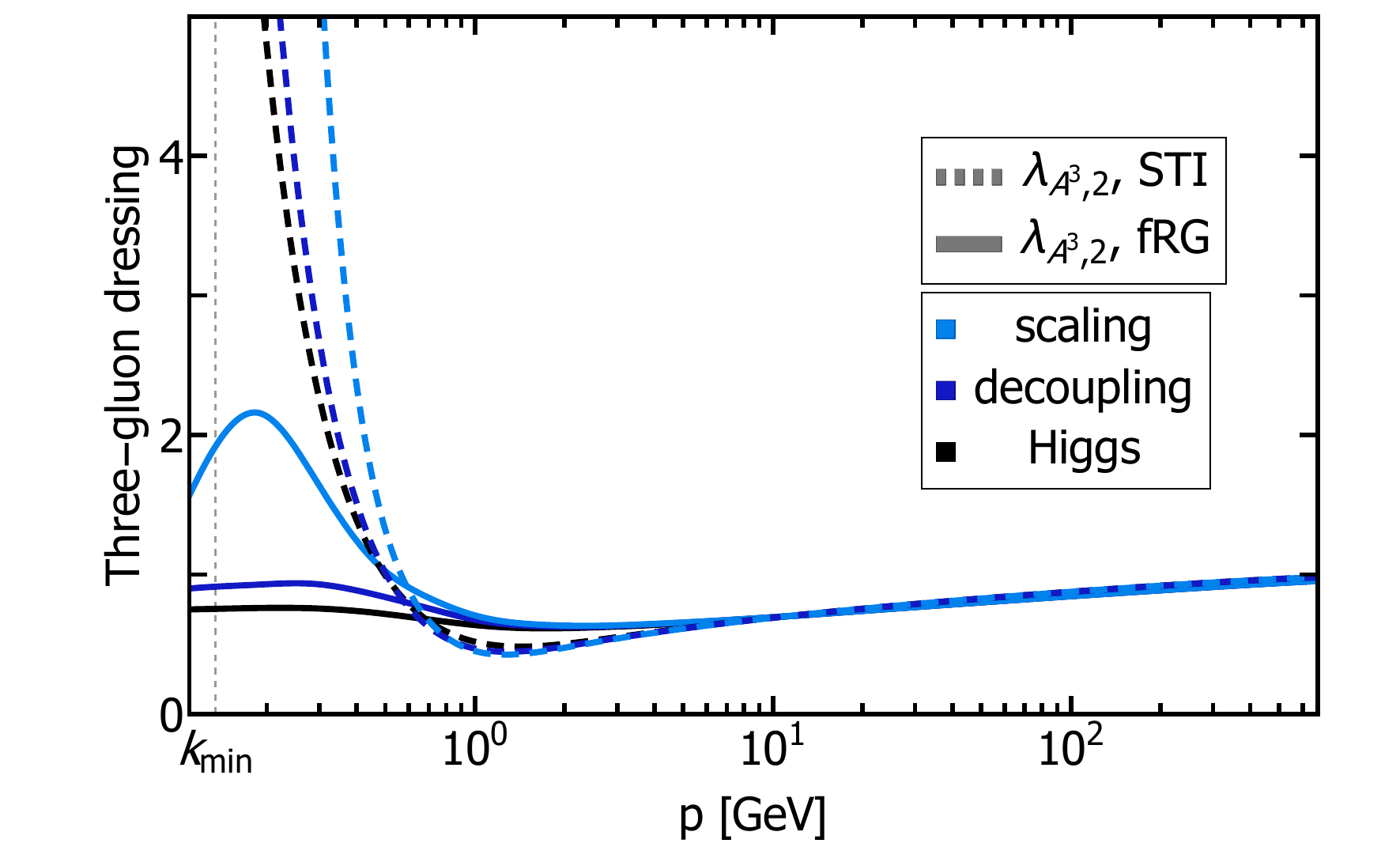}
		\caption{Absolute value of individual contributions: fRG (solid lines), STI (dashed lines). \hspace*{\fill}}
		\label{fig:three_gluon_msti_abs}
	\end{subfigure}%
	\hspace{0.05\textwidth}
	\begin{subfigure}[t]{0.45\textwidth}
		\centering
		\includegraphics[width=\linewidth]{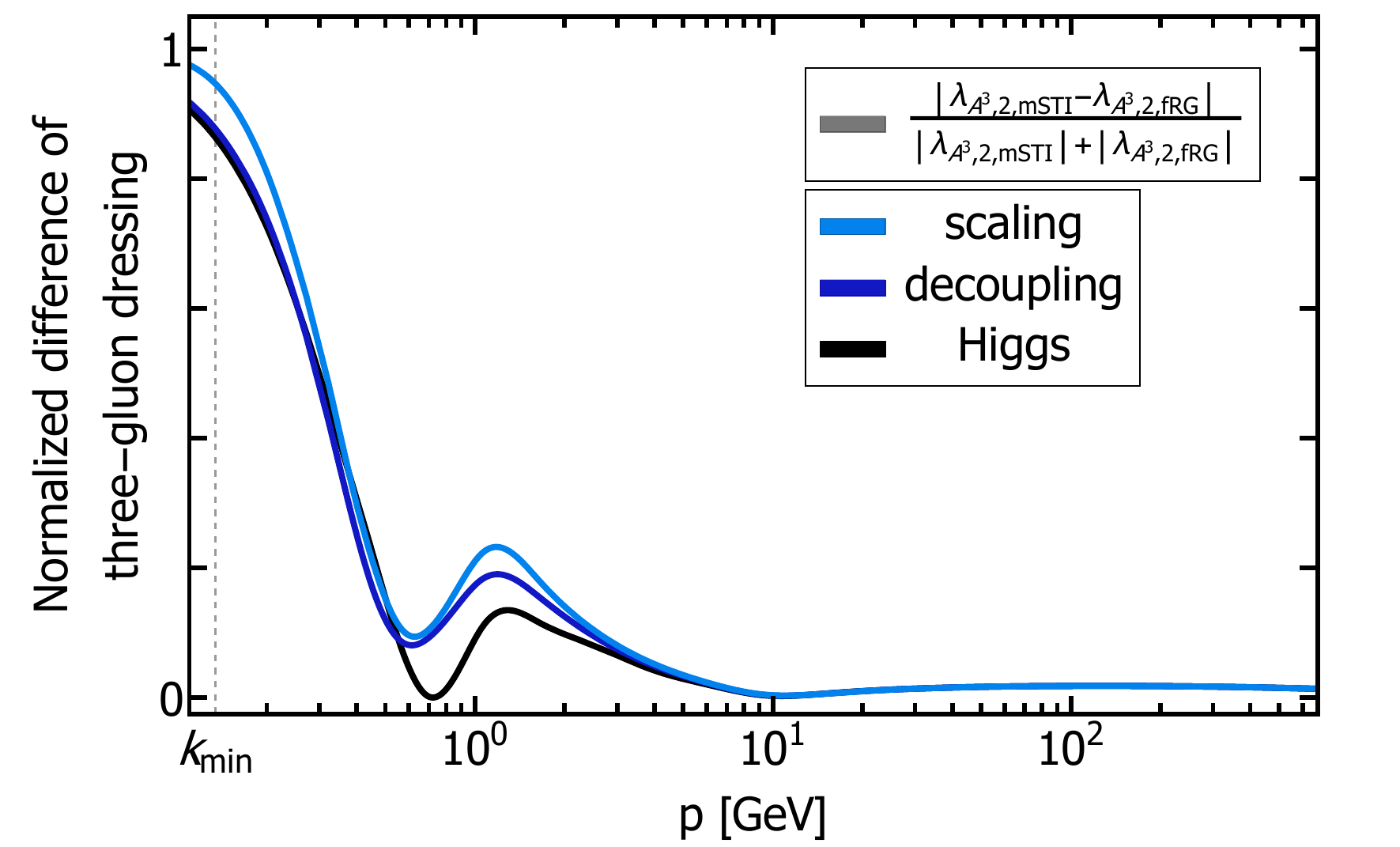}
		\caption{Normalized difference.}
		\label{fig:three_gluon_mSTI_relative}
	\end{subfigure}
	\caption{Results for the longitudinal three-gluon dressing $\lambda_{A^3,2}(p)$ as a function of the momentum $p$. The colors differentiate IR scenarios.\hspace*{\fill}}
	\label{fig:YM:theree_gluon_msti_rel}
\end{figure*}
%

%%%%%%%%%%%%%%%%%%%%%%%%%%%%%%%%%%%
\subsection{Discussion of Different Solutions}%%%%%%%%%%%%%%%
\label{sec:YM:Discussiondec}%%%%%%%%%%%%%
%%%%%%%%%%%%%%%%%%%%%%%%%%%%%%%%%%%

In this section, we present a comparison of the mSTIs for different types of solutions, i.e. the scaling, decoupling, and a Higgs-type solution, that are highlighted amongst the range of solutions in \Cref{fig:gluon_mass_tuning}.

Studying the normalized difference of the transverse andf longitudinal gluon mass, see \Cref{fig:YM:massMeasuredec}, where the longitudinal mass is obtained from the mSTI and the effective transverse mass from the fRG, one can see that all three solutions agree equally well up until $p \approx 20\,$GeV. There, the Higgs-type solution starts to deviate. It is expected that for a more massive Higgs-type solution, the deviation starts at an even larger momentum scale $p$. Vice versa, the scaling solution starts to deviate at a smaller scale than the decoupling solution. 

The ghost-gluon mSTI is fulfilled for all three types of solutions, the small deviations shown in \Cref{fig:YM:MassPlot} simply depict the numerical precision in the computation.

Generally, we do not expect the mSTIs to be fulfilled for scales $p \ll 1\,$GeV, since non-classical vertices and tensor structures that were not taken into account in our truncation, contribute significantly in this regime. We furthermore expect IR cutoff effects contributing below this scale, being within one order of magnitude of $k_{\mathrm{min}}$.

However, we can clearly see the expected divergence in the gluon two-point, \Cref{fig:YM:MassPlot}, and three-point mSTI, \Cref{fig:YM:theree_gluon_msti_rel}, for momenta $p \ll 1\,$GeV.

Comparing the longitudinal three-gluon dressing  $\lambda_{A^3,2}$ from the fRG and from the STI for different types of solutions, one can see that the scaling solution yields the best qualitative agreement of the dressings.

%%%%%%%%%%%%%%%%%%%%%%%%%%%%%%%%%%%
\section{Conclusion}%%%%%%%%%%%%%%%
\label{sec:YM:Conclusion}%%%%%%%%%%%%%
%%%%%%%%%%%%%%%%%%%%%%%%%%%%%%%%%%%
We studied the gauge consistency of functional approaches to Yang-Mills within the Landau gauge. For this purpose, we resolved the transverse and longitudinal sector by solving the corresponding fRG flow equations for the dressings of transverse and longitudinal propagators and vertices in a sufficiently advanced truncation. To quantify the violation of gauge consistency in such a set-up, we also computed longitudinal dressings with the associated modified Slavnov-Taylor identities. The results agree within numerical accuracy for momentum scale $p\gtrsim  1\,\mathrm{GeV}$. We found that the agreement of STI dressings and fRG dressings differ for the solutions branches, i.e. for scaling, decoupling and Higgs type solutions: in general, the scaling solution shows the best agreement. In turn,  the longitudinal fRG dressings for decoupling and Higgs-type solutions start to deviate from the ST dressings for successively larger momenta, the farer away they are from scaling. Interestingly, the longitudinal three-gluon vertex dressing is an exception, as there the deviation of fRG and STI dressings happens at roughly the same scale. This structure deserves further investigation, one of the reasons being that the STIs used in the present work are the standard Landau gauge STIs and the decoupling and Higgs-type solutions triggered here are effectively induced by an explicit mass parameter as in the Curci-Ferrari (CF) model. This suggests to repeat the present investigation on the basis of the CF BRST transformations.

In summary, these comparisons provide us with a tool to control the gauge consistency in truncations of the quantum effective action. In our opinion, the present level of gauge consistency supports the reliability of such truncations in applications of non-perturbative functional methods to Yang-Mills theory.

Monitoring the mSTIs in such a manner provides additional, much needed, guidance to improve on truncations. Alternatively, they can be used as constraints to determine correlation functions. 

This work provides the foundation for an advanced study thereof in QCD, where the gauge consistency of the coupled self-consistent quark-gluon system poses an intricate problem. We hope to report on this in the near future.

%%%%%%%%%%%%%%%%%%%%%%%%%%
\section*{Acknowledgments}
%%%%%%%%%%%%%%%%%%%%%%%%%%
 We thank G.~Eichmann, J.~Horak, J.~Papavassiliou and U.~Reinosa for discussions. This work is done within the fQCD-collaboration~\cite{fQCD} and we thank the members for discussion and collaborations on related projects. This work is supported by EMMI, and is part of
and supported by the DFG Collaborative Research Centre "SFB 1225 (ISOQUANT)". It is also supported by Germany’s Excellence Strategy EXC 2181/1 - 390900948 (the Heidelberg STRUCTURES Excellence Cluster).
N.~W.~is additionally supported by the Hessian collaborative research cluster ELEMENTS and by the DFG Collaborative Research Centre "CRC-TR 211 (Strong-interaction matter under extreme conditions)".

%%%%%%%%%%%%%%%%%%%%%%%%%%%%%%%%%%%%%%%%%%%%%%%%%%%%%%%%%%%%%%%%%%%%%%%%
\appendix
%%%%%%%%%%%%%%%%%%%%%%%%%%%%%%%%%%%
\section{Numerical Implementation}
\label{app:YM:Implementation}%%%%%%%%%%%%%
%%%%%%%%%%%%%%%%%%%%%%%%%%%%%%%%%%%
The fRG and mSTI equations were derived using \textit{QMeS} \cite{Pawlowski:2021tkk, github:QMeS}, a Mathematica package for the derivation of symbolic functional equations. After projecting onto the respective tensor structures, the equations were traced with \textit{FormTracer} \cite{Cyrol:2016zqb, github:FormTracer}. The resulting momentum-dependent integral-differential and integral equations were solved in Mathematica 12.0.

%%%%%%%%%%%%%%%%%%%%%%%%%%%%%%%%%%%
\section{Additional Details on the fRG Computation}
\label{app:YM:FRGYM}%%%%%%%%%%%%%
%%%%%%%%%%%%%%%%%%%%%%%%%%%%%%%%%%%

%%%%%%%%%%%%%%%%%%%%%%%%%%%%%%%%%%
\subsection{Projecting onto Tensor Structures}
\label{app:YM:tensors}%%%%%%%%%%%%%
%%%%%%%%%%%%%%%%%%%%%%%%%%%%%%%%%%%
On the level of two-point functions, we have two quantities that contribute to the transverse sector, the gluon and the ghost two-point function. Their tensor structures are,
\begin{align}
\tau^{ab}_{AA,\mu\nu}(p) &= \delta^{ab}\left(\Pi^\perp_{\mu\nu}(p) p^2+ \Pi^\parallel_{\mu\nu}(p) p^2\right)\,,\nonumber\\[10pt]
\tau^{ab}_{c\bar{c}}(p) &= - \delta^{ab}p^2\,.
\end{align}
For the derivation of their respective fRG equations we trace the diagrammatic equations with the projection operators,
\begin{align}
\mathcal{P}^{ab}_{c\bar{c}}(p) &= \tau_{c\bar{c}}^{ab}(p)\,,\nonumber\\[10pt]
\mathcal{P}^{\perp,ab}_{AA,\mu\nu}(p) &= \Pi^\perp_{\mu\bar{\mu}}(p)\Pi^\perp_{\nu\bar{\nu}}(p) \tau^{ab}_{AA,\bar{\mu}\bar{\nu}}(p)\,,\nonumber\\[10pt]
\mathcal{P}^{\parallel,ab}_{AA,\mu\nu}(p) &= \Pi^\parallel_{\mu\bar{\mu}}(p)\Pi^\parallel_{\nu\bar{\nu}}(p) \tau^{ab}_{AA,\bar{\mu}\bar{\nu}}(p)\,.
\end{align}
for the ghost and transverse gluon propagator dressing, and for the longitudinal gluon two-point respectively.

The tensor structures of the ghost-gluon vertex can be written as,
\begin{align}
\tau^{abc}_{A\bar{c}c,\mu}(p,q) &= i f^{abc}\left(\Pi^{\perp}_{\mu \nu}(p)+ \Pi^{\parallel}_{\mu \nu}(p)\right)q_\nu\,.%\nonumber\\[10pt]
%&= i f^{abc}\left(q_\mu+p_\mu\right)\,.
\end{align}
\begin{figure}[b]
	\centering
	\includegraphics[width=0.98\linewidth]{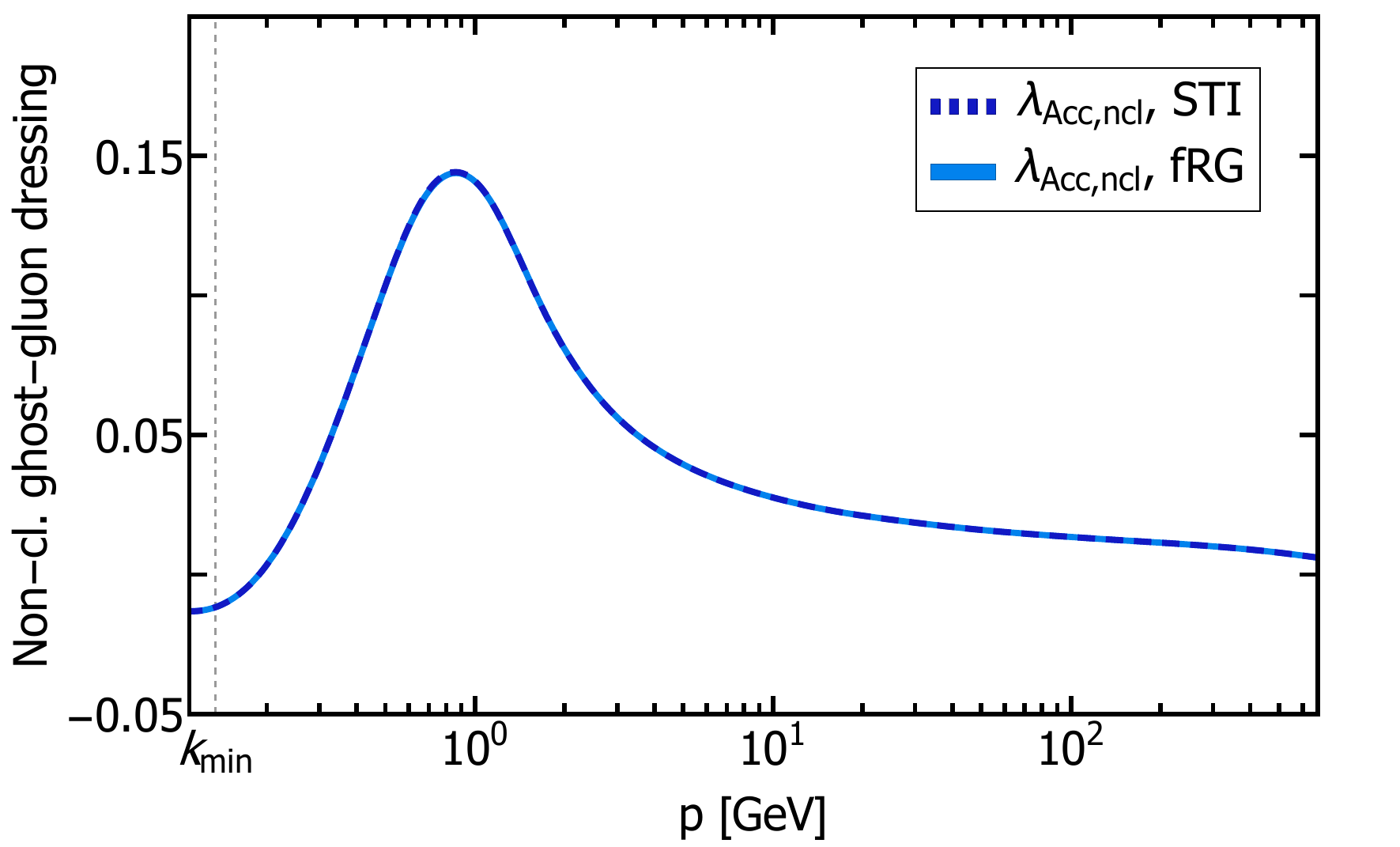}
	\caption{Non-classical ghost-gluon dressing from the fRG (solid blue) and STI (dashed dark blue).}
	\label{fig:YM:accnonclcouplingsfrg}
\end{figure}
The fRG equations for the transverse and longitidunal ghost-gluon dressing are obtained by contracting the equation from \Cref{fig:YM:flowequations} with,
\begin{align}
\mathcal{P}^{abc}_{A\bar{c}c,\mu}(p,q) &= \Pi^\perp_{\mu\bar{\mu}}(p)\tau^{ abc}_{A\bar{c}c,\bar{\mu}}(p,q)\,,\nonumber\\[10pt]
\mathcal{P}^{\parallel,abc}_{A\bar{c}c,\mu}(p,q) &= \Pi^\parallel_{\mu\bar{\mu}}(p)\tau^{abc}_{A\bar{c}c,\bar{\mu}}(p,q)\,.
\end{align}
From the longitudinal and transverse/classical ghost-gluon dressing one can compute the non-classical vertex dressing \labelcref{eq:YM:nonclDressing}. The result is shown in \Cref{fig:YM:accnonclcouplingsfrg}. One can see that it is approximately zero for perturbative momenta.

\begin{figure*}[t]
	\includegraphics[width=.45\textwidth]{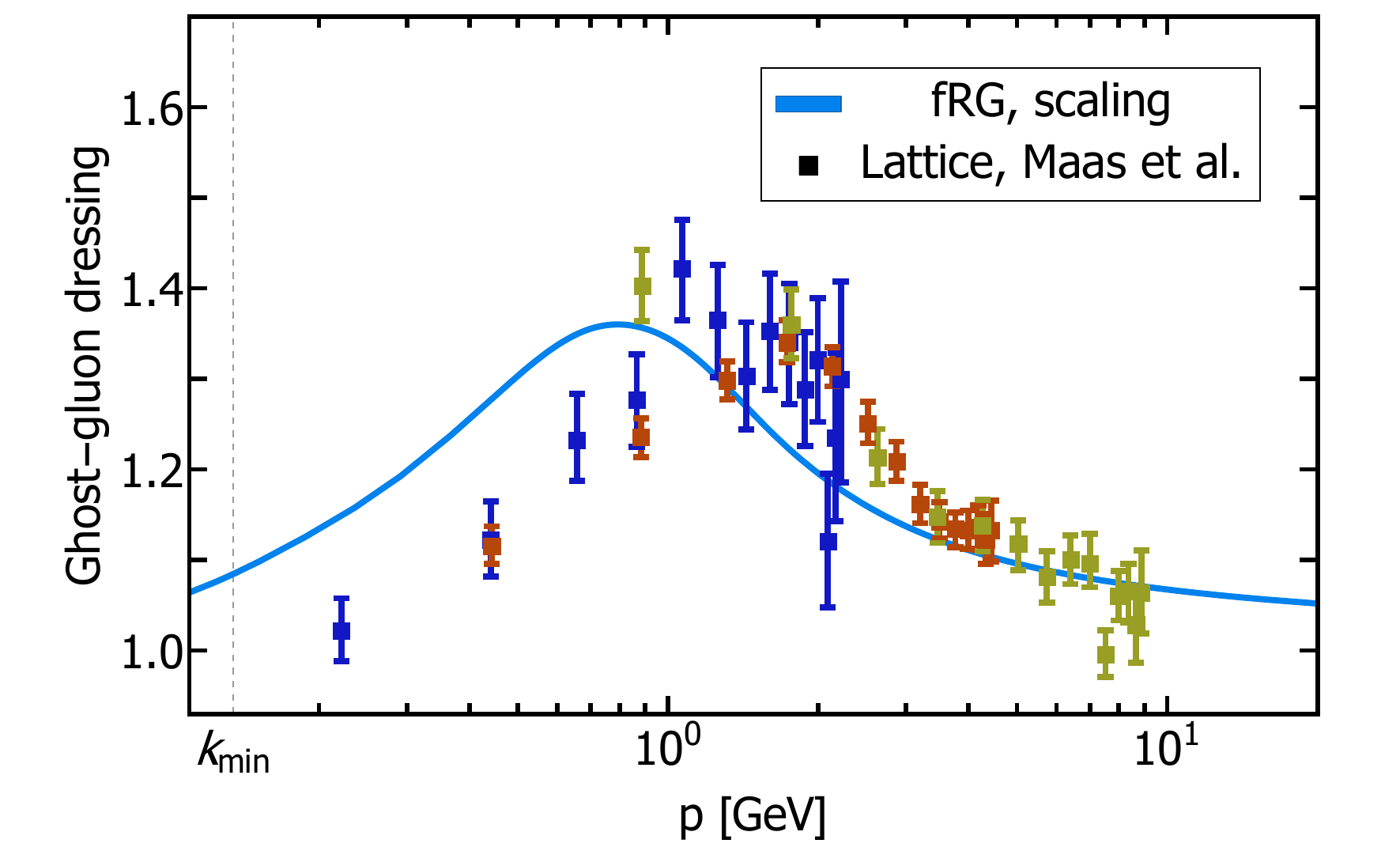}
	\includegraphics[width=.45\textwidth]{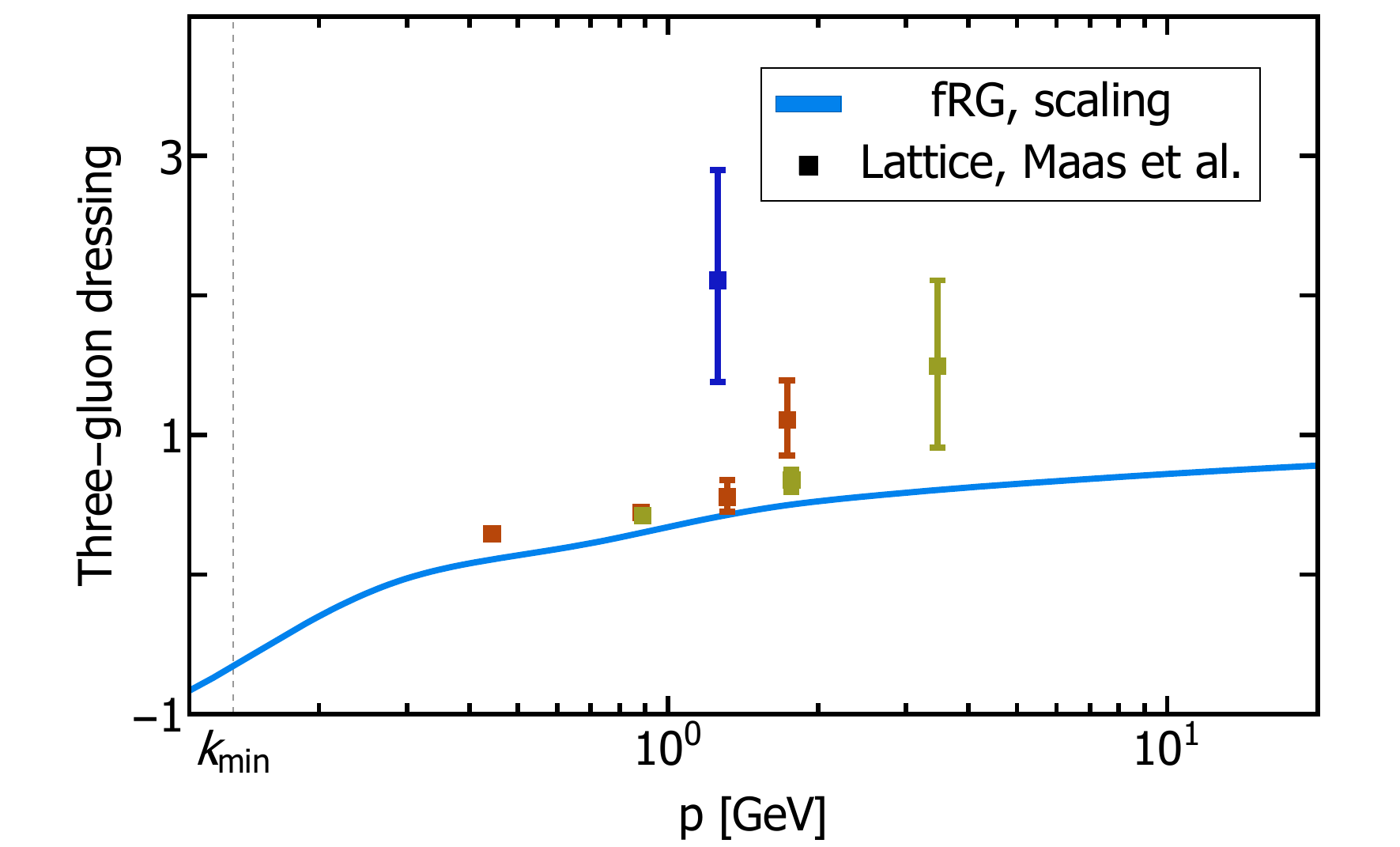}
	\caption{Transverse ghost-gluon and three-gluon dressings (blue lines) in comparison to $SU(2)$ lattice results \cite{Cucchieri:2006tf, Cucchieri:2008qm, Maas:2016pr} with $N=32^4$, $\beta = \{2.13,\,2.39,\,2.60\}$ and $a^{-1} = \{0.8\,\text{GeV},\, 1.6\,\text{GeV},\, 3.2\,\text{GeV}\}$ for the dark blue/rusty red/ocher points.\hspace*{\fill}}
	\label{fig:YM:latticedressings}
\end{figure*}

As a tensor basis for the three- and four-gluon vertex, we apply transverse and longitudinal projections of the classical tensor structures,
\begin{align}
&\tau_{A^3,\text{cl},\mu\nu\rho}^{abc}(p,q) = -i f^{abc}\nonumber\\[4pt] &\left(\delta_{\mu\nu}(p-q)_\rho+\delta_{\nu\rho}(2q+p)+\delta_{\rho\mu}(-2p-q)\right)\,,
\end{align}
and
\begin{align}
\tau_{A^4,\text{cl},\mu\nu\rho\sigma}^{abcd}(p,q,r) &= f^{abe}f^{cde}\left(\delta_{\mu\rho}\delta_{\nu\sigma}-\delta_{\mu\sigma}\delta_{\nu\rho}\right)\nonumber\\[4pt]
&+f^{ace}f^{bde}\left(\delta_{\mu\nu}\delta_{\rho\sigma}-\delta_{\mu\sigma}\delta_{\nu\rho}\right)\nonumber\\[4pt]
&+f^{ade}f^{bce}\left(\delta_{\mu\nu}\delta_{\rho\sigma}-\delta_{\mu\rho}\delta_{\nu\sigma}\right)\,.
\end{align}

We obtain the transverse and longitudinal three- and four-gluon vertex dressings by contracting the full vertices with the transverse longitudinal projection operators \labelcref{eq:YM:ProjOps} applied to the classical tensor structures. The resulting projection operators are,
\begin{align}
\mathcal{P}^{abc}_{A^3,\mu\nu\rho}(p,q) = &\Pi^\perp_{\mu \bar{\mu}}(p)\Pi^\perp_{\nu \bar{\nu}}(q)\Pi^\perp_{\rho \bar{\rho}}(-p-q)\nonumber\\[4pt]
&\tau_{A^3,\text{cl},\bar{\mu}\bar{\nu}\bar{\rho}}^{abc}(p,q)\,,\nonumber\\[10pt]
\mathcal{P}^{abc}_{A^3,1,\mu\nu\rho}(p,q) = &\Pi^\perp_{\mu \bar{\mu}}(p)\Pi^\perp_{\nu \bar{\nu}}(q)\Pi^\parallel_{\rho \bar{\rho}}(-p-q)\nonumber\\[4pt]
&\tau_{A^3,\text{cl},\bar{\mu}\bar{\nu}\bar{\rho}}^{abc}(p,q)\,,\nonumber\\[10pt]
\mathcal{P}^{abc}_{A^3,2,\mu\nu\rho}(p,q) = &\Pi^\perp_{\mu \bar{\mu}}(p)\Pi^\parallel_{\nu \bar{\nu}}(q)\Pi^\parallel_{\rho \bar{\rho}}(-p-q)\nonumber\\[4pt]
&\tau_{A^3,\text{cl},\bar{\mu}\bar{\nu}\bar{\rho}}^{abc}(p,q)\,,
\end{align}
and,
\begin{align}
&\mathcal{P}^{abcd}_{A^4,\mu\nu\rho\sigma}(p,q,r) =\Pi^\perp_{\mu \bar{\mu}}(p)\Pi^\perp_{\nu \bar{\nu}}(q)\Pi^\perp_{\rho \bar{\rho}}(r)\nonumber\\[4pt]
&\Pi^\perp_{\sigma \bar{\sigma}}(-p-q-r)\tau_{A^4,\text{cl},\bar{\mu}\bar{\nu}\bar{\rho}\bar{\sigma}}^{abcd}(p,q,r)\,,\nonumber\\[10pt]
&\mathcal{P}^{abcd}_{A^4,1,\mu\nu\rho\sigma}(p,q,r) =\Pi^\perp_{\mu \bar{\mu}}(p)\Pi^\perp_{\nu \bar{\nu}}(q)\Pi^\perp_{\rho \bar{\rho}}(r)\nonumber\\[4pt]
&\Pi^\parallel_{\sigma \bar{\sigma}}(-p-q-r)\tau_{A^4,\text{cl},\bar{\mu}\bar{\nu}\bar{\rho}\bar{\sigma}}^{abcd}(p,q,r)\,,\nonumber\\[10pt]
&\mathcal{P}^{abcd}_{A^4,2,\mu\nu\rho\sigma}(p,q,r) =\Pi^\perp_{\mu \bar{\mu}}(p)\Pi^\perp_{\nu \bar{\nu}}(q)\Pi^\parallel_{\rho \bar{\rho}}(r)\nonumber\\[4pt]
&\Pi^\parallel_{\sigma\bar{\sigma}}(-p-q-r)\tau_{A^4,\text{cl},\bar{\mu}\bar{\nu}\bar{\rho}\bar{\sigma}}^{abcd}(p,q,r)\,.
\label{eq:YM:longProjgluonic}
\end{align}

The tensor structures of the BRST-projected two-point vertices are given as, 
\begin{align}
&\tau^{ab}_{AQ_{\bar{c}},\mu}(p) = -i\frac{p_\mu}{\xi}\delta^{ab}\,,\nonumber\\[10pt]
&\tau^{ab}_{cQ_{A},\mu}(p) = -i p_\mu\delta^{ab}\,,
\end{align}
and of the BRST three-point functions as,
\begin{align}
\tau^{abc}_{ccQ_c}(p,q) =& -f^{abc}\,,\nonumber\\[10pt]
\tau^{abc}_{AcQ_A,\mu\nu}(p,q)  =& -f^{abc}\left(\Pi^\perp_{\mu\nu}(p)+\Pi^\parallel_{\mu\nu}(p)\right)\,.
\end{align}
Analogously, we can construct the respective projection operators,
\begin{align}
&\mathcal{P}^{ab}_{AQ_{\bar{c}},\mu}(p) = \tau^{ab}_{AQ_{\bar{c}},\mu}(p)\,,\nonumber\\[10pt]
&\mathcal{P}^{ab}_{cQ_{A},\mu}(p) = \tau^{ab}_{cQ_{A},\mu}(p)\,,
\end{align}
and,
\begin{align}
\mathcal{P}^{abc}_{ccQ_c}(p,q) =& \tau^{abc}_{ccQ_c}(p,q)\,,\nonumber\\[10pt]
\mathcal{P}^{abc}_{AcQ_A,\mu\nu}(p,q)  =& \Pi^\perp_{\mu\bar{\mu}}(p)\tau^{abc}_{AcQ_A,\bar{\mu}\nu}(p,q)\,,\nonumber\\[10pt]
\mathcal{P}^{abc}_{AcQ_A,1,\mu\nu}(p,q)  =& \Pi^\parallel_{\mu\bar{\mu}}(p)\tau^{abc}_{AcQ_A,\bar{\mu}\nu}(p,q)\,.
\end{align}
The relation between these tensor structures and the pure Yang-Mills ones is further elaborated in \Cref{app:YM:shiftsym}.

%%%%%%%%%%%%%%%%%%%%%%%%%%%%%%%%%%%
\subsection{Tuning of Initial Parameters}
\label{app:YM:tuning}
%%%%%%%%%%%%%%%%%%%%%%%%%%%%%%%%%%
The constant initial conditions for the dressings and the transverse gluon mass at the UV cut-off scale $\Lambda = 672.6\,\text{GeV}$ for our approximation of the scaling solution were chosen as,
\begin{align}
\lambda_{A\bar{c}c,\Lambda} &= \lambda_{A\bar{c}c, 1,\Lambda} = \lambda_{Q_ccc,\Lambda}=  1\nonumber \\
\lambda_{A^3,\Lambda} &= \lambda_{A^3,1,\Lambda} = \lambda_{A^3,2,\Lambda} =  0.981\nonumber \\
\lambda_{A^4,\Lambda} &=  0.981\nonumber \\
m^2_{\perp,\Lambda} &= -3708.5272\,\text{GeV}^2\, .
\end{align}

%%%%%%%%%%%%%%%%%%%%%%%%%%%%%%%%%%%
\subsection{Regulators and Gluon Mass}
\label{app:YM:reg}
%%%%%%%%%%%%%%%%%%%%%%%%%%%%%%%%%%
The transverse gluon dressing $Z_A(p)$ contains the gluon mass gap and thus diverges below $\sim 1\,$GeV for $k\rightarrow0$. For the numerical computations we parameterize the gluon two-point function as,
\begin{align}\nonumber 
\Gamma^{ab}_{AA,\mu\nu}(p) =&\, \Pi^\perp_{\mu\nu}(p)\, Z_T(p) \left(p^2+m_T^2\right)\,\delta^{ab}\\[1ex]
&\hspace{1cm}+\Pi^\parallel_{\mu\nu}(p) \Gamma^{\parallel}_{AA}(p)\,\delta^{ab}\,, 
\end{align}
where the longitudinal part $\Gamma^{\parallel}_{AA}$ contains the gauge fixing term $p^2/\xi$. We obtain the effective transverse and longitudinal gluon mass from,
\begin{align}\label{eq:mPerpPar}
m_{\perp,k}^2 = Z_T(0)\,m_T^2\,,\qquad 
m_{\parallel,k}^2 =\Gamma^{\parallel}_{AA,k}(0)\,.
\end{align}

The longitudinal gluon mass from the mSTI however is extracted from,
\begin{align}
m_{\parallel,k,\text{mSTI}}^2=\Gamma^{\parallel}_{AA,k,\text{mSTI}}(p=4k_{\mathrm{min}})\,,
\end{align}
where the minimal momentum was chosen sufficiently large to avoid any IR cutoff effects.
\begin{figure*}[t]
	\includegraphics[width=.45\textwidth]{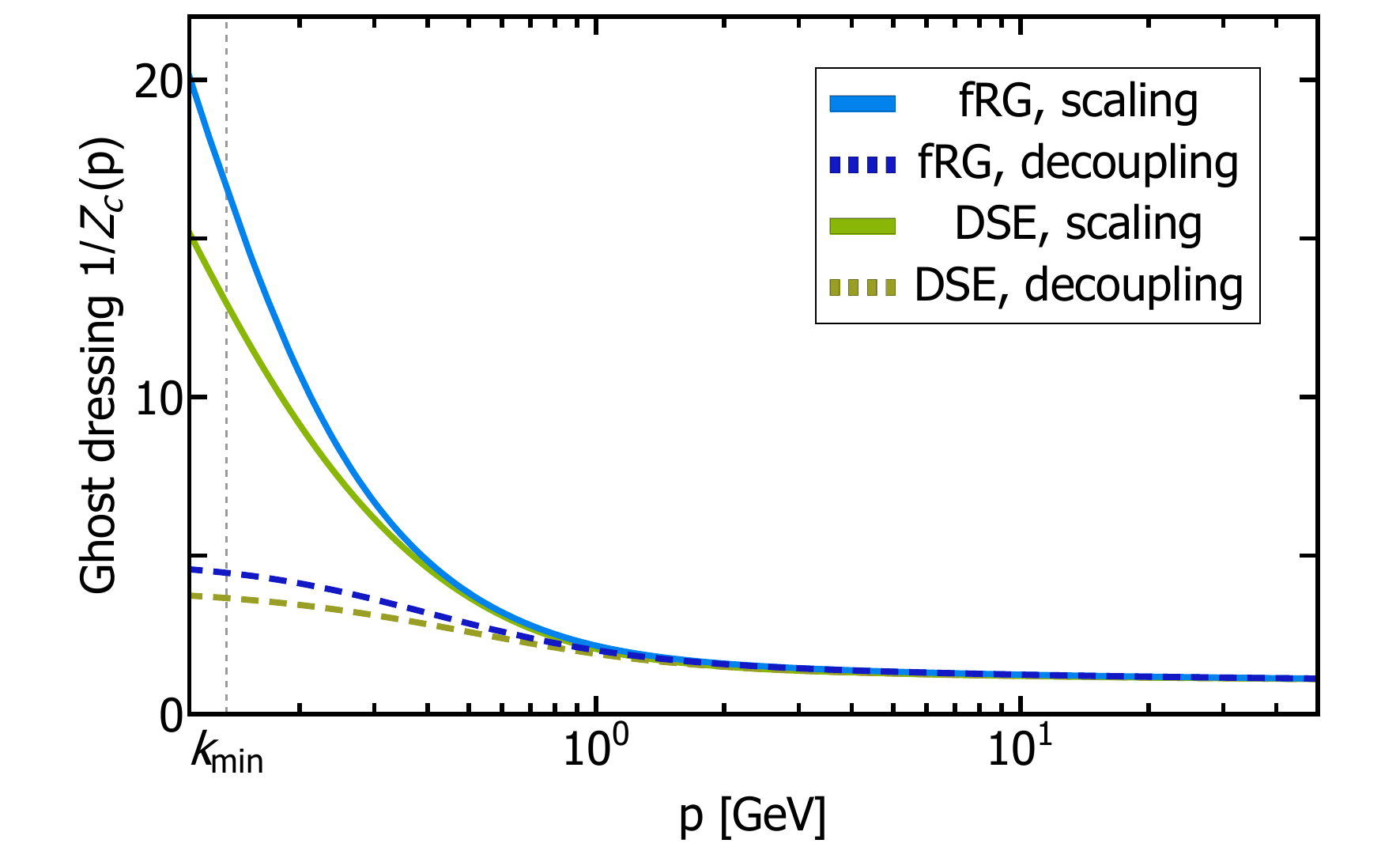}
	\includegraphics[width=.45\textwidth]{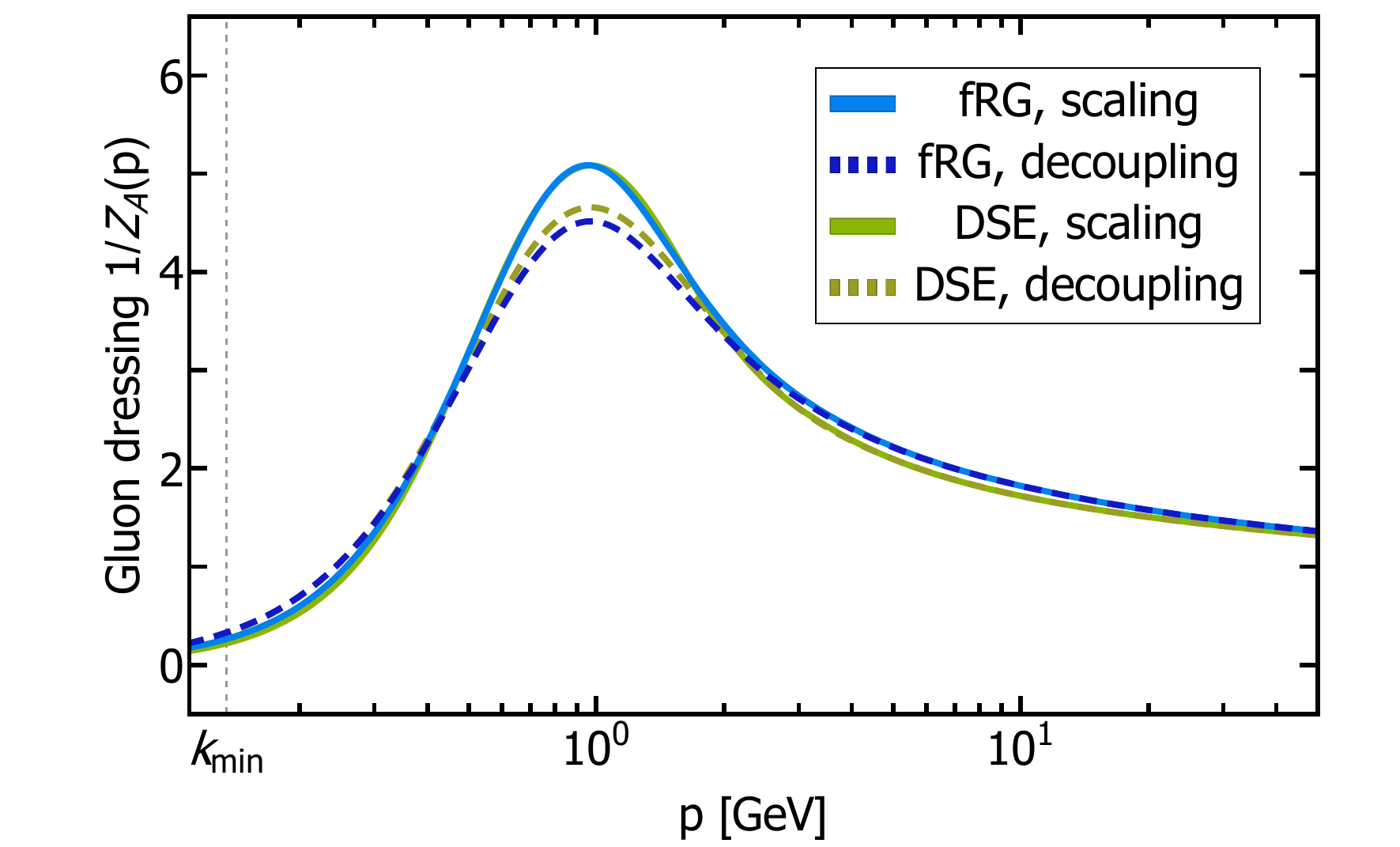}
	\caption{Ghost and gluon propagator dressings: Results from the fRG (blue) in comparison to the DSE~\cite{Huber:2020keu} (green), for the scaling (solid) and decoupling (dashed) solution. The DSE dressings were rescaled such that they agree with the corresponding fRG dressings at $p = \Lambda = 672.6\,$GeV.\hspace*{\fill}}
	\label{fig:YM:dsepropdressings}
\end{figure*}
The effective parameter for the true scaling mass was extracted by extrapolation of $m^2_{\perp,\mathrm{kmin}}(m^2_{\perp,\Lambda})$.
Since we are dealing with scaling, the relation between the effective mass and the UV mass parameter follows

\begin{align}
\label{eq:fit_function_mass}
	m^2_{\perp,\mathrm{kmin}} = x_0\, (m^2_{\perp,\Lambda} - m^2_\mathrm{ref})^\alpha
\, .
\end{align}
Fitting the six left most values, we obtain
\begin{align}
	x_0 &= 1.08\cdot10^{-4} \nonumber \\
	\alpha &= -0.097 \nonumber \\
	m^2_\mathrm{ref} &= -0.00812 \Lambda^2 = 1.15\, m^2_{\perp,\Lambda,\,\text{1-loop}}
\, ,
\end{align}
with the one-loop mass defined in \labelcref{eq:YM:oneloopmass}.

For the ghost and gluon regulator we use respectively,
\begin{align}
(R_A)^{ab}_{\mu\nu}(p) =&\,\delta^{ab}  r(p^2/k^2)\nonumber\\[4pt]
&\left(\Pi^\perp_{\mu \nu}(p)\bar{Z}_T(p)\, p^2+\Pi^\parallel_{\mu \nu}(p)\Gamma^{\parallel}_{AA}(p)\right) \,,\nonumber\\[10pt]
(R_c)^{ab}(p) &= \delta^{ab} p^2 Z_c(p) r(p^2/k^2)\,,
\label{eq:YM:regs}
\end{align}
where k is the RG scale and we parametrise the transverse gluon two-point as,
\begin{align}
\Gamma_{AA,\mu\nu}^{\perp,ab}(p) &= \Pi_{\mu\nu}^\perp(p) Z_T(p)(p^2+m_T^2)\delta^{ab}\\[4pt]
&=\Pi_{\mu\nu}^\perp(p)\left(\bar{Z}_T(p)(p^2+\bar{m}_T^2+k^2)-k^2\right)\delta^{ab}\,.\nonumber
\end{align}

\begin{figure*}[t]
	\includegraphics[width=.45\textwidth]{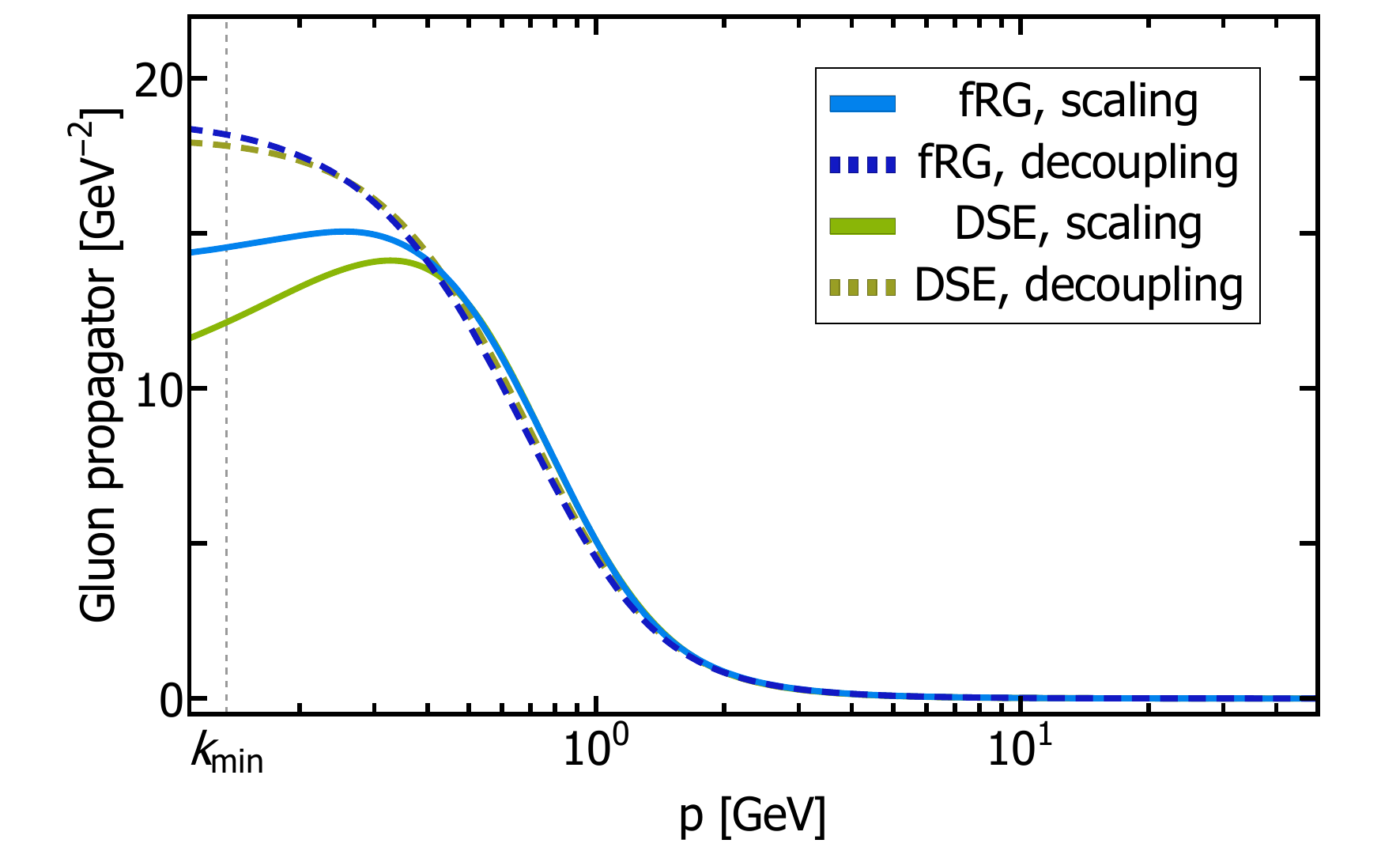}
	\includegraphics[width=.45\textwidth]{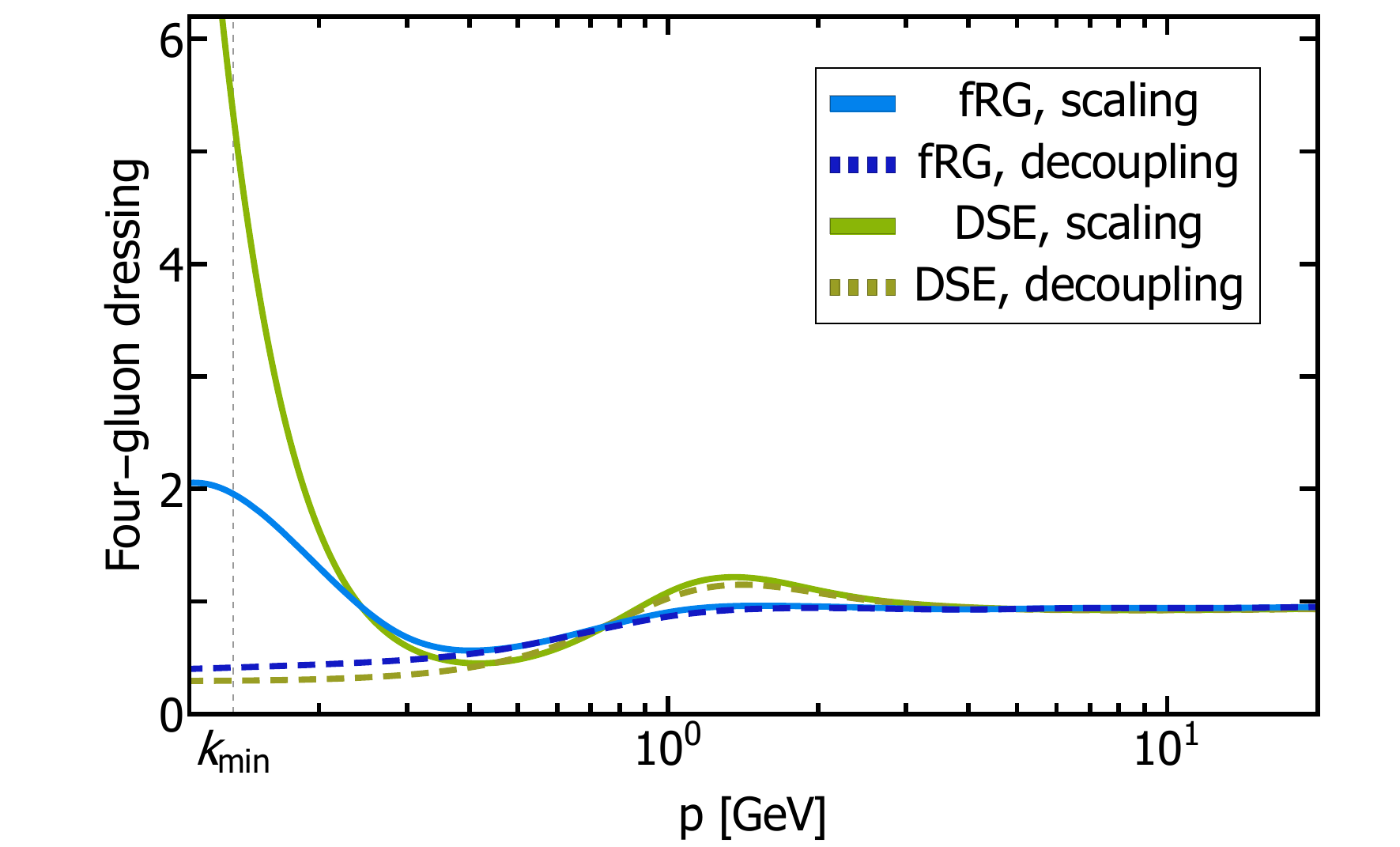}
	\caption{Gluon propagator and four-gluon dressing: Results from the fRG (blue) in comparison to the DSE~\cite{Huber:2020keu} (green), for the scaling (solid) and decoupling (dashed) solution. The DSE dressings were rescaled such that they agree with the corresponding fRG dressings at $p = \Lambda = 672.6\,$GeV.\hspace*{\fill}}
	\label{fig:YM:dseprop}
\end{figure*}

The dressing $\bar{Z}_T$ guarantees a well-defined regulator in the infrared.
The shape function is defined as an exponential,
\begin{align}
r(p^2/k^2) = \frac{e^{-p^2/k^2}}{1-e^{-p^2/k^2}}\,.
\label{eq:YM:regshape}
\end{align}

\subsection{Extraction of Scale}
\label{app:YM:scalesetting}
%%%%%%%%%%%%%%%%%%%%%%%%%%%%%%%%%%
We set the momentum scale by positioning the maximum of the gluon dressing $1/Z_A(p)$ at the lattice scale $p\approx 0.955 \,$GeV from \cite{Sternbeck:2006cg}.

When comparing our propagators and dressings to the aforementioned lattice results one also has to adjust the global normalisation, $Z_{\text{Lat}}^{-1}(p) \rightarrow y Z_{\text{Lat}}^{-1}(p)$. This is done via minimizing
\begin{align}
N_{Z_A}(y) = \sum_{i} \frac{\Delta x_i}{\Delta E_i^2} \left[\left( Z_A^{-1}(p_i)-y Z_{\text{Lat}}^{-1}(p_i)\right)^2\right.\nonumber\\[4pt]
\left.+\left( \partial_p Z_A^{-1}(p_i)- y\partial_p Z_{\text{Lat}}^{-1}(p_i)\right)^2\right]
\end{align}
in the region $0.8\,$GeV$\,\leq p_i \leq 4\,$GeV, where the lattice input $\Delta x_i$ denotes the distances to the next point, $\Delta E_i^2$ the statistical error of that point and $Z_{\text{Lat}}^{-1}(p)$ is the gluon dressing obtained from the respective lattice computation \cite{Sternbeck:2006cg} and \cite{Aguilar:2021okw}. The normalisation procedure for the ghost dressing is done analogously in the fitting range $1.5\,$GeV$\,\leq p_i \leq 6\,$GeV.

\subsection{Scaling Exponents}
\label{app:YM:scalingexponents}
%%%%%%%%%%%%%%%%%%%%%%%%%%%%%%%%%%
The scaling exponents, c.f. \labelcref{eq:YM:scalingexponentsscaling}, obtained from our transverse scaling results in the regime $0.135\,$GeV$\,\leq p \leq 0.471\,$GeV are
\begin{align}
\kappa_{gl} = 0.535\nonumber\\[4pt]
\kappa_{gh} = 0.541
\label{eq:YM:scalingexponents}
\end{align}
%%%%%%%%%%%%%%%%%%%%%%%%%%%%%%%%%%%
\subsection{BRST Projected Vertices}
\label{app:YM:shiftsym}%%%%%%%%%%%%%
%%%%%%%%%%%%%%%%%%%%%%%%%%%%%%%%%%%
%%%%%%%%%%%%%%%%%%%%%%%%%%%%%
The effective action exhibits shift symmetry in the anti-ghost. This symmetry leads to the identities,
\begin{align}
\partial \cdot \frac{\delta \Gamma}{\delta Q_{A}} &=\frac{\delta \Gamma}{\delta \bar{c}}\,,\nonumber\\[10pt]
\frac{\delta \Gamma}{\delta Q_{\bar{c}}} &= \frac{1}{\xi}\partial\cdot A\,.
\end{align}
Fourier transforming both gives,
\begin{align}
-iq_\nu \Gamma_{Q_A,\nu}(q) &= \Gamma_{\bar{c}}(q)\,,\nonumber\\[10pt]
\Gamma_{Q_{\bar{c}}}(-p) &= \frac{1}{\xi}ip_\mu A_\mu(p)\,.
\end{align}
The first identity relates,
\begin{align}
iq_\nu \Gamma^{adb}_{AcQ_A,\mu\nu}(p,-p-q) &= \Gamma^{abd}_{A\bar{c}c,\mu}(p,q)\,,\nonumber\\[10pt]
i p_\mu\Gamma^{ba}_{cQ_A,\mu} (-p) &= \Gamma^{ab}_{\bar{c}c} (p)\,.
\label{eq:YM:shiftsymid}
\end{align}
Generally we can write down the full tensor basis,
\begin{align}
\Gamma^{adb}_{AcQ_A,\mu\nu}(p,-p-q) = &f^{abd}\left(\delta_{\mu\nu}\lambda_1+p_\mu p_\nu\lambda_2 +q_\mu q_\nu \lambda_3\right.\nonumber\\[4pt]
&\left.+p_\mu q_\nu \lambda_4 + q_\mu p_\nu \lambda_5\right)\,.
\end{align}
The shift symmetry identity \labelcref{eq:YM:shiftsymid} then relates,
\begin{align}
\lambda_{A \bar{c} c} =&\, \lambda_1+q^2\lambda_3+pq \lambda_5\nonumber\\[10pt]
\lambda_{A \bar{c} c, ncl} =&\, pq \lambda_2 +q^2 \lambda_4\nonumber\\[10pt]
\lambda_{A \bar{c} c,\parallel} =&\, \lambda_1+q^2\lambda_3+pq \lambda_5\nonumber\\[4pt]
&-2pq \lambda_2 -2 q^2 \lambda_4\,.
\end{align}

\begin{figure*}[t]
	\includegraphics[width=.45\textwidth]{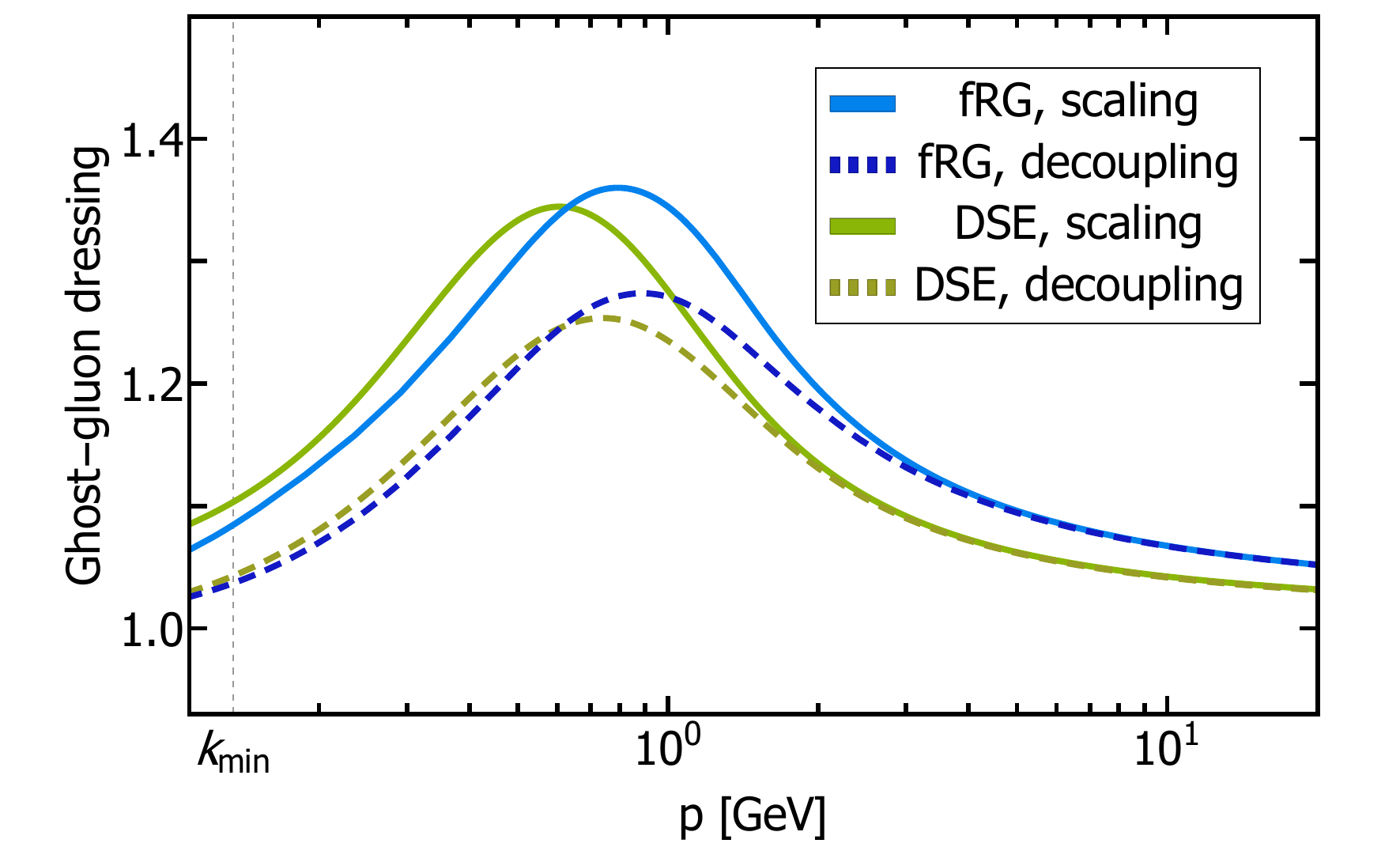}
	\includegraphics[width=.45\textwidth]{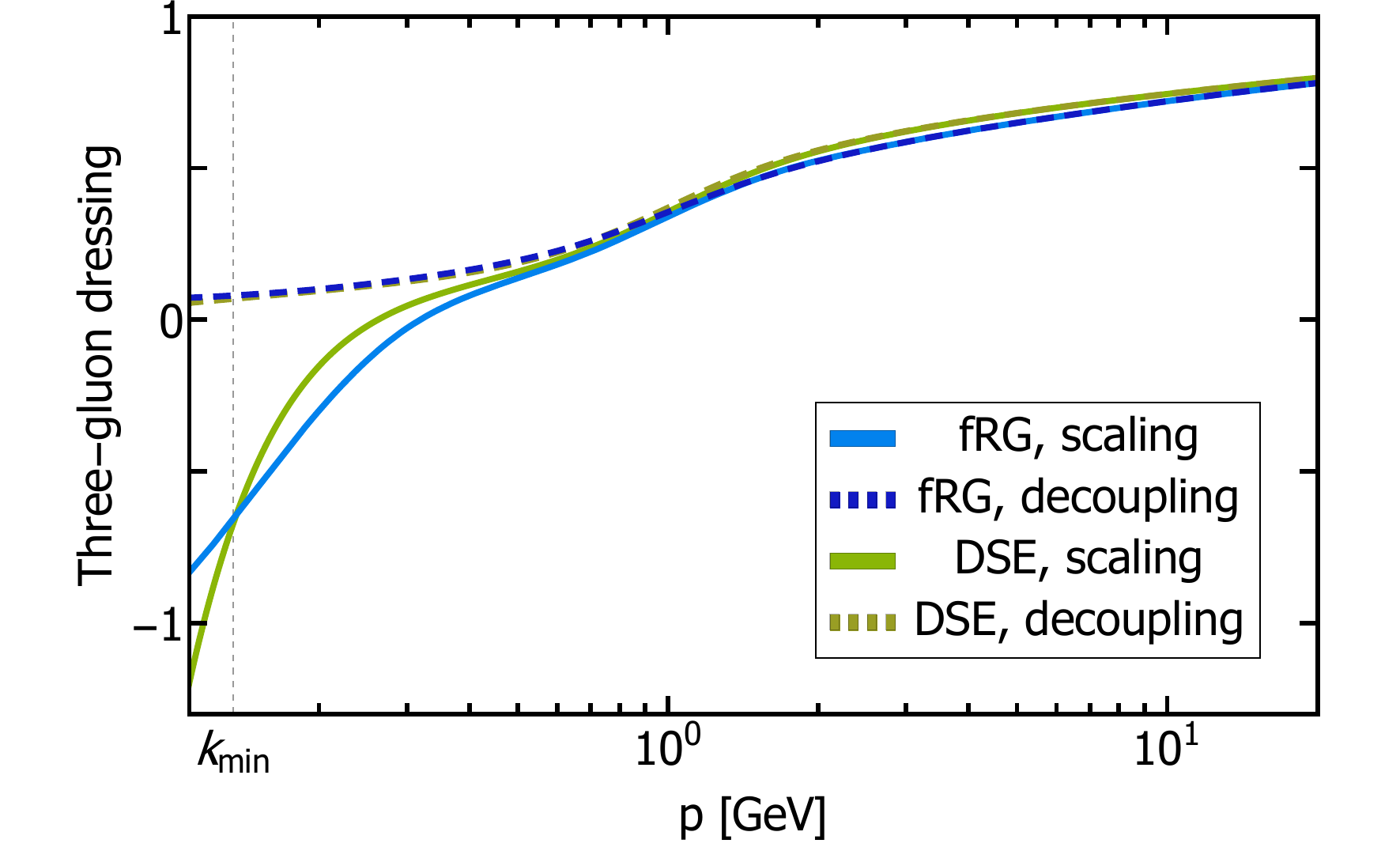}
	\caption{Transverse ghost-gluon and three-gluon dressings: Results from the fRG (blue) in comparison to the DSE~\cite{Huber:2020keu} (green), for the scaling (solid) and decoupling (dashed) solution. The DSE dressings were rescaled such that they agree with the corresponding fRG dressings at $p = \Lambda = 672.6\,$GeV.\hspace*{\fill}}
	\label{fig:YM:dsedressings}
\end{figure*}

The diagrams contributing to the flow of the $AcQ_A$- and $cQ_A$-dressings are depicted in \Cref{fig:YM:flowequations}. One can see that the identity \labelcref{eq:YM:shiftsymid} is already diagrammatically fulfilled for the fRG equations. Thus they are exact identities. Therefore applying them in the STIs is valid and the exactness of the identities guarantees gauge parameter independence of the three-gluon STI. However the STIs not only depend on these specific combinations of dressings, therefore a computation of all five from the fRG is nonetheless necessary. 

We however simplify the approximation by identifying,
\begin{align}
\lambda_{AcQ_A,1}(p,-p-q) &= \lambda_{A\bar{c}c,1}(p,q)\,,\nonumber\\[10pt]
\lambda_{AcQ_A}(p,-p-q) &= \lambda_{A\bar{c}c}(p,q)\,,\nonumber\\[10pt]
Z_{cQ_A}(p) &= Z_{c}(p)\,,
\end{align}
where the index $1$ indicates a longitudinally projected gluon.
This approximation should not lead to qualitative differences.

%%%%%%%%%%%%%%%%%%%%%%%%%%%%%%%%%%%
\section{Additional Details on the mSTI}
\label{app:YM:mSTI}%%%%%%%%%%%%%
%%%%%%%%%%%%%%%%%%%%%%%%%%%%%%%%%%%

%%%%%%%%%%%%%%%%%%%%%%%%%%%%%%%%%%%
\subsection{Longitudinal Gluon Two-Point Function and Gluon Mass}
\label{app:YM:twopointmSTI}
%%%%%%%%%%%%%%%%%%%%%%%%%%%%%%%%%%%
To obtain the mSTI of the longitudinal gluon two-point function one takes the functional derivatives $\delta^2/(\delta A_{\mu}^a \delta c^b)$ of the mSTI \labelcref{eq:YM:mSTI} projects both sides of the equation with,
\begin{align}
\mathcal{Q}^{ab}_{AA,\mu} = p_\nu \tau^{ab}_{AA,\nu\mu}(p)\,.
\end{align}
After normalisation, one obtains the STI
\begin{align}
\Gamma^{\parallel}_{AA,\text{reg}}(p) =0\,.
\end{align}
We can compare the effective one-loop, transverse and longitudinal, mass running from the fRG with the one-loop longitudinal mass from the STI. The different diagrams contributing to the equations are depicted in \Cref{fig:YM:flowequations} and \Cref{fig:YM:mSTIequations}. Inserting undressed propagators and vertices with a gauge coupling $g$ and the regulator shape function $r(q^2)\equiv r(q^2/k^2)$, see \labelcref{eq:YM:regshape}, one obtains for the effective transverse gluon mass from the fRG,
\begin{align}
\dot{m}^2_{\perp,\text{1-loop}} = \int dz\, dq\,& \frac{q^3\sqrt{1-z^2}}{4\pi^3}\frac{g^2 q^2\dot{r}(q^2)}{q^4(1+r(q^2))^3}\nonumber\\[4pt]
&\left(3-9z^2+r(q^2)(-7+z^2)\right)
\label{eq:YM:onelooptransmass}
\end{align}
where $z= \frac{p\cdot q}{\left| p\right| \left|q\right|}$. Performing the loop integration, one arrives at the one-loop running,
\begin{align}
\dot{m}^2_{\perp,\text{1-loop}} = k\partial_k m^2_\perp = -\frac{9g^2}{64\pi^2}k^2\,.
\end{align}
Solving the differential equation yields,
\begin{align}
m^2_{\perp,\text{1-loop}} = -\frac{9g^2}{128\pi^2}k^2\,.
\label{eq:YM:oneloopmass}
\end{align}
For the longitudinal gluon mass from the fRG one obtains,
\begin{align}
\dot{m}^2_{\parallel,\text{1-loop}} = \int dz dq &\frac{q^3 \sqrt{1-z^2}}{4\pi^3}\frac{g^2 q^2 \dot{r}(q^2)}{q^4(1+r(q^2))^3}\nonumber\\[4pt]
&\left(2-9z^2+r(q^2)(2+z^2)\right)\,.
\end{align}
Even though the longitudinal vertices contribute differently, one arrives at the same one-loop running as for the effective transverse mass and thus the longitudinal mass is,
\begin{align}
m^2_{\parallel,\text{1-loop}} = -\frac{9g^2}{128\pi^2}k^2\,.
\end{align}
For the derivation of the one-loop mass from the mSTI, one has to take the zero momentum limit $p\rightarrow0$ of the equation for $\Gamma^{AA}_\parallel(p)$ with the rule of l'Hospital,
\begin{align}
\lim\limits_{p\rightarrow0}\frac{f(p)}{g(p)} = \lim\limits_{p\rightarrow0}\frac{f'(p)}{g'(p)}\,.
\end{align}
One then obtains the equation,
\begin{align}
m^2_{\parallel,\text{1-loop}} = &\int dz dq \frac{q^3\sqrt{1-z^2}}{4\pi^3}\frac{g^2 q^2 r(q^2)}{q^4(1+r(q^2))^3}\nonumber\\[4pt]
&\left((-2+9z^2)(1+r(q^2))+8q^2z^2 \partial_{q}r(q^2)\right)\,.
\end{align}
Integrating this, one arrives at the same one-loop mass as from the fRG,
\begin{align}
m^2_{\parallel,\text{1-loop}} = -\frac{9g^2}{128\pi^2}k^2\,.
\end{align}
Although all three equations contain different contributions and diagrams, they yield the same effective gluonic mass at one-loop level.

%%%%%%%%%%%%%%%%%%%%%%%%%%%%%%%%%%%
\subsection{Ghost-Gluon Vertex mSTI}
\label{app:YM:AccmSTI}
%%%%%%%%%%%%%%%%%%%%%%%%%%%%%%%%%%%

We take the functional derivatives $\delta^3/(\delta \bar{c}^a\delta c^b\delta c^c)$ of \labelcref{eq:YM:mSTI}. To project onto the dressings we trace the equation with $f^{abc}$. Thus the projection operator is, 
\begin{align}
\mathcal{Q}^{abc}_{A\bar{c}c}(p,q) = p_\mu \tau^{abc}_{A\bar{c}c,\mu}(p,q)\,.
\end{align}
After normalisation, the STI for the non-classical ghost-gluon dressing at the symmetric point is given as,
\begin{align}
\lambda_{Q_ccc}(p)-\lambda_{A\bar{c}c}(p)+2\lambda_{A \bar{c} c,noncl}(p)=0\,.
\end{align}
From this we can derive:
\begin{align}
\lambda_{A \bar{c} c,ncl}(p) &= \frac{1}{2}\left(\lambda_{A\bar{c}c}(p)-\lambda_{Q_ccc}(p)\right)\,,\nonumber\\[10pt]
\lambda_{A \bar{c} c}(p) &= \lambda_{A \bar{c} c}(p)\,,\nonumber\\[10pt]
\lambda_{A \bar{c} c,1}(p) &= \lambda_{A \bar{c} c}(p)-2\lambda_{A \bar{c} c,ncl}(p)=\lambda_{Q_ccc}(p)\,.
\end{align}
A comparison of the non-classical ghost-gluon dressing from the mSTI and from the fRG is shown in \Cref{fig:YM:accnonclcouplingsfrg}.

%%%%%%%%%%%%%%%%%%%%%%%%%%%%%%%%%%%%%%%%%
\subsection{Three-Gluon Vertex mSTI}
%%%%%%%%%%%%%%%%%%%%%%%%%%%%%%%%%%%%%%%%%
One can derive the three-gluon mSTI by taking the functional derivatives $\frac{\delta^3}{\delta A_\mu^a \delta A_\nu^b \delta c^c}$. For the projection onto the mSTI for the three-gluon dressings with one and two longitudinal legs, $\lambda_{A^3,1}(p)$ and $\lambda_{AAA,2}(p)$, one uses,
\begin{align}
\mathcal{Q}^{abc}_{A^3,1,\mu\nu}(p,q) = &\, \Pi^\perp_{\mu \bar{\mu}}(p)\Pi^\perp_{\nu \bar{\nu}}(q)(-p-q)_{\bar{\rho}}\nonumber\\[4pt] &\tau^{abc}_{A^3,cl,\bar{\mu}\bar{\nu}\bar{\rho}}(p,q)\,,\nonumber\\[10pt]
\mathcal{Q}^{abc}_{A^3,2,\mu\nu}(p,q) = &\,\Pi^\perp_{\mu \bar{\mu}}(p)\Pi^\parallel_{\nu \bar{\nu}}(q)(-p-q)_{\bar{\rho}}\nonumber\\[4pt] &\tau^{abc}_{A^3,cl,\bar{\mu}\bar{\nu}\bar{\rho}}(p,q)\,.
\end{align}
Since the classical tensor structure has no overlap with the first projection operator, there exists no mSTI for $\lambda_{A^3,1}(p)$ in our truncation, see also \labelcref{eq:YM:AAA1nonrunning}, for the equivalent case in the fRG equations.

%%%%%%%%%%%%%%%%%%%%%%%%%%%%%%%%%%%%%%%%
\subsection{Four-Gluon Vertex mSTI}
%%%%%%%%%%%%%%%%%%%%%%%%%%%%%%%%%%%%%%%%
We take the field derivatives $\frac{\delta^3}{\delta A^a_\mu(p4)A^b_\nu(p3)A^c_\rho(p2)}\frac{\delta}{\delta c^d(p1)}$ of \labelcref{eq:YM:mSTI}. The projection operators to obtain mSTI equations for the longitudinal four-gluon dressings $\lambda_{A^4,1}(p)$ and $\lambda_{A^4,2}(p)$ are given as,
\begin{align}
&\mathcal{Q}^{abcd}_{A^4,1,\mu\nu\rho}(p,q,r) = \Pi^\perp_{\mu \bar{\mu}}(p)\Pi^\perp_{\nu \bar{\nu}}(q)\Pi^\perp_{\rho \bar{\rho}}(r)\nonumber\\[4pt]
&(-p-q-r)_{\bar{\sigma}}\tau^{abcd}_{A^4,cl,\bar{\mu}\bar{\nu}\bar{\rho}\bar{\sigma}}(p,q,r)\,,\nonumber\\[10pt]
&\mathcal{Q}^{abcd}_{A^4,2,\mu\nu\rho}(p,q,r) = \Pi^\perp_{\mu \bar{\mu}}(p)\Pi^\perp_{\nu \bar{\nu}}(q)\Pi^\parallel_{\rho \bar{\rho}}(r)\nonumber\\[4pt] &(-p-q-r)_{\bar{\sigma}}\tau^{abcd}_{A^4,cl,\bar{\mu}\bar{\nu}\bar{\rho}\bar{\sigma}}(p,q,r)\,.
\end{align}

%%%%%%%%%%%%%%%%%%%%%%%%%%%%%%%%%%%
\section{Diagrammatic mSTIs and fRG Equations}
\label{app:YM:Diags}%%%%%%%%%%%%%
\begin{figure}[H]
	\centering
	\includegraphics[width=0.75\textwidth]{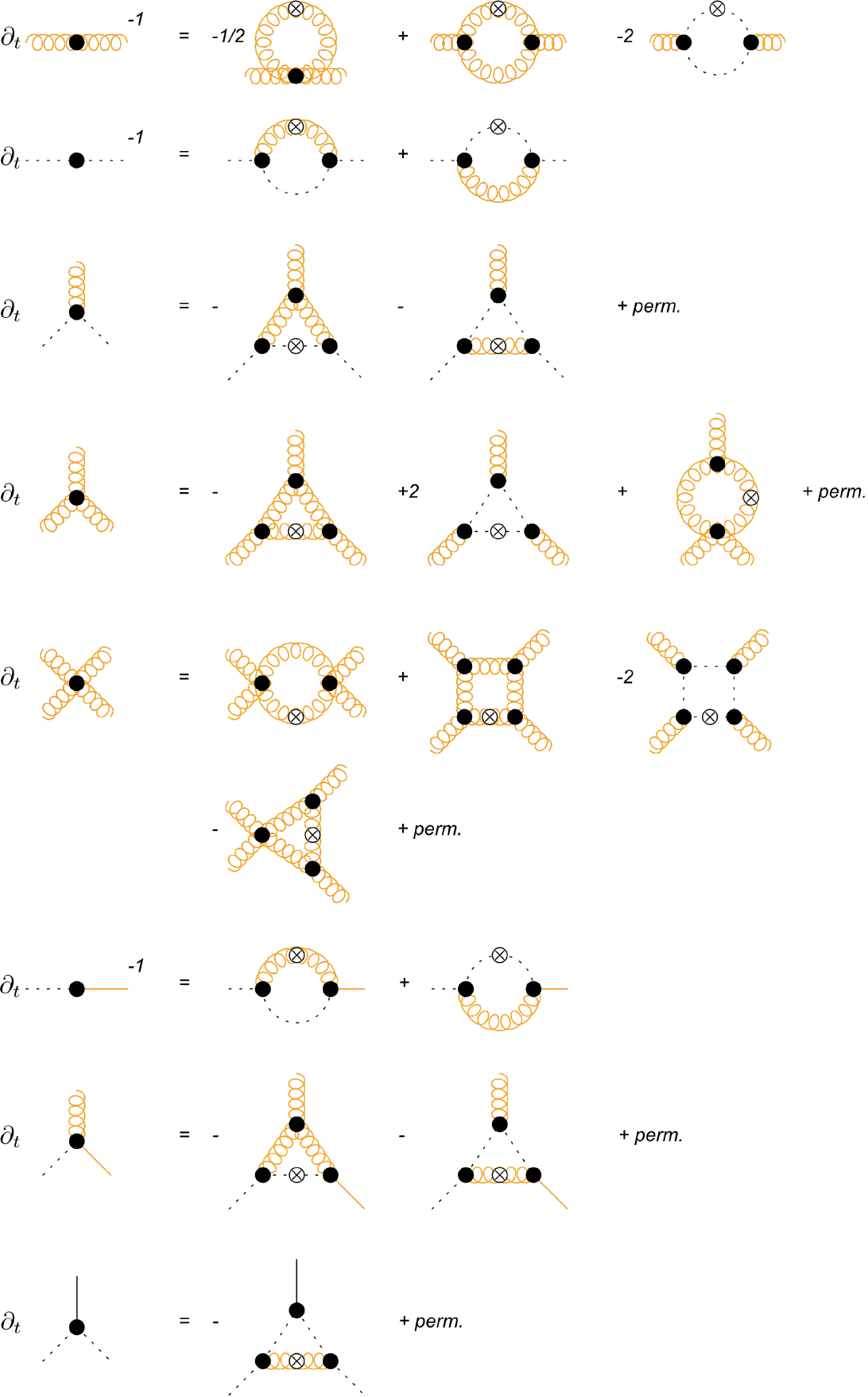}
	\caption{Diagrammatic representation of the flow equations of propagators and vertices. The dashed lines represent the fully dressed ghost, curly orange lines the fully dressed gluon, solid black lines to fully dressed $Q_c$ and solid orange lines to the fully dressed $Q_A$ BRST charge. The permutations include permutations of external legs as well as the position of the regulator derivative insertions which is indicated by the crossed circle. The power $-1$ indicates a full two-point function.\hspace*{\fill}}
	\label{fig:YM:flowequations}
\end{figure}

\begin{figure}[H]
	\centering
	\includegraphics[width=0.9\textwidth]{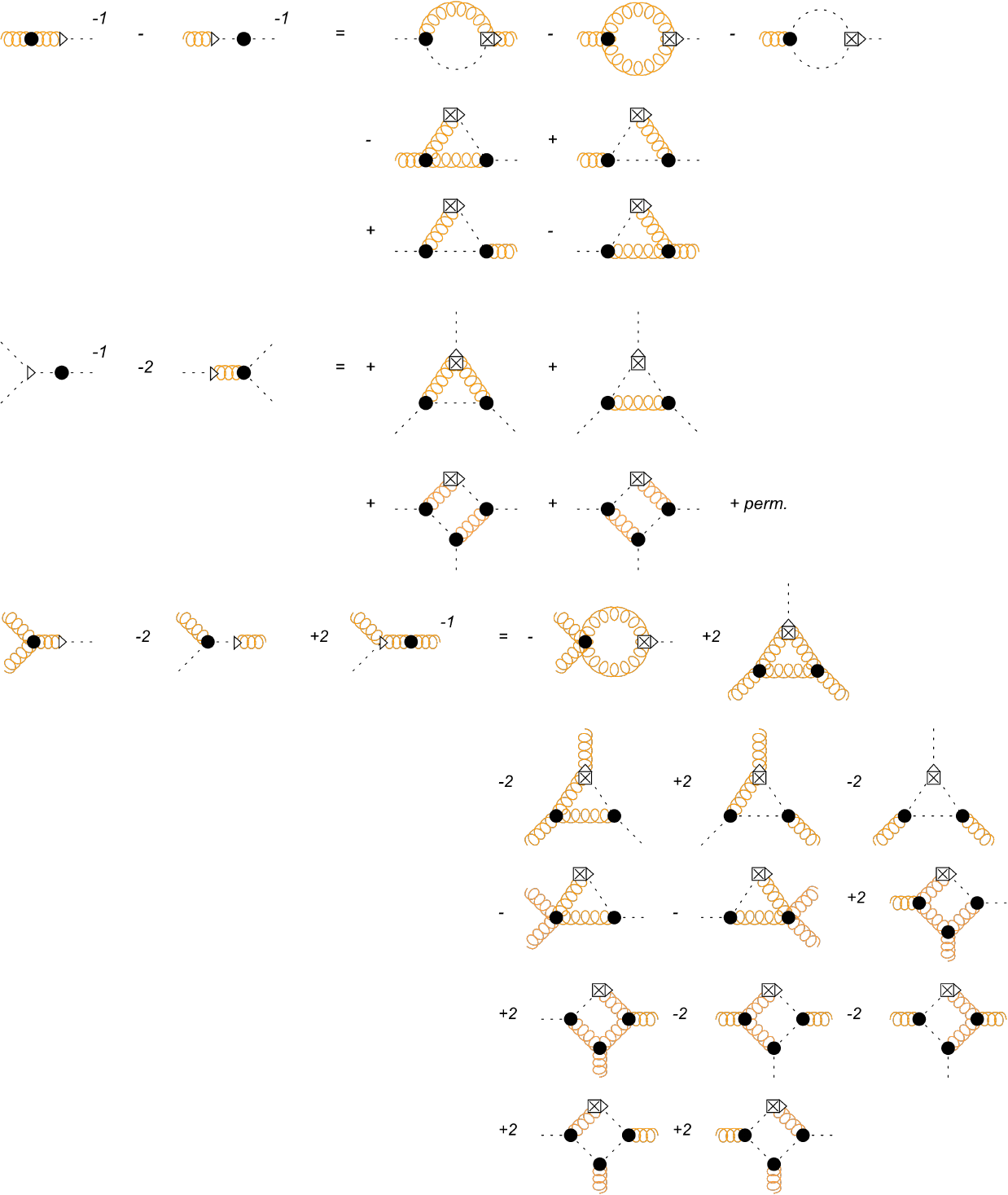}
	\caption{Diagrammatic representation of the modified Slavnov-Taylor identities for the longitudinal vertices. 
		 The dashed lines represent the fully dressed ghost, curly orange lines the fully dressed gluon. The black circles represent full vertices and the crossed square corresponds to a regulator and the triangle to a BRST vertex. The power $-1$ indicates a full two-point function. The permutations include permutations of external ghost legs.\hspace*{\fill}}
	\label{fig:YM:mSTIequations}
\end{figure}

%%%%%%%%%%%%%%%%%%%%%%%%%%%%%%%%%%%
The diagrammatic fRG equations and mSTIs within the truncation that was used throughout this work are depicted in \Cref{fig:YM:flowequations} and \Cref{fig:YM:mSTIequations}. \vfill
\onecolumngrid

\newpage

\newpage
\twocolumngrid

\bibliography{bib_master}

\end{document}